\pdfoutput=1
\documentclass[cernpreprint, atlasdraft=false, texlive=2023, UKenglish, texmf, orcidlogo]{atlasdoc}
 
\usepackage{multirow}
\usepackage{bm}
\usepackage{siunitx}
\usepackage{atlaspackage}
\usepackage{atlasbiblatex}
 
\usepackage[BSM=true,jetetmiss=true,process=true]{atlasphysics}

\graphicspath{{logos/}{figures/}}
 
\usepackage{ANA-EXOT-2020-25-PAPER-defs}

\AtlasTitle{Search for heavy neutral Higgs bosons decaying into a top quark pair in 140~fb$^{-1}$ of proton--proton collision data at $\sqrt{s}=13$~TeV with the ATLAS detector}

\AtlasAbstract{
A search for heavy pseudo-scalar ($A$) and scalar ($H$) Higgs bosons decaying into a top-quark pair ($t\bar{t}$) has been performed with 140~fb$^{-1}$ of
proton--proton collision data collected by the ATLAS experiment at the Large Hadron Collider at a centre-of-mass energy of $\sqrt{s}=13$~TeV.
Interference effects between the signal process and Standard Model (SM) $t\bar{t}$ production are taken into account.
Final states with exactly one or exactly two electrons or muons are considered.
No significant deviation from the SM prediction is observed.
The results of the search are interpreted in the context of a two-Higgs-doublet model (2HDM) of type II in the alignment limit
with mass-degenerate pseudo-scalar and scalar Higgs bosons ($m_A = m_H$)
and the hMSSM parameterisation of the minimal supersymmetric extension of the Standard Model.
Ratios of the two vacuum expectation values, $\tan\beta$, smaller than 3.49 (3.16) are excluded at 95\% confidence level for $m_A = m_H = 400$\,GeV in the 2HDM (hMSSM).
Masses up to 1240~GeV are excluded for the lowest tested $\tan\beta$ value of 0.4 in the 2HDM.
In the hMSSM, masses up to 950~GeV are excluded for $\tan\beta=1.0$.
In addition, generic exclusion limits are derived separately for single scalar and pseudo-scalar states for different choices of their mass and total width.
}
 
\AtlasRefCode{EXOT-2020-25}
 
\PreprintIdNumber{CERN-EP-2024-090}

\arXivId{2404.18986}

\AtlasJournal{JHEP}
\AtlasJournalRef{JHEP 08 (2024) 013}
\AtlasDOI{https://doi.org/10.1007/JHEP08(2024)013}

\hypersetup{pdftitle={ATLAS document},pdfauthor={The ATLAS Collaboration}}
 
\begin{document}
 
\maketitle

\section{Introduction}

New massive scalar or pseudo-scalar states with strong couplings to the top quark are predicted in numerous extensions of the Standard Model (SM) of particle physics.
These include models with an extended Higgs sector, such as
two-Higgs-doublet models (2HDMs)~\cite{PhysRevD.8.1226,Branco_2012,Inoue:1982pi,Gunion:2002zf}, which postulate the existence of a second complex Higgs doublet.
These 2HDMs are motivated, for example, by supersymmetry~\cite{Golfand:1971iw,Volkov:1973ix,Wess:1974tw,Wess:1974jb,Ferrara:1974pu,Salam:1974ig} and axion models~\cite{Kim:1986ax}.
Further examples of models with an extended Higgs sector include models predicting a new electroweak singlet~\cite{Huitu:2018gbc},
a combination of singlet and doublet fields~\cite{Muhlleitner:2017dkd,Moortgat-Pick:2023lhi}, or three-Higgs-doublet models (3HDMs)~\cite{Dey:2023exa}.
Additional scalar or pseudo-scalar states could also provide a portal to dark matter, acting as a mediator between the SM and the dark sector~\cite{Buckley:2014fba,Bauer:2017ota,Afik:2018fjk}.
 
In this paper, a search for massive scalar ($H$) and pseudo-scalar ($A$) states decaying into a top–antitop quark pair (\ttbar) is presented.
The significant interference between the signal process and the dominant, irreducible background from SM \ttbar production is taken into account.
The results of this search are interpreted in a range of representative benchmark models, in particular 2HDM-based models.
 
\begin{figure}[t]
\centering
\subfloat[]{
\includegraphics[width=0.3\textwidth]{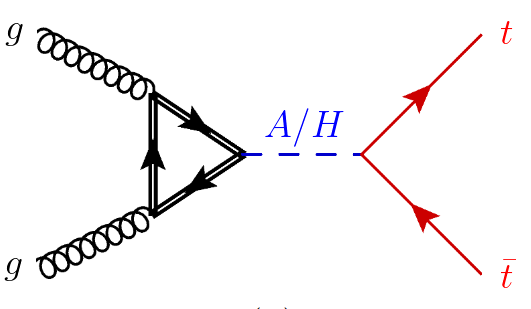}
}
\subfloat[]{
\includegraphics[width=0.25\textwidth]{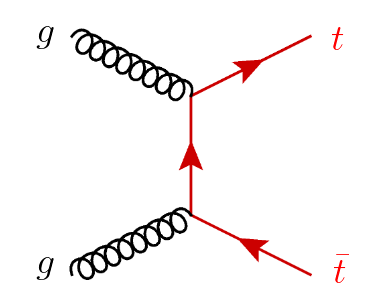}
}\\
\caption{Diagram for (a) the resonant production of a pseudo-scalar or scalar from gluon-gluon ($gg$) initial states via a fermion loop with subsequent decay into \ttbar at leading order,
$gg\rightarrow A/H \rightarrow \ttbar$,
and (b) the interfering diagram for $gg$ induced \ttbar production via the strong force at tree level.}
\label{fig:feyn}
\end{figure}

A 2HDM model is assumed with a CP-conserving potential with a softly broken
$Z_2$ symmetry~\cite{Gunion:2002zf}. After electroweak symmetry breaking the model contains five Higgs bosons:
a lighter CP-even boson, $h$, a heavier CP-even boson, $H$, a CP-odd boson, $A$, and two charged bosons, $H^{\pm}$.
It is assumed that the 125~GeV Higgs boson discovered by ATLAS and CMS~\cite{HIGG-2012-27,CMS-HIG-12-028} corresponds to the lighter CP-even state, $h$.
The 2HDM coupling structure is chosen to be of type-II~\cite{Gunion:2002zf}.
In this case, the parameter governing the fermionic decays of the neutral
scalar and pseudo-scalar states is the ratio of the vacuum expectation
values of the two Higgs doublets, $\tan \beta$.
According to precision measurements of the Higgs boson by ATLAS and CMS, the parameters of the 2HDM must be close to the alignment limit ($\cos(\beta - \alpha) = 0$),
where $\alpha$ denotes the mixing angle between the two doublets.
A prominent realisation of such a 2HDM is the minimal supersymmetric extension of the SM (MSSM)~\cite{Fayet:1976et,Fayet:1977yc}.
Type-II 2HDMs are also predicted in models of dark matter (DM), such as the
2HDM+$a$~\cite{Bauer:2017ota}, in which an additional pseudo-scalar $a$,
mediating the interaction between the SM- and the
DM-sector, is introduced, in addition to the five Higgs bosons of the 2HDM.
For large $\tan \beta$, stringent limits on the $A/H$ masses exist, the most stringent ones obtained from searches targeting the decays $A/H \rightarrow\tau\tau$~\cite{HDBS-2018-46,CMS-HIG-17-020}.
For small values of $\tan \beta$ the decay $A/H \rightarrow \ttbar$ dominates.
The present limits are much weaker than for large $\tan\beta$,
with the most stringent constraints derived from searches targeting heavy scalars decaying into a pair of Higgs bosons,
$H \rightarrow hh$~\cite{HDBS-2018-58,CMS-HIG-20-014}, from searches for decays of charged Higgs bosons to a top and a bottom quark, $H^\pm \rightarrow tb$~\cite{HDBS-2018-51,CMS-HIG-18-015},
and from searches targeting the \ttbar associated production of $A/H$ with subsequent decay into \ttbar, leading to a 4-top ($\ttbar\ttbar$) final state~\cite{ATLAS-CONF-2024-002,EXOT-2019-26}.
The dominant production mode for a heavy $H$ or $A$ is, like for the SM Higgs, via gluon-gluon fusion, which proceeds dominantly via a top-quark loop.
Due to the top quarks in the loop, the matrix element acquires a complex
phase which leads to the fact that the matrix element at the mass pole is no longer purely imaginary.
This, in turn, leads to a complicated interference pattern with the irreducible background $gg \rightarrow \ttbar$.
Because of this interference pattern the peak at the $A$, $H$ mass can
disappear completely and be replaced by a broad peak at lower masses and a dip around the resonance mass.
In Figure~\ref{fig:feyn}, the leading-order Feynman graphs for the signal
process and for the interfering part of the SM \ttbar background are shown.
 
A search for $A/H \rightarrow \ttbar$ taking the interference into account was first presented by ATLAS on 20.3~\ifb of 8~TeV data~\cite{EXOT-2016-04}
and later by CMS on 36~\ifb of 13~TeV data~\cite{CMS-HIG-17-027}.
Neither of these two searches observed a significant deviation from the SM expectation.
In this paper, a search for $gg \rightarrow A/H \rightarrow \ttbar$ on the
full dataset collected with the ATLAS~\cite{PERF-2007-01} detector during
LHC~\cite{Evans:2008zzb} Run 2,
amounting to 140~\ifb of $pp$ collision data at $\sqrt{s}=13$~TeV, is presented.
The search targets events in which one top quark decays leptonically and the other hadronically (1-lepton channel),
as well as events in which both top quarks decay leptonically (2-lepton channel).
Here, leptonic top-quark decays are defined as all top-quark decays resulting in an electron ($e$) or muon ($\mu$) in the final state,
either directly from the $W$-boson decay or via a $W$ boson that decays into a leptonically decaying $\tau$-lepton.
In the 1-lepton channel, separate analysis strategies targeting resolved and merged hadronic top-quark decays are used.
In the latter case, the merged top-quark decay is reconstructed using large
jets with a variable radius parameter $R$ that are re-clustered from calibrated small-$R$ jets.
The dominant background in both channels arises from SM \ttbar production, which is estimated using simulated events that are corrected to high-precision predictions calculated
at next-to-next-to-leading order (NNLO) in QCD and NLO in the electroweak (EW) interaction~\cite{Czakon:2017wor}.
 
This paper is structured as follows. In Section \ref{sec:models}, the
benchmark models used for the interpretation of the results of this search are introduced.
In Sections \ref{sec:detector} and \ref{sec:datamc}, the ATLAS detector as well as the data and simulated event samples are described.
The event selection, categorisation, reconstruction of observables, the estimation of background processes and systematic uncertainties are described in Sections \ref{sec:selection} to \ref{sec:unc}.
The statistical model for the interference analysis is discussed in Section~\ref{sec:stats}. Finally, the results are presented and discussed in Section~\ref{sec:results}.


\section{Theoretical framework}
\label{sec:models}

In this paper, a CP-conserving type-II 2HDM with a softly broken $Z_2$ symmetry~\cite{Gunion:2002zf} is considered as a benchmark model.
The alignment and decoupling limit are assumed, i.e. $m_h = 125$~\GeV, $v=246$~\GeV, and $\cos(\beta-\alpha)=0$, where $v$ denotes the electroweak vacuum expectation value.
The parameter $m_{12}$ of the $Z_2$ breaking term of the potential is taken to be $m_{12}^2 = m_A^2 \tanb/(1 + \tanb^2)$~\cite{Kanemura:2004mg,EXOT-2016-04}.
In this model, the production cross-sections and widths of $A$ and $H$, as well as their signal shapes, are uniquely determined by \tanb and the masses $m_A$ and $m_H$.
In the alignment limit, the scalar $h$ is SM-like and the scalar $H$ does not couple to gauge bosons.
The coupling relative to the SM Higgs coupling for the pseudo-scalar $A$ is multiplied by $\tan \beta$ for down type quarks and by $1/\tan \beta$ for up-type quarks.
For the scalar $H$, the coupling to up-type quarks receives in addition a  minus sign \cite{ATL-PHYS-PUB-2013-016}.
In this paper, the scalar and pseudo-scalar states are assumed to be mass degenerate, i.e. $m_H = m_A$.
 
As a second benchmark model, the Higgs sector of the hMSSM~\cite{Djouadi:2013uqa} is considered, which constitutes a special instance of a type-II 2HDM.
In this specific parameterisation of the MSSM,
the lighter scalar $h$ is identified with the Higgs boson ($m_h = 125$~GeV) discovered in 2012.
This choice fixes the dominant radiative corrections that enter the MSSM~\cite{Djouadi:2015jea}.
As a result, the hMSSM can be fully described by only two free parameters, $m_A$ and \tanb.
In particular, in this case, the mass of the heavier scalar $H$ depends on $m_A$.
For $m_A \approx 2 m_t$, the $H$ is around 40\,GeV heavier than the $A$, while the mass difference reduces to $\Delta m \sim 10\,$GeV at $m_A \approx 1\,$TeV.
The couplings of the $A$ and the $H$ are to a good approximation the same as in the alignment limit.
 
A 2HDM-based model of dark matter, referred to as 2HDM+$a$, is also considered in this publication.
In addition to the extra Higgs bosons introduced by the 2HDM, the 2HDM+$a$ includes a fermionic DM particle $\chi$ and a pseudo-scalar mediator $a$ with Yukawa-like couplings
to both the SM fermions and the Dirac DM particle $\chi$.
The mediator mixes with the pseudo-scalar $A$ of the 2HDM sector with a mixing angle $\theta$.
The 2HDM+$a$ is recommended by the LHC Dark Matter Working Group as it is a simple, ultra-violet-complete (UV-complete), gauge-invariant,
and renormalisable benchmark model of DM~\cite{2HDMWGproxi}. A range of different benchmark scenarios are recommended to and explored by the LHC experiments~\cite{2HDMWGproxi}.
In all the benchmark scenarios, the scalar and pseudo-scalar states are assumed to be mass-degenerate: $m_H = m_A$.
The most stringent constraints to date on the 2HDM+$a$ are summarised in Ref.~\cite{EXOT-2018-64}.
 
Finally, a simplified generic scenario is considered in which only the interference pattern of either a scalar or a pseudo-scalar are assumed to appear in the spectra of sensitive variables.
The full spectrum in the presence of signal-background interference can be described in terms of the contributions from the pure (resonant) signal $S$, the interference term $I$,
and the total background $B$, which is dominated by SM \ttbar production, $B_{\ttbar}$. The interference pattern $S+I$, which describes the deviation from the background-only hypothesis
in the presence of a signal process with signal-background interference, is obtained by subtracting the SM \ttbar background from the inclusive process $S+I+B_{\ttbar}$,
see Section~\ref{sec:datamc} for details.
In the simplified generic scenario, only the coupling to \ttbar is considered, both in the production loop and the decays.
The Yukawa coupling to \ttbar for a (pseudo-)scalar of a given mass and width is scaled by the coupling modifier $g_{A/H\ttbar}$,
which is a free parameter in this scenario, along with the mass $m_{A/H}$ and the width $\Gamma_{A/H}$.
This benchmark scenario differs from the other, 2HDM-based, benchmark models considered in this publication, in which $\Gamma_{A/H}$ is a function of $m_{A/H}$ and $\tan\beta$.
In a type-II 2HDM, the coupling modifier is inversely proportional to $\tan\beta$: $g_{A\ttbar} = 1/\tanb$ and $g_{H\ttbar} = -1/\tanb$.
In the generic scenario, the interference pattern $(S+I)_{g_{A/H\ttbar}^2}$ for a given value of the coupling modifier $g_{A/H\ttbar}$
is derived from the pure-signal ($S \equiv S(m_{A/H},\Gamma_{A/H})$)
and the signal-plus-interference ($S+I \equiv (S+I)(m_{A/H},\Gamma_{A/H})$) templates,
both obtained for a given choice of $m_{A/H}$ and $\Gamma_{A/H}$, as follows:
\begin{eqnarray}\nonumber
(S+I)_{g_{A/H\ttbar}^2} &=& g_{A/H\ttbar}^4 \,\cdot\, S \, + \, g_{A/H\ttbar}^2 \,\cdot\, I \\
&=& (g_{A/H\ttbar}^4 - g_{A/H\ttbar}^2) \,\cdot\, S \, + \, g_{A/H\ttbar}^2 \,\cdot\, (S+I).
\label{eq:generic}
\end{eqnarray}
This scenario allows for the results of the search to be re-interpreted in the context of models beyond the 2HDM-based ones discussed in this paper
by providing constraints on a wider set of interference patterns obtained from the well-defined $S$ and $S+I$ templates for a given mass and width
via variations of the coupling modifier according to Eq.~(\ref{eq:generic}).
This includes models which, unlike 2HDM-based models, predict the existence of a single new pseudo-scalar or single scalar,
such as models predicting a single heavy axion-like particle coupling to the top quark~\cite{Esser:2023fdo}.
It also includes models in which the width is not only a function of $m_{A/H}$ and $g_{A/H\ttbar}$ but additionally depends on other model assumptions,
such as contributions from new, heavier particles to the production loop.
 
In Figure~\ref{fig:SI_truth}, two sets of interference patterns for the production and decay into \ttbar of a single pseudo-scalar with mass $m_A = 500$~GeV
and a single scalar with mass $m_H = 800$~GeV are shown for different values of \tanb in a type-II 2HDM in the alignment limit.
The interference patterns are shown as a function of the \ttbar invariant mass, \mttbar, calculated at parton level before the emission of final-state radiation (FSR).
The SM \ttbar background contribution has been subtracted in both cases to yield the signal-plus-interference ($S+I$) distribution only.
For all scenarios, a deficit of events compared to the SM expectation is predicted in the \mttbar region around the signal mass, while an excess of events is predicted for lower values of \mttbar.
Higher values of \tanb correspond to smaller relative widths
$\Gamma_{A/H}/m_{A/H}$ and hence to a narrower peak-dip structure.
In addition, the interference patterns obtained for the generic scenario for a single pseudo-scalar and scalar, respectively, with width $\Gamma_{A/H}/m_{A/H}=10\%$
are shown in Figure~\ref{fig:SI_truth_mod_ind} and for various values of the coupling modifier $g_{A/H\ttbar}$.
For small values of the coupling modifier, the interference pattern exhibits a peak-dip structure.
For coupling values $g_{A/H\ttbar} > 1$, the resonant contribution $S$, which scales like $g_{A/H\ttbar}^4$, dominates over the interference contribution $I$,
which scales like $g_{A/H\ttbar}^2$. Hence for large values of the coupling modifier, the interference pattern can exhibit a peak-peak structure.
 
\begin{figure}[t]
\centering
\subfloat[]{
\includegraphics[width=0.5\textwidth]{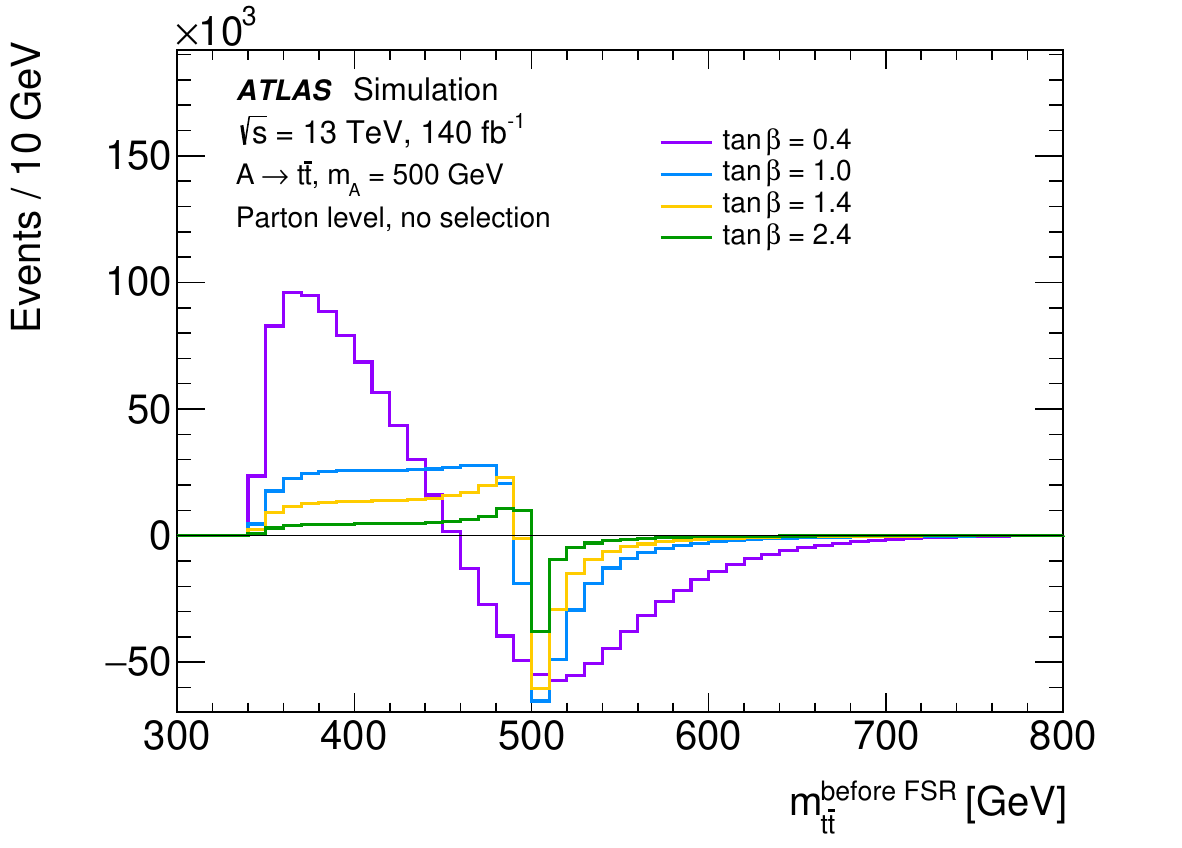}
}
\subfloat[]{
\includegraphics[width=0.5\textwidth]{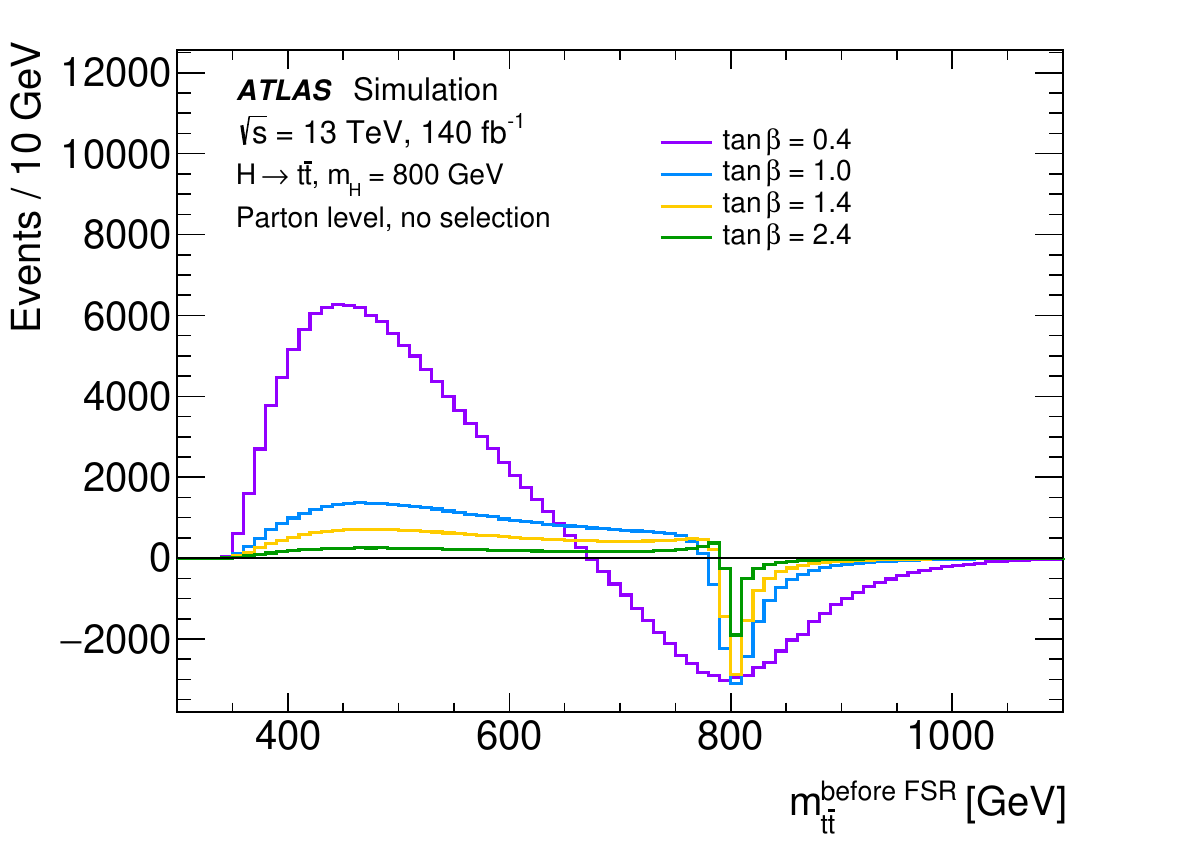}
}\\
\caption{Signal-plus-interference distributions in $m_{t\bar{t}}$ at parton level before FSR for (a) a single pseudo-scalar $A$ with mass $m_A = 500$~GeV
and (b) a single scalar $H$ with mass $m_H = 800$~GeV for various values of $\tan\beta$ in a type-II 2HDM in the alignment limit.
\label{fig:SI_truth}}
\end{figure}
 
\begin{figure}[t]
\centering
\subfloat[]{
\includegraphics[width=0.5\textwidth]{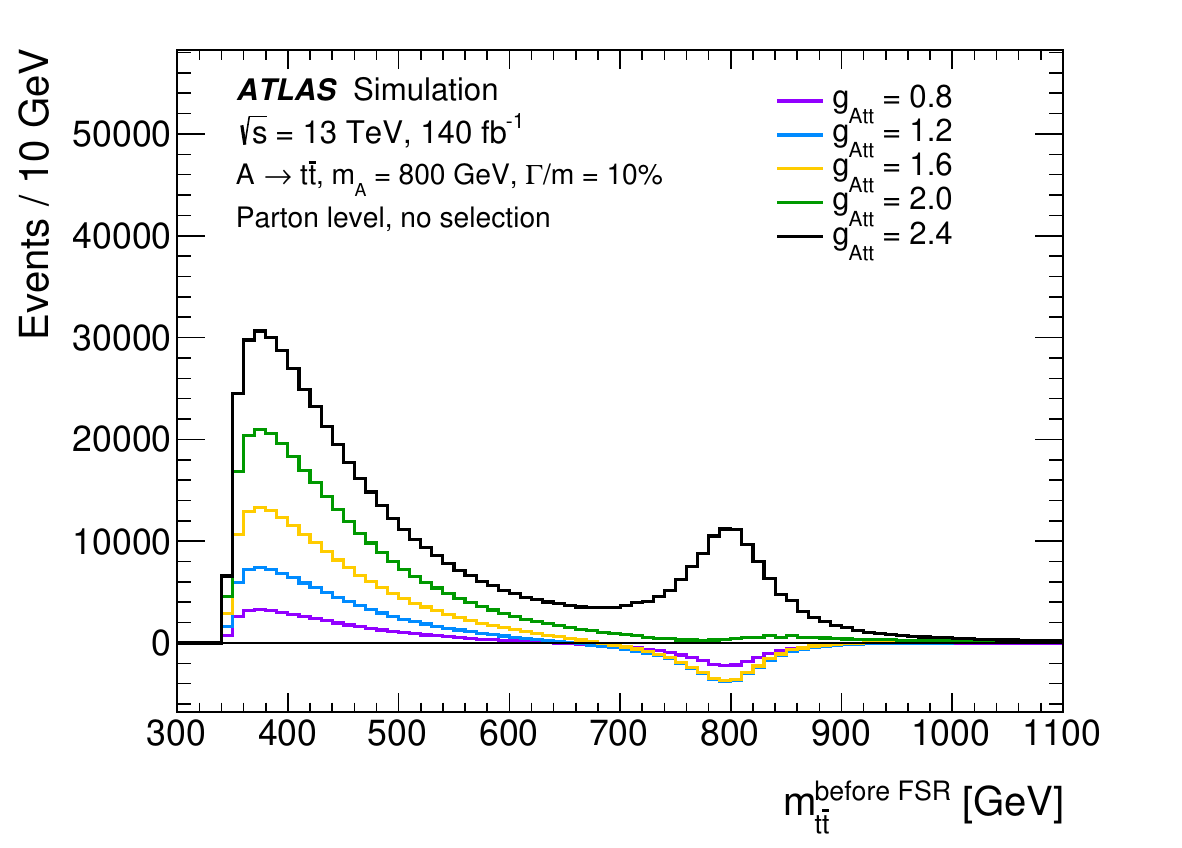}
}
\subfloat[]{
\includegraphics[width=0.5\textwidth]{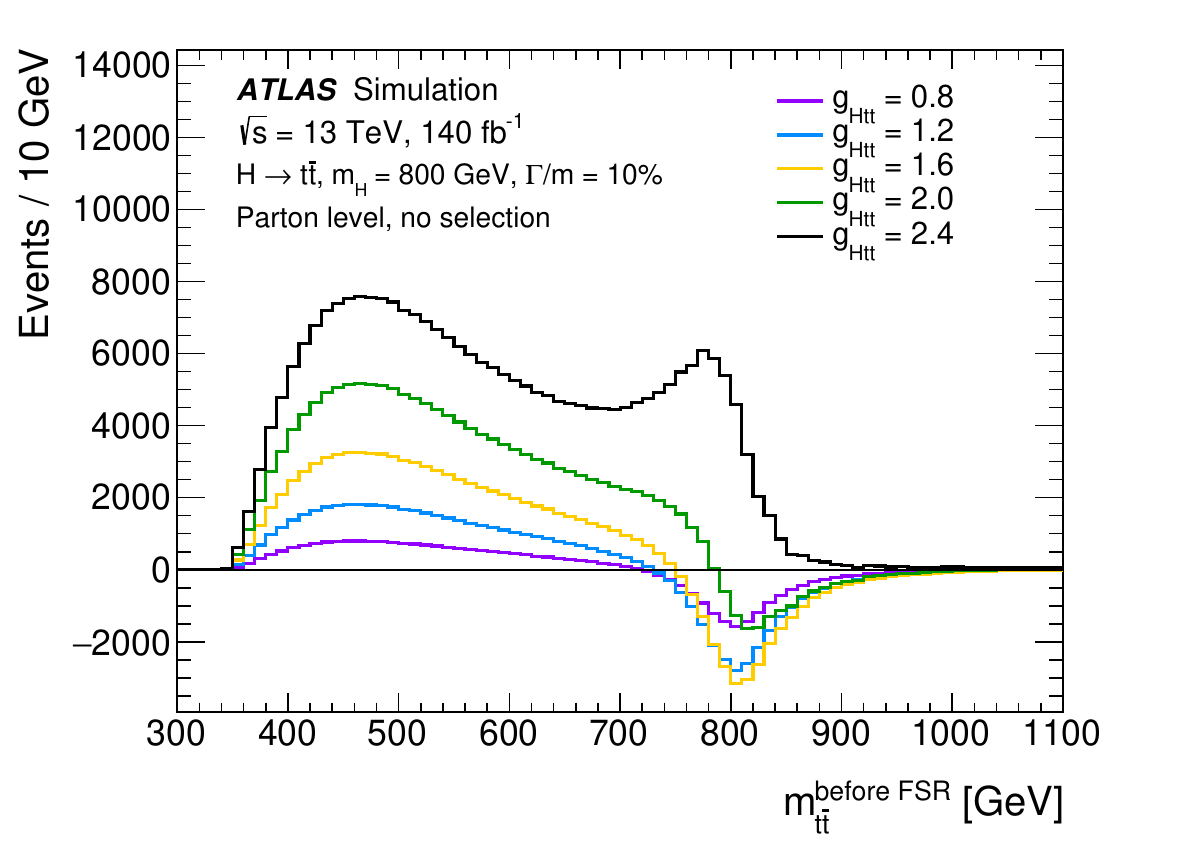}
}\\
\caption{Signal-plus-interference distributions in $m_{t\bar{t}}$ at parton level before FSR for (a) a single pseudo-scalar $A$ with mass $m_A = 800$~GeV
and for (b) a single scalar $H$ with mass $m_H = 800$~GeV, both with relative width $\Gamma_{A/H}/m_{A/H}=10\%$. The distributions are shown for different values of the
coupling modifier $g_{A/H\ttbar}$ in the generic benchmark scenario
(see text for details).
\label{fig:SI_truth_mod_ind}}
\end{figure}


\section{ATLAS detector}
\label{sec:detector}

The ATLAS detector~\cite{PERF-2007-01} is a multipurpose particle detector with a forward--backward symmetric cylindrical
geometry and a near 4$\pi$ coverage in solid angle.\footnote{
ATLAS uses a right-handed coordinate system with its origin at the nominal interaction point in the centre of the detector.
The positive $x$-axis is defined by the direction from the interaction point to the centre of the LHC ring,
with the positive $y$-axis pointing upwards, while the beam direction defines the $z$-axis.
Cylindrical coordinates $(r,\phi)$ are used in the transverse plane, $\phi$ being the azimuthal angle around the $z$-axis.
The pseudorapidity $\eta$ is defined in terms of the polar angle $\theta$ by $\eta=-\ln\tan(\theta/2)$,
while the rapidity $y$ is defined as $y = 0.5 \ln[(E + p_z)/(E-p_z)]$, where $E$ denotes the energy and $p_z$ the component of the momentum along the beam direction.
The angular distance $\Delta R$ is defined as $\sqrt{(\Delta y)^2 + (\Delta \phi)^2}$.
}
It consists of an inner tracking detector surrounded by a thin superconducting solenoid providing a 2 T axial magnetic field,
electromagnetic and hadron calorimeters, and a muon spectrometer.
The inner tracking detector (ID) consists of silicon pixel and silicon microstrip detectors covering the pseudorapidity range $|\eta| < 2.5$,
and a surrounding transition radiation tracking detector that enhances electron identification in the range $|\eta| < 2.0$.
A new inner pixel layer, the insertable B-layer~\cite{ATLAS-TDR-2010-19, PIX-2018-001}, was added at a mean radius of 3.3~cm during the period between Run~1 and Run~2 of the LHC.
Lead/liquid-argon (LAr) sampling calorimeters provide electromagnetic (EM) energy measurements with high granularity in the region $|\eta|<3.2$.
A steel/scintillator-tile hadron calorimeter covers the central pseudorapidity range ($|\eta| < 1.7$).
The endcap and forward regions ($1.5<|\eta|<4.9$) of the hadron calorimeter are made of LAr active layers with either copper or tungsten
as the absorber material.
The muon spectrometer (MS) surrounds the calorimeters and is based on three large superconducting air-core toroidal magnets with eight coils each.
The field integral of the toroids ranges between 2.0 and 6.0 Tm across most of the detector.
Three layers of high-precision tracking chambers provide coverage in the range $|\eta|<2.7$, while dedicated fast chambers allow triggering in the region $|\eta|<2.4$.
The ATLAS trigger system consists of a hardware-based level-1 trigger followed by a software-based high-level trigger~\cite{TRIG-2016-01}.
The level-1 trigger uses a subset of the detector information to accept events at a rate below 100 kHz,
while the software-based trigger reduces the accepted event rate to 1 kHz on average depending on the data-taking conditions.
An extensive software suite~\cite{ATL-SOFT-PUB-2021-001,} is used in the reconstruction and analysis of real and simulated data,
in detector operations, and in the trigger and data acquisition systems of the experiment.


\section{Data and simulated event samples}
\label{sec:datamc}

In this search, data from $pp$ collisions at $\sqrt{s}=13$~TeV corresponding to an integrated luminosity of 140~fb$^{-1}$,
collected in the years 2015 to 2018 with the ATLAS detector, are analysed.
The uncertainty in the combined 2015--2018 integrated luminosity is 0.83\%~\cite{DAPR-2021-01},
obtained using the LUCID-2 detector~\cite{LUCID2} for the primary luminosity measurements.
All detector subsystems were required to be operational during data taking.
The average number of interactions per bunch crossing (pile-up) in this dataset is \averagemu$=33.7$.
 
Candidate events are selected using single-muon~\cite{TRIG-2018-01} and single-electron~\cite{TRIG-2018-05} triggers.
These triggers require a muon (electron) with transverse momentum \pt (transverse energy \et) above a certain threshold
and passing certain quality and, for some triggers, lepton isolation requirements.
The unprescaled triggers with the lowest \pT (\et) threshold require $\pT > 26$~\GeV\ for muons (electrons) in 2016--2018
and include a lepton isolation requirement that is not applied for triggers with higher thresholds.
The trigger efficiency is mostly constant in the transverse momentum for leptons with $\pT > 28$~GeV.
 
Monte Carlo (MC) simulated event samples are used to model the signal and background processes, including the interference effects between the signal and SM \ttbar production.
The ATLAS simulation infrastructure~\cite{SOFT-2010-01} was used to simulate the detector and its response.
Signal and nominal background samples as well as several alternative background samples used to assess systematic uncertainties
were produced with a detailed $\GEANT$~\cite{Agostinelli:2002hh} detector simulation.
A faster simulation based on a parameterisation of the calorimeter response and $\GEANT$ for the other detector systems~\cite{SOFT-2010-01} was used for the remaining alternative background samples.
Pile-up effects were modelled by overlaying minimum-bias events simulated using the soft QCD processes of \PYTHIA{8.186}~\cite{Sjostrand:2007gs}
with the \NNPDF[2.3lo] set of parton distribution functions (PDFs)~\cite{Ball:2012cx} and the A3~\cite{ATL-PHYS-PUB-2016-017} set of tuned parameters (tune).
The pile-up profiles match the ones of each dataset between 2015 and 2018.
The same offline reconstruction methods used for data were applied to the simulated event samples.
Corrections were applied to the simulated events in order to match the selection efficiencies, energy and mass scales and resolutions of reconstructed simulated particles to those measured in data control samples.
An overview of the generator choices for the signal and background processes can be found in Table~\ref{tab:samples:generators}, with further details given in the following.
 
The dominant and irreducible background process is SM \ttbar production.
Smaller background components arise from single vector-boson ($W$,$Z$) production in association with hadronic jets, referred to as $W$/$Z$+jets,
single-top-quark, and multijet production.
In the 2-lepton channel, another small background component arises from processes with at least one fake or non-prompt lepton that
satisfies the lepton identification and isolation criteria applied in this search.
These are mostly \ttbar production with one top quark decaying hadronically and the other leptonically, as well as single-top, and $W$+jets production.
This background component is referred to as \textit{fakes} in the following.
Minor backgrounds from \ttbar+$V$, \ttbar+$H$ and diboson ($WW$, $ZZ$, $WZ$) production are also considered.
All background components, with the exception of multijet production in the 1-lepton channel (Section~\ref{subsec:multijet_1l}), are modelled using MC simulation.
Data-driven corrections are applied to the MC-based estimates of the $W$+jets background in the 1-lepton channel (Section~\ref{subsec:wjets_1l})
and the $Z$+jets background in the 2-lepton channel (Section~\ref{subsec:zjets_2l}) to correct the $W$+jets normalisation and $Z$+jets shape, respectively.
The modelling of the fakes background is also validated in data (Section~\ref{subsec:fakes_2l}).
 
SM \ttbar production was generated at next-to-leading order (NLO) accuracy in QCD
using \POWHEGBOX[v2]~\cite{Nason:2004rx,Frixione:2007nw,Frixione:2007vw,Alioli:2010xd,Campbell:2014kua} with the \NNPDF[3.0nlo]~\cite{Ball:2014uwa} PDF set
and the \hdamp parameter set to 1.5\,\mtop~\cite{ATL-PHYS-PUB-2016-020}.\footnote{
The \hdamp parameter controls the transverse momentum, \pT, of the first additional emission beyond the leading-order Feynman
diagram in the parton shower and therefore regulates the high-\pT emission against which the \ttbar system recoils.
}
The functional form of the renormalisation and factorisation scales was set to the default scale $\sqrt{m_{\textrm{top}}^2 + p_{\textrm T}^2}$.
The top-quark mass was set to \mtop=172.5~GeV. The top quarks were decayed in \POWHEGBOX, thus preserving their spin correlations.
The matrix-element (ME) generator is interfaced with \PYTHIA[8.230]~\cite{Sjostrand:2014zea} to model the parton shower (PS), hadronisation, and the underlying event
with parameters set according to the A14 tune~\cite{ATL-PHYS-PUB-2014-021} and using the \NNPDF[2.3lo] PDF set.
The decays of bottom and charm hadrons were modelled with \EVTGEN[1.6.0]~\cite{Lange:2001uf}.
 
Alternative SM \ttbar\ samples obtained with different generator choices and settings are used to estimate systematic uncertainties related to the modelling of this main background component.
The details can be found in Section~\ref{subsec:unc_mod}.
 
The top-quark kinematics in the nominal and all alternative SM \ttbar samples were corrected to more accurate differential predictions calculated
at NNLO-QCD+NLO-EW accuracy for a top-quark mass value of \mbox{\mtop=173.3~GeV~\cite{Czakon:2017wor}}.
The difference between this \mtop value and the value of 172.5 GeV used in the NLO MC generation is covered by the \mtop uncertainty, which is taken into account in the final fit
and is found to only have minor impact on the fit result.
The corrections were applied as event-by-event weights.
The weights were obtained via an iterative recursive reweighting procedure, referred to as \textit{NNLO reweighting} in the following,
in which the calculated and generated binned differential distributions of the top and anti-top quark \pT as well as that of the \ttbar invariant mass, \mttbar, were compared.
All variables were obtained at the parton level before final-state radiation.
In each step, weights were obtained as the ratio of the bin contents of the calculated parton-level distribution and
the corresponding distribution obtained from the given MC sample, already corrected by the previous steps of the reweighting procedure.
The weights were then applied to the generated sample.
In the case of the top and anti-top quark \pT distributions, in order to avoid introducing an artificial asymmetry between the top and anti-top kinematic distributions,
the weight applied to the sample was the geometric average of the two weights derived from comparing separately the top and anti-top quark \pT distributions.
Corrections were first derived based on \mttbar, then on the $\pt(t)$ and $\pt(\bar{t})$ distributions, and then again on the $\pt(t)$ and $\pt(\bar{t})$ distributions.
The procedure was iterated three times to achieve a good agreement between the reweighted and calculated target distributions in all three variables.
The systematic uncertainties related to this reweighting procedure are described in Section~\ref{subsec:unc_mod}.
 
All reweighted \ttbar samples were additionally normalised to the cross-section prediction at NNLO
in QCD including the resummation of next-to-next-to-leading logarithmic (NNLL) soft-gluon terms calculated using
\TOPpp[2.0]~\cite{Beneke:2011mq,Cacciari:2011hy,Baernreuther:2012ws,Czakon:2012zr,Czakon:2012pz,Czakon:2013goa,Czakon:2011xx}.
For $pp$ collisions at a centre-of-mass energy of \rts~=~\SI{13}{\TeV}, this cross-section corresponds to
$\sigma(\ttbar)_{\textrm{NNLO+NNLL}} = \mathrm{814\pm76~fb}$ using a top-quark mass of $\mtop = 173.3\,\GeV$.
The uncertainties in the cross-section due to the PDF and \alphas were calculated using the \PDFforLHC[15] prescription~\cite{Butterworth:2015oua}
with the \MSTW[nnlo]~\cite{Martin:2009iq,Martin:2009bu}, \CT[10nnlo]~\cite{Lai:2010vv,Gao:2013xoa}
and \NNPDF[2.3ffn] PDF sets in the five-flavour scheme, and were added in quadrature to the effect of the scale uncertainty.
 
Single-top production in the $Wt$-channel~\cite{Re:2010bp}, which constitutes the main single-top contribution in this analysis,
was generated with \POWHEGBOX[v2] and using the \NNPDF[3.0nlo] PDF set.
The overlap between \ttbar and $Wt$ production was treated within the diagram removal (DR) scheme~\cite{Frixione:2008yi}
and the renormalisation and factorisation scales were set to $H_T/2$,
where $H_T$ denotes the scalar sum of the transverse momenta of all final-state particles in the event.
The ME generator was interfaced with \PYTHIA[8.307]~\cite{Bierlich:2022pfr} with parameters set according to the A14 tune and using the \NNPDF[2.3lo] PDF set.
\EVTGEN[1.7.0] was used to decay bottom and charm hadrons.
Single-top production in the $s$-channel~\cite{Alioli:2009je}, like SM \ttbar production,
were generated with \POWHEGBOX[v2] at NLO in QCD using the five-flavour scheme and the \NNPDF[3.0nlo] PDF set.
Single-top production in the $t$-channel was generated with \POWHEGBOX[v1]~\cite{Frederix:2012dh}.
This generator uses the four-flavour scheme for the NLO matrix element calculations together with the four-flavour PDF set \NNPDF[3.04f].
For this process, the top-quark decays were simulated using \MADSPIN~\cite{Artoisenet:2012st}, preserving all spin correlations.
For both processes, the ME generator was interfaced with \PYTHIA[8.230] with parameters set according to the A14 tune and using the \NNPDF[2.3lo] PDF set.
\EVTGEN[1.6.0] was used to decay bottom and charm hadrons.
The respective samples were normalised to the theoretical cross-sections for $Wt$-channel~\cite{Kidonakis:2010ux}, calculated at NNLO+NNLL accuracy, and for
$s$-channel~\cite{Kidonakis:2010tc}, and $t$-channel~\cite{Kidonakis:2011wy} production, both calculated at NLO accuracy.
 
The background from $t\bar{t}V$ ($V=W,Z$) production was generated at NLO accuracy in QCD with \MGNLO[2.3.3]~\cite{Alwall_2014} with the \NNPDF[3.0nlo] PDF set,
interfaced with \PYTHIA[8.210] using the A14 set of tuned parameters and the \NNPDF[2.3lo] PDF set for parton showering, hadronisation, and underlying event.
\EVTGEN[1.2.0] was used for the decay of bottom and charm hadrons.
The samples were normalised to the theoretical cross-section calculated at NLO in QCD~\cite{LHCHiggsCrossSectionWorkingGroup:2016ypw}.
The background from \ttbar+$h$ production was generated at NLO accuracy in QCD with \POWHEGBOX[v1] with the \NNPDF[3.0nlo] PDF set,
interfaced with \PYTHIA[8.230] using the A14 set of tuned parameters and the \NNPDF[2.3lo] PDF set for parton showering,
hadronisation, and underlying event. \EVTGEN[1.6.0] was used to decay bottom and charm hadrons.
The samples were normalised to the theoretical cross-section calculated at NLO QCD and NLO EW accuracies~\cite{LHCHiggsCrossSectionWorkingGroup:2016ypw}.
 
The production of a single $W$ boson in association with hadronic jets ($W$+jets)
was simulated with \SHERPA[2.2.11]~\cite{Bothmann:2019yzt},
while $Z$+jets production was simulated with \SHERPA[2.2.11] (\SHERPA[2.2.1]) in the 1-lepton (2-lepton) channel.
NLO MEs were used for up to two partons for all samples,
and LO MEs for up to five (four) partons for the \SHERPA[2.2.11] (\SHERPA[2.2.1]) samples.
The MEs were calculated with the Comix~\cite{Gleisberg:2008fv} and \OPENLOOPS~\cite{Buccioni:2019sur,Cascioli:2011va,Denner:2016kdg} libraries.
They were matched with the \SHERPA parton shower~\cite{Schumann:2007mg} using the \MEPSatNLO prescription~\cite{Hoeche:2011fd,Hoeche:2012yf,Catani:2001cc,Hoeche:2009rj}
using the set of tuned parameters developed by the \SHERPA authors.
The \NNPDF[3.0nnlo] set of PDFs was used for all $V$+jets samples.
The $W$+jets and $Z$+jets samples were normalised to the theoretical cross-sections calculated at NLO accuracy in QCD~\cite{Catani:2009sm}.
An additional data-driven normalisation correction was derived for the $W$+jets background in the 1-lepton channel.
 
Diboson events with fully leptonic and semileptonic decays were simulated using \SHERPA[2.2.1] and \SHERPA[2.2.2]~\cite{ATL-PHYS-PUB-2017-005}, respectively.
The simulation includes off-shell effects and Higgs boson contributions, where appropriate.
Events were generated using matrix elements at NLO accuracy in QCD for up to one additional parton and at LO accuracy for up to three additional parton emissions.
Samples for the loop-induced processes $gg \to VV$ were generated using LO-accurate matrix elements for up to one additional parton emission.
The matrix element calculations were matched and merged with the \SHERPA parton shower based on the Catani--Seymour dipole factorisation~\cite{Gleisberg:2008fv,Schumann:2007mg}
using the MEPS@NLO prescription.
The virtual QCD corrections were provided by the \OPENLOOPS library.
The \NNPDF[3.0nnlo] set of PDFs was used, along with the dedicated set of tuned parton-shower parameters developed by the \SHERPA authors.
The cross-sections from the generator were used for sample normalisation.
 
\begin{table}[t]
\centering
\caption{List of ME generators and the order of the strong coupling constant in the perturbative calculation, PDF sets,
shower generator and tune for the different signal and background processes.}
\label{tab:samples:generators}
\resizebox{\linewidth}{!}{\begin{tabular}{lccccc}
\toprule
Process 			& ME generator 		& ME order 		& PDF set	& PS and hadronisation 	& UE tune   \\
\midrule
Signal				& \MGNLO[2.6.7]     	& LO 			& \NNPDF[3.0nlo]     & \PYTHIA[8.244]    	& A14       \\
\ttbar			        & \POWHEGBOX[v2] 	& NLO, rew. to NNLO + NLO EW 			& \NNPDF[3.0nlo] 	& \PYTHIA[8.230]	& A14 	    \\
Single top ($Wt$)	        & \POWHEGBOX[v2] 	& NLO 			& \NNPDF[3.0nlo] 	& \PYTHIA[8.235]	& A14 	    \\
Single top ($s$)	        & \POWHEGBOX[v2] 	& NLO 			& \NNPDF[3.0nlo] 	& \PYTHIA[8.230]	& A14 	    \\
Single top ($t$)	        & \POWHEGBOX[v2] 	& NLO 			& \NNPDF[3.0nlo4f] 	& \PYTHIA[8.230]	& A14 	    \\
Diboson 			& \SHERPA[2.2.1/2.2.2]   & \MEPSatNLO	        & \NNPDF[3.0nnlo]    & \SHERPA		& internal  \\
$W$+jets			& \SHERPA[2.2.11]     	& \MEPSatNLO	        & \NNPDF[3.0nnlo]	& \SHERPA 		& internal  \\
$Z$+jets			& \SHERPA[2.2.1/2.2.11] & \MEPSatNLO	        & \NNPDF[3.0nnlo]	& \SHERPA 		& internal  \\
\ttbar+$V$			& \MGNLO[2.3.3] 	& NLO 			& \NNPDF[3.0nlo]     & \PYTHIA[8.210]	& A14 	    \\
\ttbar+$h$			& \POWHEGBOX[v1] 	& NLO 			& \NNPDF[3.0nlo]     & \PYTHIA[8.230]	& A14 	    \\
\bottomrule
\end{tabular}}
\end{table}
 
The signal process $gg \rightarrow A/H \rightarrow t\bar{t}$ and its interference with the background from SM $t\bar{t}$ production
were simulated with the \MADGRAPH~\cite{Alwall:2014hca} v2.6.7 generator with the model of Ref.~\cite{HiggsEffFormFac},
which implements $A/H$ production through loop-induced gluon–gluon fusion with loop contributions from top and bottom quarks at leading order in QCD.
Additional, non-resonant contributions at the same order~\cite{Moretti:2012mq} or NLO contributions~\cite{BuarqueFranzosi:2017jrj} were not considered in the simulation.
The latter are partially accounted for via the multiplicative correction factors described below.
The parton luminosities were modelled using the \NNPDF[3.0nlo] PDF set.
The factorisation and renormalisation scales were set to $\frac{1}{2} \sum_{i=1}^N\sqrt{m_i^2 + p_{T,i}^2}$,
where $N$ denotes the number of final-state partons,
and the top-quark mass was set to 173.3~GeV to be consistent
with the top-quark mass value that was used in the NNLO-QCD+NLO-EW predictions to which the generated SM \ttbar sample was corrected.
The widths of the (pseudo-)scalar states for a given signal hypothesis were calculated with \textsc{2HDMC}~\cite{Eriksson:2009ws}~v1.8.0 and used as an input to the ME generation.
The top-quark decays were simulated using \MADSPIN to preserve all spin correlations.
\PYTHIA[8.244] was used to model the PS, hadronisation, and the underlying event
with parameters set according to the A14 tune and using the \NNPDF[2.3lo] PDF set.
The decays of bottom and charm hadrons were modelled with \EVTGEN[1.6.0].
 
For the statistical interpretation, distributions of the \ttbar invariant mass and related variables in data are compared
to a combination of the expected distributions from all background processes $B$,
the pure signal process $S$, and the signal-plus-interference component $S+I$ for a given signal hypothesis.
Details can be found in Eq.~(\ref{eq:LH}) in Section~\ref{sec:stats}, which illustrates the parameterisation of the likelihood function in terms of the $S$ and $S+I$ components.
The $S+I$ contributions therefore need to be obtained from the inclusive process $S+I+B_{\ttbar}$ that \MADGRAPH produces by default.
This was achieved by modifying the \MADGRAPH software to remove the pure (LO) SM \ttbar process from the inclusive $S+I+B_{\ttbar}$ process at the ME level on an event-by-event basis.
The modified \MADGRAPH code was used and validated in the context of Ref.~\cite{EXOT-2016-04}.
The $S+I$ events obtained with the modified software can have positive or negative weights and the overall binned distribution typically exhibits several bins with negative yields
in the dip region.
 
Event samples for both the $S$ and $S+I$ components for different signal hypotheses (i.e. different values of $(m_{A/H},\tanb)$ or $(m_{A/H},\Gamma_{A/H})$ in the generic model)
were obtained from a set of signal samples $S$ after the detector simulation
by applying an event-by-event reweighting. This reweighting substantially reduces the computing time required to model all tested signal hypotheses.
The weight is the ratio of the \MADGRAPH matrix elements, calculated from the four-momenta of the incoming gluons and outgoing top quarks of the generated event
with the new and the old values of $(m_{A/H},\tanb)$, respectively.
 
Two sets of input samples were generated for signal processes involving scalar and pseudo-scalar states, respectively.
All $S$ and $S+I$ distributions for signal hypotheses with a (pseudo-)scalar were obtained from the set of input samples generated for a (pseudo-)scalar.
For the input samples, $\tanb=0.4$ and $m_{A/H}$ ranging from 400~GeV to 1400~GeV were chosen to obtain a good coverage of the \mttbar spectrum and related kinematic distributions.
The reweighting procedure was validated by comparing the particle-level \mttbar distributions obtained via reweighting to the equivalent distributions generated directly with \MADGRAPH.
The reweighted and generated distributions were found to agree within statistical uncertainties.
Only signal hypotheses with $\tanb\geq 0.4$ were considered to ensure the perturbativity of the top-quark Yukawa coupling~\cite{Branco_2012}.
Signal hypotheses with $m_{A/H} < 400$~GeV were not considered as these require an accurate modelling of $A/H$ decays to off-shell top quarks as well as
the inclusion of higher-order contributions to the signal process that are not included in the model of Ref.~\cite{HiggsEffFormFac}.
The $S$ and $S+I$ distributions for the two states can be trivially added in order to obtain the interference patterns predicted in signal models involving both states, such as the 2HDM and hMSSM,
because the scalar and pseudo-scalar represent two orthogonal CP eigenstates that do not interfere with each other.
 
Multiplicative correction factors $k_S$ were used in the 2HDM, hMSSM, and the simplified generic scenario interpretations
to correct the generated signal ($S$) cross-section to the value calculated at partial NNLO precision in QCD
with \textsc{SusHi}~v1.7.0~\cite{Harlander:2012pb,Harlander:2002wh,Harlander:2003ai,Aglietti:2004nj,Bonciani:2010ms,Harlander:2005rq}.
The values of $k_S$ range from $3-4$ for low values of $m_{A/H}$ and \tanb to around $1-2$ for high values of $m_{A/H}$ and \tanb.
A multiplicative correction factor $k_I$ is also applied to the interference term $I$. It is defined as $k_I = \sqrt{k_S\cdot k_B^{\mathrm{LO}}}$~\cite{Hespel:2016qaf},
where $k_B^{\mathrm{LO}}=2.07$ is the ratio of the \ttbar cross-section calculated at NNLO+NNLL precision and the cross-section for LO \ttbar production
with \MADGRAPH using settings consistent with those for $S$ and $S+I$ production.
These correction factors $k_S$ and $k_I$ are applied to the $S$ and $I$ templates, respectively, obtained via the event-by-event reweighting,
where the $I$ template is obtained by subtracting the $S$ template from the $S+I$ template.
This correction thus not only affects the normalisation but also the shape of the $S+I$ template, providing an approximate correction of the interference pattern to higher orders in QCD.
No correction factors were applied for the 2HDM+$a$ due to the fact that higher-order cross-section predictions were not available for the process $gg\rightarrow a/A \rightarrow \ttbar$,
which depends on the mixing of the two pseudo-scalar particles $a$ and $A$.


\section{Event reconstruction}
\label{sec:objects}

Common event-quality criteria and object definitions are applied for both analysis channels,
including standard data-quality requirements to select data events
with the detector in good operating condition~\cite{DAPR-2018-01}.
In addition, in each analysis channel, dedicated event selection criteria,
which are specific to the objects and kinematics of interest in those final states,
are applied as described in \Sect{\ref{sec:selection}}.
 
Events are required to have at least one reconstructed $pp$ interaction vertex
with a minimum of two associated tracks with transverse momenta $\pT > 0.5$~GeV.
The primary vertex is defined as the vertex with the highest sum of squared transverse momenta of associated tracks~\cite{ATL-PHYS-PUB-2015-026}.
A set of baseline quality criteria are applied to reject events with non-collision backgrounds or detector noise~\cite{ATLAS-CONF-2015-029}.
Two levels of object identification requirements are defined for charged leptons and jets: baseline and signal.
Baseline leptons and jets are selected with looser identification criteria,
and are used in computing the missing transverse momentum as well as in resolving possible reconstruction ambiguities.
Signal leptons and jets are a subset of the baseline objects, with tighter quality requirements which are used to define the search regions.
Isolation criteria are used to discriminate between signal leptons and leptons arising from semileptonic heavy-flavour decays or jets misidentified as leptons.
 
Jets are reconstructed from particle-flow objects~\cite{PERF-2015-09,JETM-2018-05} using the anti-$k_t$ algorithm~\cite{Cacciari:2008gp,Fastjet} with a radius parameter $R=0.4$.
The particle-flow algorithm combines information about ID tracks and energy deposits in the calorimeters to form the input for jet reconstruction.
An energy calibration is applied to both the input calorimeter clusters~\cite{PERF-2014-07} and the final reconstructed jets~\cite{JETM-2018-05}.
Additionally, a pile-up subtraction procedure~\cite{JETM-2018-05} is applied along with a global sequential calibration to account for flavour dependencies.
To suppress jets arising from pile-up, a jet-vertex-tagging (JVT) technique using a multivariate likelihood~\cite{PERF-2014-03} is applied to jets with $\pt < 60$~GeV and $|\eta|<2.4$,
ensuring that selected jets are matched to the primary vertex.
Baseline jets are selected by requiring $\pt>20\,\GeV$ and $|\eta|<4.5$.
Signal jets are selected via the tighter requirements $\pt>30$~\GeV\ and $|\eta|<2.5$.
 
The selected and calibrated jets are used as inputs for jet reclustering~\cite{ATLAS-CONF-2017-062} using the anti-$k_t$ algorithm
with a variable radius parameter~\cite{ATL-PHYS-PUB-2016-013}.
These reclustered jets are referred to as \textit{large-$VR$ jets}, and are used as proxies of the hadronically decaying top quark in the 1-lepton channel.
The effective radius of these jets is inversely proportional to the jet \pT, according to the relation $R_{\mathrm{eff}}\simeq \rho/\pT^{\mathrm{jet}}$.
The parameter $\rho$ is chosen to be 600~GeV, a value found to be optimal for the reconstruction of boosted hadronically decaying top quarks~\cite{ATL-PHYS-PUB-2016-013}.
The maximum and minimum radius of the large-$VR$ jets is set to 1.5 and 0.4, respectively.
The calibration corrections and uncertainties for the reclustered large-$VR$ jets are inherited from the input jets~\cite{ATLAS-CONF-2017-062}.
A trimming procedure~\cite{Krohn:2009th} is applied to reclustered large-$VR$ jets which removes all the associated small-$R$ jets
that have a \pT below 5\% of the \pT of the reclustered jet to suppress gluon radiation from the primary vertex and mitigate pile-up effects.
The large-$VR$ jets are required to have $\pT > 200$~GeV, $|\eta|<2.0$, a jet mass $m>100$~GeV, and at least two constituent jets.
 
Small-$R$ jets in the range $|\eta| < 2.5$ are identified as containing a $b$-hadron (henceforth called $b$-tagged) using the `DL1r' algorithm~\cite{FTAG-2019-07}.
This algorithm is based on a multivariate classification technique with a DNN combining information from the impact parameters of tracks
and topological properties of secondary and tertiary decay vertices reconstructed from the tracks associated to the jet.
The $b$-tagged jets are selected in this analysis using a working point corresponding to an efficiency of 77\% for identifying true $b$-jets in simulated SM \ttbar~events.
This working point corresponds to a rejection factor of 6 for charm and of 134 for light-flavour jets.
Correction factors are applied to the simulated event samples to compensate for differences between data and simulation in the $b$-tagging efficiency for $b$-, $c$-, and light-flavour jets.
The correction for $b$-jets is derived from \ttbar events with final states containing two leptons,
and the corrections are consistent with unity with uncertainties at the level of a few percent over most of the jet \pT range.
 
Muon candidates are reconstructed from matching tracks in the ID and MS, refined through a global fit which uses the hits from both subdetectors~\cite{PERF-2015-10}.
Baseline muons must have $\pt>10$~\GeV\ and $|\eta|<2.5$, and satisfy a set of medium identification criteria.
Additionally, the longitudinal impact parameter is required to satisfy $|z_0\sin\theta| <0.5$~mm.
Signal muons are required to have $\pt>25$~GeV and additionally satisfy the following requirement on the transverse impact parameter $d_0$ and its uncertainty $\sigma_{d_0}$:
$|d_0/\sigma_{d_0}|<3$.
Signal muons are required to be isolated~\cite{MUON-2018-03} using the requirement that the sum of the transverse momenta of the tracks within a variable-radius cone
around the muon direction, excluding the muon track, be less than 6\% of the transverse momentum of the muon.
The track isolation cone size is given by the minimum of $R = 10\;\mathrm{GeV}/p_T^{\mu}$ and $R=0.3$, where $p_T^{\mu}$ is the muon \pt.
Thus, the cone radius increases with decreasing $p_T^{\mu}$ up to a maximum of 0.3.
 
Electron candidates are reconstructed from energy deposits in the electromagnetic calorimeter matched to a charged-particle track in the ID~\cite{PERF-2017-01}.
The track is required to be matched to the primary vertex by imposing the requirement $|z_0 \sin\theta| < 0.5$~mm.
Electron candidates are required to be within $|\eta|<2.47$, excluding the transition region between the barrel and endcap calorimeters ($1.37 < |\eta| < 1.52$).
Baseline electrons are required to satisfy $\pt>10$~\GeV\ and fulfil loose identification criteria,
using a likelihood-based discriminant that combines information about tracks in the ID and energy deposits in the calorimeter system~\cite{PERF-2017-01}.
The number of hits in the innermost pixel layer is used to discriminate between electrons and converted photons.
Signal electrons are required to also satisfy $\pt>25$~\GeV\ and the tight likelihood identification criteria~\cite{PERF-2017-01}.
Additionally, signal electrons must fulfil $|d_0/\sigma(d_0)|<5$.
The same variable-cone isolation requirement as for muons is imposed on signal electrons, with the exception that the maximum cone radius is set to 0.2.
 
An overlap removal procedure is applied to resolve the reconstruction ambiguities between electrons, muons, and jets.
First, if an electron shares the same ID track with another electron, the electron with the lower transverse momentum is discarded.
Electrons sharing the same track with a muon candidate are rejected as they are assumed to be a falsely reconstructed photon from bremsstrahlung.
Next, jets are rejected if they lie within $\Delta R = 0.2$ of an electron.
Similarly, jets are rejected if they are within $\Delta R = 0.2$ of a muon
if the jet has fewer than three associated tracks or the muon is matched to the jet through ghost association~\cite{Cacciari:2008gn}.
Next, to reduce the background contributions due to muons from heavy-flavour decays inside jets,
muons are discarded if they are separated from the nearest jet by $\Delta R < 0.04 + 10\;\mathrm{GeV}/p_T^{\mu}$.
For electrons, an electron-in-jet subtraction method is used to suppress backgrounds from leptonic heavy-flavour decays inside jets
while at the same time maintaining a high reconstruction efficiency for electrons from highly collimated leptonic decays of boosted top quarks,
which typically result in an electron close to a $b$-jet.
If an electron is within $\Delta R < 0.4$ of a jet, the electron \pt\ is subtracted from the jet \pt.
If the transverse momentum of the resulting modified jet fails the requirement $\pt(\mathrm{jet})>25$~GeV,
the jet is removed as it can be assumed that the jet was actually caused by the electron. The selected electron is kept in this case.
If the modified jet passes the jet \pt\ cut, the $\Delta R$ between the electron and the modified jet is recalculated.
If $\Delta R(e, \mathrm{mod.\,jet}) > 0.2$, then both the electron and jet are kept.
Otherwise, it is assumed that the electron resulted from a heavy-flavour decay.
In this case, the electron is removed and the original jet is kept.
 
The missing transverse momentum \ptmiss, with magnitude \met, is calculated as the negative vectorial sum of the transverse momenta
of all baseline reconstructed objects (electrons, muons, jets and photons~\cite{EGAM-2018-01}) and a soft term.
The soft term includes all tracks associated with the primary vertex but not matched to any reconstructed physics object.
Tracks not associated with the primary vertex are not considered in the \met\ calculation,
improving the \met\ resolution by suppressing the effect of pile-up~\cite{ATLAS-CONF-2018-023}.


\section{Event selection and categorisation}
\label{sec:selection}

Events are required to have fired one of the single-electron or single-muon triggers.
For all events the pre-selection detailed in Section~\ref{sec:objects} is applied.
The further selection depends on the number of leptons in the event.
 
\subsection{Event selection and categorisation in the 1-lepton channel}
\label{subsec:selection_1l}
 
Events are required to contain exactly one signal electron or muon with a minimum transerse momentum of 28~\GeV.
This threshold is chosen to ensure that events are selected from the trigger efficiency plateau, avoiding the turn-on region.
The electron or muon is also required to be matched, within $\Delta R < 0.15$, to the respective trigger object.
Events with a second signal electron or muon (with transverse momentum larger than 25~\GeV, Section~\ref{sec:objects}) are vetoed
to ensure orthogonality of the 1-lepton and 2-lepton channels.
 
The detector signature of a \ttbar decay in the 1-lepton channel involves the presence of a substantial amount of
missing transverse momentum from the leptonically decaying $W$ boson.
In order to suppress background from strong multijet production that can give rise to (typically smaller) amounts of \MET via
leptonic decays in heavy-flavour jets, mis-measured jet energies, etc.,
the requirement $\met > 20$~\GeV\ is imposed.
Additionally, events are required to fulfil $\met+\mwt > 60$~\GeV\,
where the transverse mass of the selected lepton and the \ptmiss, referred to as the $W$ transverse mass, or \mwt, is defined as:
\[
\mwt=\sqrt{2\pt^{\ell}\met(1-\cos\Delta\phi(\pT^{\ell}, \ptmiss))}.
\]
 
All events must contain at least one $b$-tagged small-$R$ jet.
 
Events are split into two categories, a merged topology where the hadronic top-quark decay is reconstructed as a single large-$VR$ jet,
and a resolved topology where the decay products of the hadronically decaying top quark are reconstructed as three small-$R$ jets.
In order to keep the signal regions orthogonal, the selection criteria for the merged topology are applied first.
The criteria for the resolved topology are applied only on the events that fail the merged-topology selection.
This preference of the merged-topology selection over the resolved-topology selection for events that would pass the selection requirements for both is based on
the superior \mttbar resolution obtained with the merged-topology reconstruction compared to the \mttbar resolution for events in the resolved-topology category (Section~\ref{subsec:reco_1l}).

Events in the merged category must contain at least one small-$R$ jet with a distance $\Delta R < 2.0$ to the selected lepton,
which is considered as the $b$-jet candidate from the leptonic top-quark decay.
If more than one jet is found within $\Delta R < 2.0$ from the selected lepton, the $b$-jet candidate is chosen
as the jet with the highest transverse momentum among all $b$-tagged candidate jets.
If none of the candidate jets is $b$-tagged, the jet with the highest transverse momentum is chosen instead.
In addition, at least one selected large-$VR$ jet must be present in the event with a distance $\Delta R > 1.5$ to the lepton
and to the selected $b$-jet candidate from the leptonic top-quark decay.
 
Only events failing the merged selection are considered for the resolved-topology selection.
A fully resolved hadronic top-quark decay is expected to result in three small-$R$ jets, one of which is a $b$-jet, in addition to a $b$-jet from the leptonic top-quark decay.
Hence, events with a resolved decay topology are required to contain at least four selected small-$R$ jets.
The \ttbar\ system is then reconstructed using a $\chi^2$ algorithm, defined in Section~\ref{subsubsec:reco_1l:resolved}, and only well-reconstructed events,
defined via the requirement $\textrm{log}_{10}(\chi^2)<0.9$, are kept for further analysis.
This requirement has a 60\% efficiency for SM \ttbar events and for signal events with a $m_{A/H}=500$~GeV.
All selection criteria are summarised in Table~\ref{tab:selection_1l}.

\begin{table}[t]
\caption{Summary of the event selection criteria for the resolved and merged signal regions in the 1-lepton channel.}
\label{tab:selection_1l}
\centering
\begin{tabular}{l | l  }
\toprule
\textbf{Selection} & \textbf{Criteria}\\
\midrule
\multicolumn{2}{c}{Common selection} \\
\midrule
Run and event cleaning & All detector components with
acceptable conditions\\
Single lepton trigger & Separate single-electron or single-muon triggers\\
Exactly one lepton & Exactly one $e$ or $\mu$ with $\pt > 28$ GeV.\\
\met & $\met > 20$ GeV \\
\met + $W$ transverse mass& $\met + \mwt > 60$ GeV \\
$b$-tagging & $\geq 1$ $b$-tagged jet \\
\midrule
\multicolumn{2}{c}{Merged-topology selection} \\
\midrule
Large-$VR$ jet & $\geq 1$ large-$VR$ jet, $\pt > 200$ GeV \\
Top tagging (hadronic decay) & Large-$VR$ jet mass consistent with $\mtop$: $m > 100$~GeV\\
Candidate $b$-jet (leptonic decay) & $\geq 1$ jet with $\Delta R ~(\ell, \text{R=0.4 jet})< 2.0$ \\
\multirow{3}{*}{Back-to-back \ttbar topology} & $\Delta R~(\text{candidate $b$-jet},\ell) < 2.0$ \\
& $\Delta R~(\text{large-}VR~\text{jet}, \text{candidate $b$-jet})> 1.5$ \\
& $\Delta R~(\text{large-}VR~\text{jet}, \ell) > 1.5$ \\
Matching of $b$-jets and top candidates & $\geq 1$ top candidate reconstructed with exactly one $b$-jet \\
\midrule
\multicolumn{2}{c}{Resolved-topology selection} \\
\midrule
Small-$R$ jets & $\geq 4$ jets, $\pt > 25$ GeV \\
Well-reconstructed \ttbar system & $\textrm{log}_{10}(\chi^2)<0.9$ \\
Matching of $b$-jets and top candidates & $\geq 1$ top candidate reconstructed with exactly one $b$-jet \\
Veto events passing merged-topology selection & \\
\bottomrule
\end{tabular}
\end{table}
 
In Figure~\ref{fig:accXeff}, the selection efficiency times acceptance, including the branching ratio for the $\ell$+jets final state,
is shown for the merged-topology category and for all signal regions combined, separately for $e$+jets and $\mu$+jets events.
For reference, the branching ratio for \ttbar to $e$+jets or $\mu$+jets final states is about 17\% for each lepton flavour, taking into account leptonic $\tau$-lepton decays~\cite{Workman:2022ynf}.
The efficiency times acceptance is dominated by the merged-topology selection for \ttbar invariant mass values $\mttbar > 600$\,GeV,
while the resolved-topology selection covers the region down to the kinematic threshold.
The distributions correspond to the case of a single pseudo-scalar $A$. The corresponding results for a single scalar $H$ are very similar.
 
\begin{figure}[t]
\centering
\includegraphics[width=0.5\textwidth]{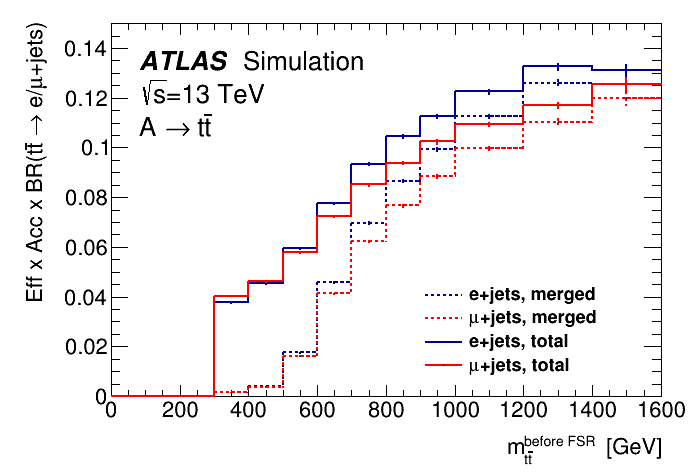}
\caption{
Selection efficiency times acceptance times $t\bar{t}$ branching ratio for the $\ell$+jets ($\ell \in {e,\mu}$) final state (Eff x Acc x BR)
as a function of the $t\bar{t}$ invariant mass at the parton level before the emission of FSR for all signal regions (solid lines)
and the merged-topology selection only (dashed lines).
The branching ratio for $t\bar{t}$ to $e$+jets or $\mu$+jets final states is about 17\% for each lepton flavour, taking into account leptonic $\tau$-lepton decays.
The distributions are obtained from all generated pure-$A$ samples in the mass range 400 -- 1400~GeV.
The error bars correspond to the statistical uncertainty in the distributions.
}
\label{fig:accXeff}
\end{figure}
 
To increase the statistical significance of the analysis, selected events passing the resolved-topology selection are classified further into separate categories
based on whether or not a $b$-tagged jet can be associated with one or both of the reconstructed top quarks.
The matching of $b$-tagged jets to the leptonic or hadronic top quark is performed by checking whether the small-$R$ jet assigned as $b$-candidate jet
to the leptonic side or one of the small-$R$ jets used for the reconstruction of the hadronic decay (based on the result of the $\chi^2$ minimisation)
are $b$-tagged. The \textit{Resolved $2b$} category contains events in which both top-quark candidates have associated $b$-tagged jets,
while the \textit{Resolved $1b$} category contains events in which only the leptonically or hadronically decaying top-quark candidate has an associated $b$-tagged jet.
The background composition in the resulting four event categories is shown in Figure~\ref{fig:sr_composition_comb}.
The highest \ttbar purity (94.2\%) is found in the Resolved $2b$ region due to the requirement of two $b$-tagged jets.
In the Resolved $1b$ and \textit{Merged} signal regions, the relative contribution of SM \ttbar production to the total SM background amounts to 74.4\% and 84.7\%, respectively.
Signal regions are not split according to the flavour of the selected lepton
because it was found that such a splitting would not increase the sensitivity of the analysis.
 
The Resolved $2b$ and Resolved $1b$ regions are each further split into five
equidistant bins of the angular variable $|\cts|$,
which provides additional discrimination between the signal process and SM backgrounds in the 1-lepton channel.
This yields a total of eleven orthogonal signal regions for the 1-lepton channel that are used in the statistical analysis of the results.
Here, $\theta^*$ denotes the angle between the momentum of the leptonically
decaying top quark in the \ttbar centre-of-mass frame
and the momentum of the reconstructed \ttbar system in the laboratory frame.
A flat \cts distribution is expected for signal events as the decays of a heavy spin-zero state would result in an isotropic distribution of the resulting top quarks.
The main background from SM \ttbar production, by contrast, is dominated by $t$-channel processes, and thus the resulting \cts distribution peaks at $\pm 1$.
No event categorisation based on the number of $b$-tagged jets or $|\cts|$ is used in the Merged category due to the fact that the sensitivity of the search at low values of \tanb is driven
by the Resolved signal regions, even at higher values of $m_{A/H}$ due to the presence of the large off-shell peak at low \mttbar
in the corresponding interference patterns (see e.g. Figure~\ref{fig:SI_truth}).
The distributions of $|\cts|$ in the Resolved $2b$ and Resolved $1b$ regions obtained after the profile-likelihood fit to the data
under the background-only hypothesis (Section~\ref{sec:stats}) are shown in Figure~\ref{fig:costhetastar:prefit:incl}.
Each $\cts$ bin corresponds to a signal region, with excellent agreement observed between the data and the SM yields in all signal regions after the fit.
 
\begin{figure}[t]
\centering
\includegraphics[width=0.45\textwidth]{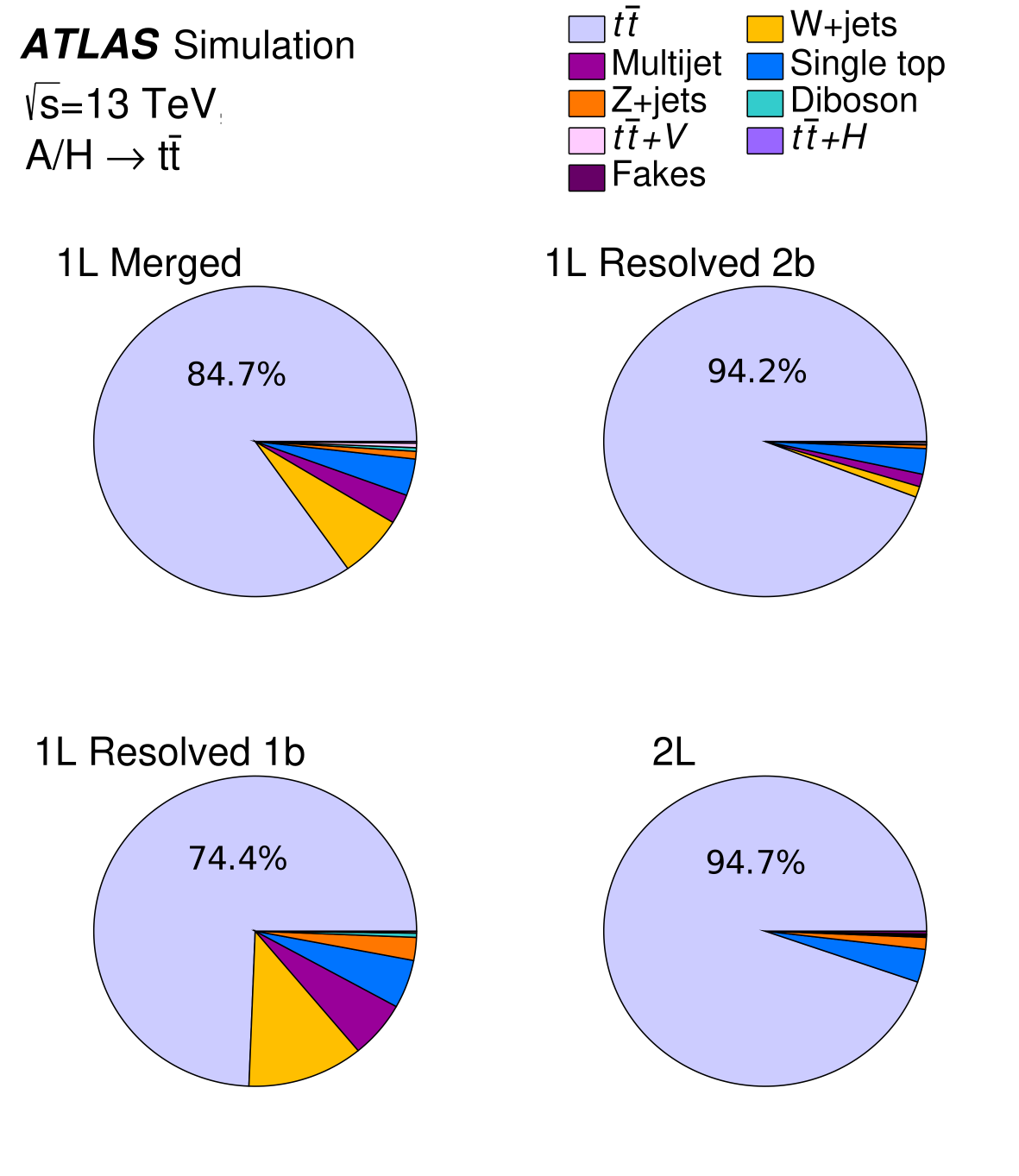}
\caption{Background composition in the Resolved $2b$ and Resolved $1b$ as well as in the Merged signal regions of the 1-lepton (1L) channel,
as well as in the signal region of the 2-lepton (2L) channel, all evaluated before the profile likelihood fit to data.
The resolved and 2-lepton signal regions are not broken down further into angular bins here for simplicity.
The background composition in the different angular bins is similar to that of the respective inclusive regions shown here.
The number within each pie chart corresponds to the $t\bar{t}$ purity in the respective region.
}
\label{fig:sr_composition_comb}
\end{figure}
 
\begin{figure}[t]
\centering
\subfloat[]{
\includegraphics[width=0.5\textwidth]{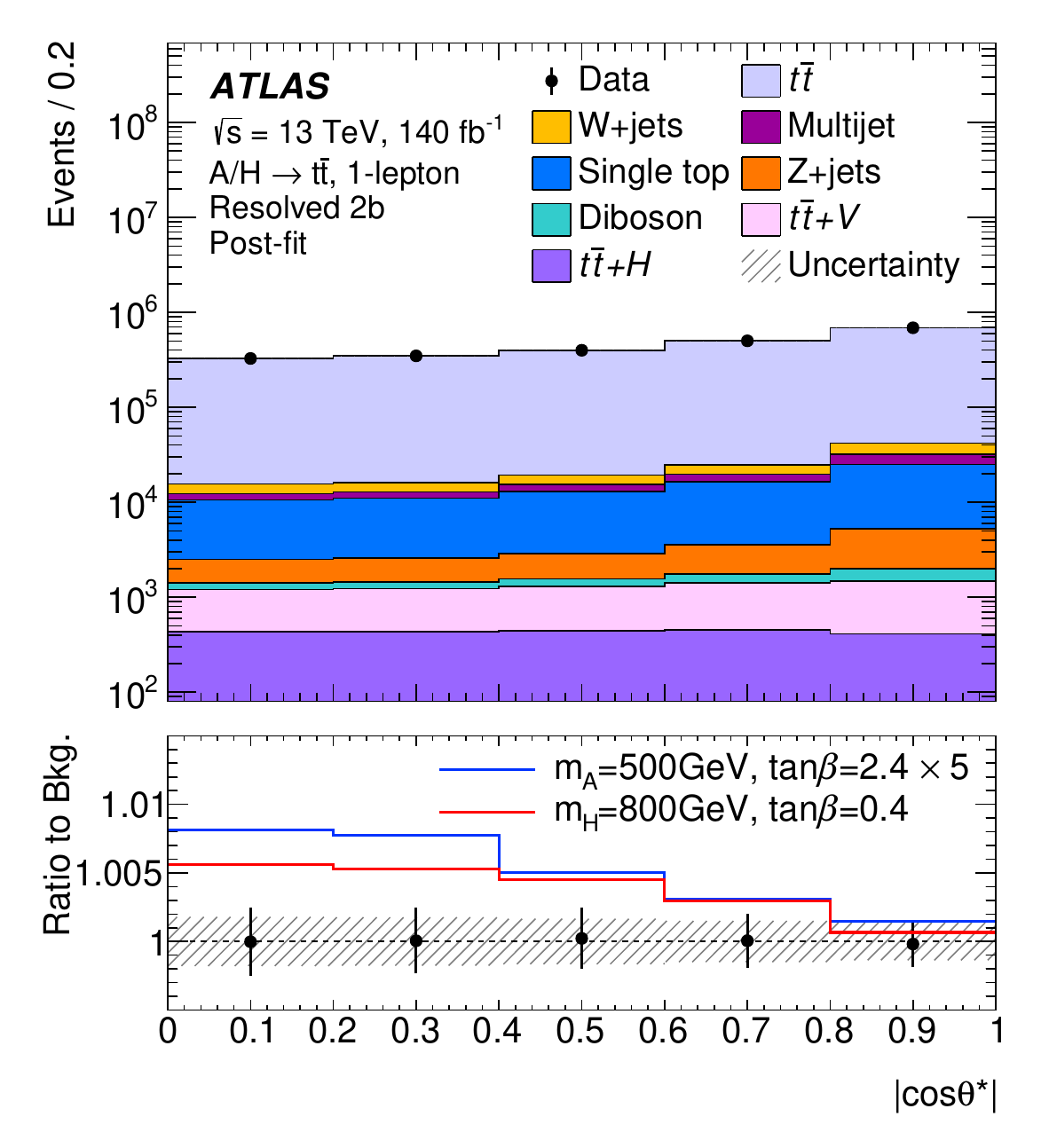}
}
\subfloat[]{
\includegraphics[width=0.5\textwidth]{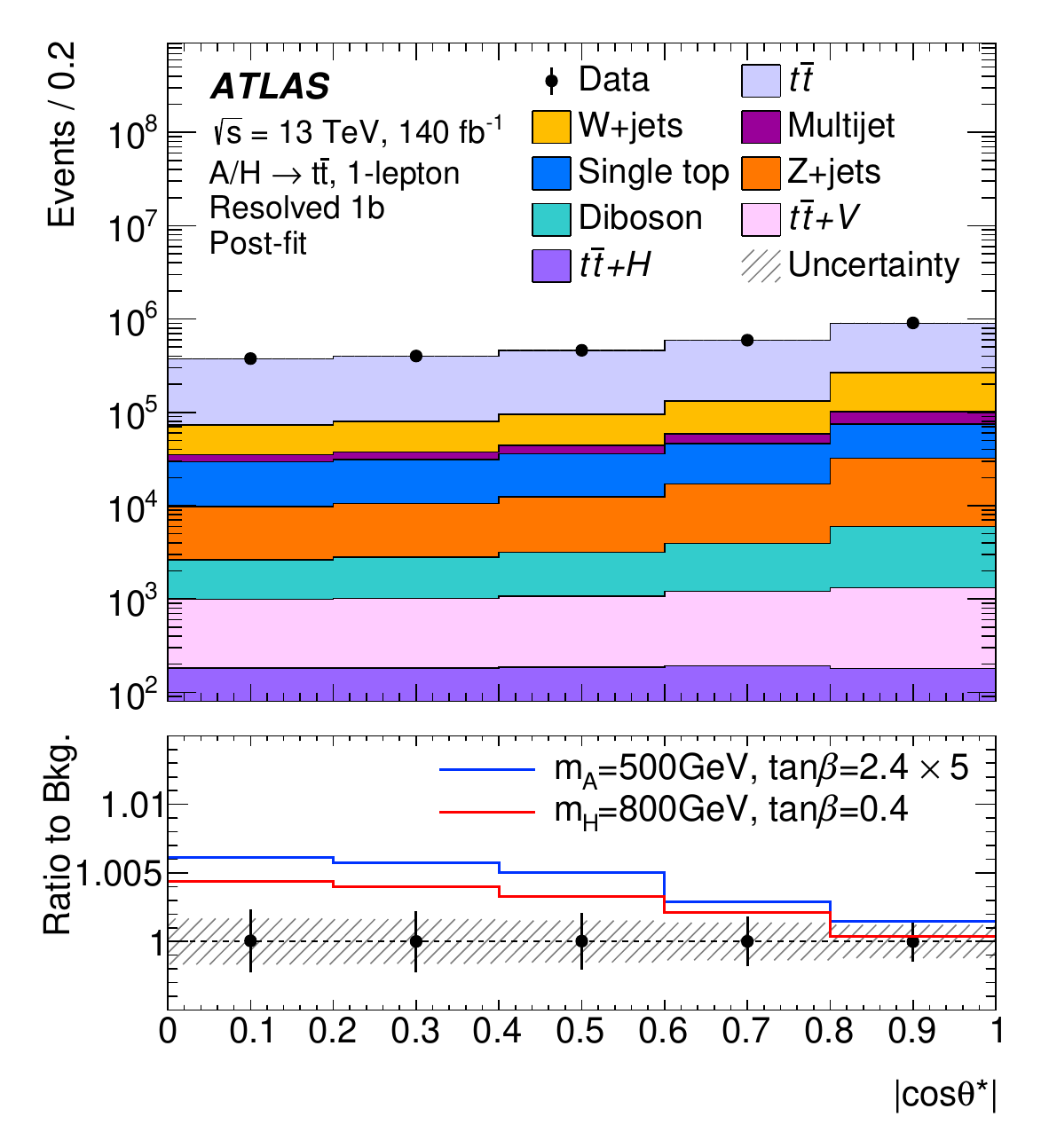}
}\\
\caption{Distributions of the reconstructed $|\cos\theta^*|$ in (a) the Resolved $2b$ and (b) the Resolved $1b$ signal regions.
Each bin corresponds to a signal region.
The expected relative deviation from the background prediction in the presence of an interference pattern, $(S+I+B)/B$,
for two representative signal hypotheses, one of them scaled by a factor of five for better visibility, is also shown in the ratio panel.
All distributions and the uncertainty bands are obtained after the profile likelihood fit to the data under the background-only hypothesis.
\label{fig:costhetastar:prefit:incl}}
\end{figure}
 
Additional control and validation regions enriched in hadronic jets falsely identified as leptons are defined
in the context of the data-driven estimate of the multijet background via the Matrix Method.
These are obtained by reverting the \met and/or \mwt requirements of the SRs and additionally loosening the lepton ID and isolation requirements.
Details are given in Section~\ref{subsec:multijet_1l}.
 
\subsection{Event selection and categorisation in the 2-lepton channel}
Candidate events for the 2-lepton channel are required to have exactly two charged leptons (electrons or muons) and at least two reconstructed jets.
At least one of these leptons is required to have $\pt>28$ GeV and to match, within $\Delta R < 0.15$, the lepton with the same flavour reconstructed by the trigger algorithm.
Additionally, at least one of the jets is required to be $b$-tagged.
Depending on the flavour of the charged leptons, three channels are defined: $ee$, $\mu\mu$ and $e\mu$.
These pre-selection requirements ensure orthogonality of the 1- and 2-lepton channels in this analysis.
Events passing the pre-selection requirements are then further classified into a number of signal and control regions.
 
Events in the signal regions are required to contain two leptons with opposite-sign electric charge (OS).
Additionally, in the $ee$ and $\mu\mu$ channels, the dilepton invariant mass, \mll, is required to be greater than 15~GeV and not within the range 81-101 GeV around the $Z$-boson mass.
To further suppress the background from $Z$+jets production, events in the
$ee$ and $\mu\mu$ channels are required to have $\met > 45$~GeV.
Signal-candidate events must also meet the requirement that the invariant masses of the lepton-$b$-jet pairs, $m_{\ell b}$,
are smaller than 150 GeV for at least one of the two possible $b$-jet to lepton assignments.
In this context, the two $b$-jets are either taken as the two $b$-tagged jets with the largest transverse momentum
or, if only one jet in the event is $b$-tagged, the $b$-tagged jet and the jet with the highest transverse momentum among the jets without a $b$-tag.
This requirement is meant to veto events where at least one of the lepton-$b$-jet pairs does not originate from a top-quark decay.
It effectively suppresses events from $tW$ production as well as \ttbar\ events in which at least one of the selected $b$-jets does not originate
from a top-quark decay but, for example, from initial-state radiation (ISR).
It has a 64\% (68\%) efficiency for SM \ttbar events with both top quarks decaying leptonically (pure-signal events with a signal mass of 1000~GeV).
 
In analogy to the 1-lepton channel event categories, the signal region of the 2-lepton channel is further split into five equidistant bins
in the azimuthal angle between the two leptons \dPhi\, to enhance the sensitivity to spin-zero states.
These five orthogonal signal regions are used in the statistical analysis of the results.
The post-fit distribution of $\dPhi$ in the 2-lepton channel is shown in Figure~\ref{fig:dPhi:prefit:incl}.
Each $\dPhi$ bin corresponds to a signal region.
Excellent agreement is observed between the data and the SM yields in all signal regions after the fit.
 
\begin{figure}[t]
\centering
\includegraphics[width=0.5\textwidth]{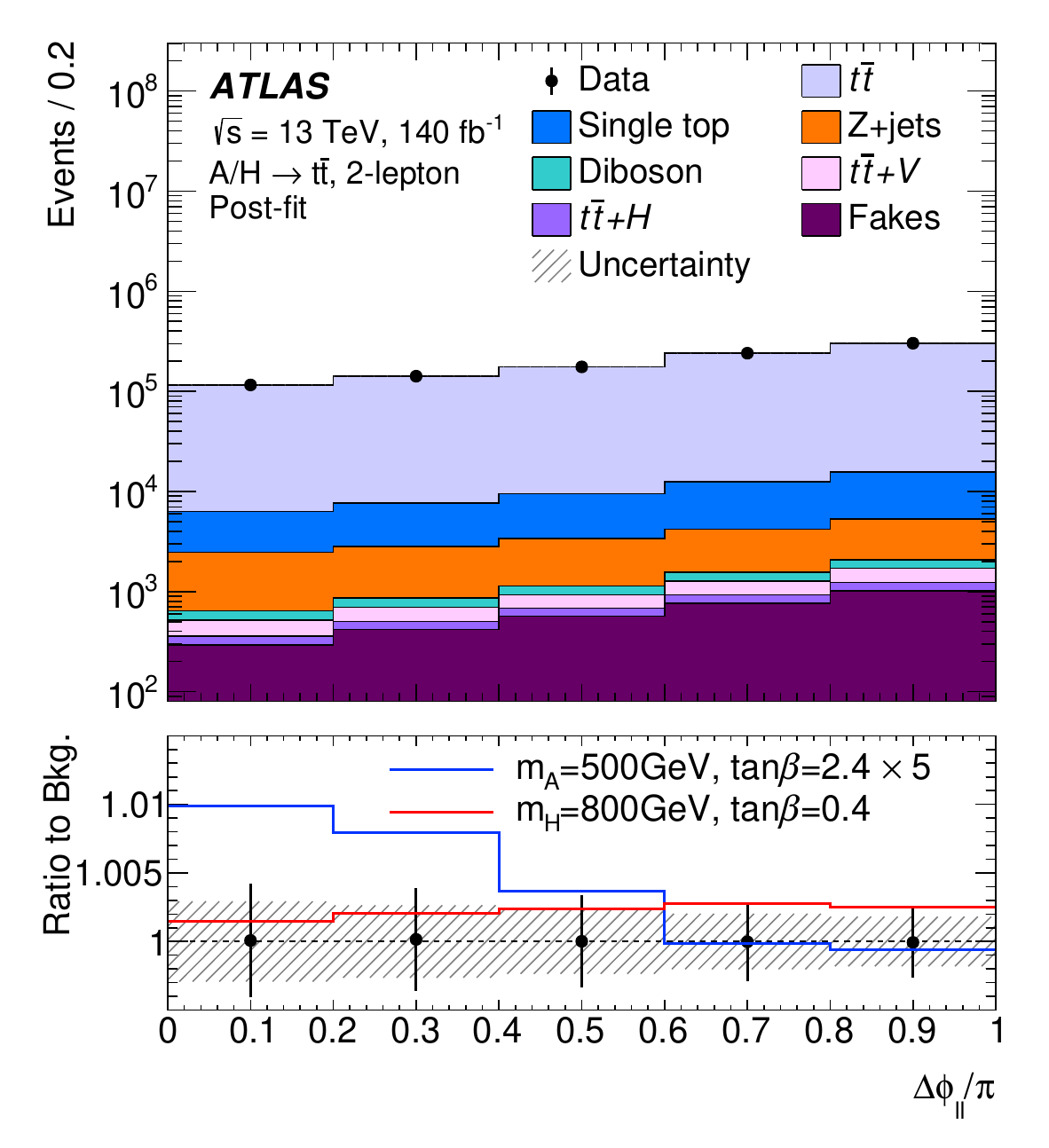}
\caption{Distribution of the reconstructed $\Delta\phi_{\ell\ell}/\pi$ after the signal selection of the 2-lepton channel for the SM backgrounds.
Each bin corresponds to a signal region.
The expected relative deviation from the background prediction in the presence of an interference pattern, $(S+I+B)/B$,
for two representative signal hypotheses, one of them scaled by a factor of five for better visibility, is also shown in the ratio panel.
All distributions and the uncertainty band are obtained after the profile likelihood fit to the data under the background-only hypothesis.
\label{fig:dPhi:prefit:incl}}
\end{figure}
 
Two orthogonal regions are used to correct and validate the MC-based modelling of smaller background components.
A control region \CRZ\ enriched in $Z$+jets events is defined by requiring events to pass the common pre-selection requirements
and additionally contain two same-flavour leptons with opposite charge and with an invariant mass consistent with leptons arising from a $Z$-boson decay, $81<\mll<101$~GeV.
A second, orthogonal control region \CRF, enriched in events with one prompt lepton and one lepton arising from a semileptonic decays of a hadron inside a jet
or a jet mis-identified as a lepton (fake lepton),
is obtained by inverting the opposite-sign requirement on the electric charges of the selected leptons and instead selecting same-sign $e\mu$ or $\mu\mu$ events.
In addition, only events in which the transverse $W$-boson masses for either the leading or sub-leading lepton is smaller than 100~GeV are considered.
This requirement reduces contributions from events containing electrons with mis-identified charge.
Events with a same-sign electron pair $ee$ are not used to control and validate the fake-lepton background
due to the large contributions from processes with two prompt electrons where one of the two electrons has its charge mis-identified.
Instead, the modelling of fake electrons is assessed in $e\mu$ events after verifying in $\mu\mu$ events that fake muons are well modelled (Section~\ref{subsec:fakes_2l}).
The event selection requirements for the signal regions are summarised in Table~\ref{tab:selection_2l}.
The background composition after the signal selection for the 2-lepton channel is also shown in Figure~\ref{fig:sr_composition_comb}.
The relative contribution of SM \ttbar production to the total SM background amounts to 94.7\% after the signal selection.

\begin{table}[t]
\caption{Summary of the event selection criteria for the signal regions in the 2-lepton channel.}
\label{tab:selection_2l}
\centering
\begin{tabular}{l | l  }
\toprule
\textbf{Selection} & \textbf{Criteria}\\
\midrule
\multicolumn{2}{c}{Common selection} \\
\midrule
Run and event cleaning & All detector components with
acceptable conditions\\
Single lepton trigger & Separate single-electron or single-muon triggers\\
Exactly two leptons & $2$ ($ee$, $\mu\mu$, $e\mu$) with $\pt > 25$ GeV. Leading one with $\pt > 28$ GeV.\\
Small-$R$ jets & $\geq 2$ jets \\
$b$-tagging & $\geq 1$ $b$-tagged jet \\
\midrule
\multicolumn{2}{c}{Signal selection} \\
\midrule
Opposite-sign leptons & $e^+e^-$, $\mu^+\mu^-$, $e^+\mu^-$, $e^-\mu^+$ \\
\met    & $\met>45$ GeV ($ee$ and $\mu\mu$ channels only) \\
Dilepton invariant mass & $\mll>15$ GeV \\
Dilepton invariant mass & $\mll<81$ GeV or $>101$ GeV ($ee$ and $\mu\mu$ channels only) \\
Lepton-plus-$b$-jet invariant mass & $m_{\ell b}<150$ GeV \\
\bottomrule
\end{tabular}
\end{table}


\section{Reconstruction of observables}

The most relevant variable to separate signal from background is the invariant mass of the $\ttbar$ system, \mttbar, which needs to be reconstructed accurately.
Additionally, angular variables sensitive to the $\ttbar$ spin correlations are used in both the 1- and 2-lepton channels to increase the sensitivity of the analysis.
These provide additional discrimination power between the signal process, in which the \ttbar system is produced in a spin-zero state,
and the background from SM \ttbar production, which involves \ttbar systems in different spin and angular momentum states.
 
\subsection{Observables in the 1-lepton channel}
\label{subsec:reco_1l}
 
In addition to the \ttbar invariant mass, \mttbar, the angular variable \cts is calculated to define the ten orthogonal signal regions for the Resolved regions,
as discussed in Section~\ref{sec:selection}.
The reconstruction of both \mttbar and \cts requires the correct identification and reconstruction of all \ttbar decay products,
including the four-momentum of the neutrino from the leptonic decay of one of the two top quarks.
In both the resolved and the merged event topologies, the transverse momentum of the neutrino is taken to be the transverse missing momentum in the event, \ptmiss.
The four-momentum component longitudinal to the beam axis $p_z^\nu$ is calculated from the kinematic constraint that the squared sum of the neutrino and
charged lepton four-momenta must yield the squared mass of the $W$-boson~\cite{TOPQ-2012-18}.
If the resulting equation has exactly one real solution, it is taken as the neutrino longitudinal momentum.
If it has no real solutions, the \ptmiss vector is rescaled and rotated in the transverse plane by the minimal amount required to obtain exactly one real solution~\cite{EXOT-2015-04}.
This procedure is based on the assumption that the lack of a real solution is caused by a mismeasurement of the transverse missing energy.
If two real solutions are found, in the resolved-topology selection, the choice is made by means of a $\chi^2$ algorithm (Section~\ref{subsubsec:reco_1l:resolved}),
while in the merged-topology selection the solution with the smallest absolute value of $p_z^{\nu}$ is chosen~\cite{EXOT-2015-04}.
 
\subsubsection{Resolved topology}
\label{subsubsec:reco_1l:resolved}
A $\chi^2$ minimisation approach is used to select the four jets from the \ttbar decay from all selected small-$R$ jets and assign them to the leptonically- and hadronically- decaying top quarks.
It is defined as follows:
\begin{eqnarray}\nonumber
\chi^2= \left[\frac{m_{jj}-m_{W_h}}{\sigma_{W_h}}\right]^2  +  \left[\frac{(m_{jjb}-m_{jj})-m_{t_h-W_h}}{\sigma_{t_h-W_h}}\right]^2 & + & \left[\frac{m_{jl\nu}-m_{t_l}}{\sigma_{t_l}}\right]^2 \\
& + & \left[\frac{(p_{T,jjb}-p_{T,jl\nu})-(p_{T,t_h}-p_{T,t_l})}{\sigma_{\mathrm{diff}p_T}}\right]^2.
\end{eqnarray}
 
The first term is a constraint requiring the mass of a pair of jets
$m_{jj}$ to be close to the $W$-boson mass. The second term constrains the
mass of a three-jet system $m_{jjb}$ to be close to the mass of the hadronically decaying top quark
but since $m_{jj}$ and $m_{jjb}$ are heavily correlated, the mass of the hadronically-decaying $W$-boson is subtracted to decouple this term from the previous one.
The third term is used to constrain the mass of the leptonically decaying top quark.
The last term constrains the magnitudes of the transverse momenta of the two top quarks to be similar,
as expected for $pp\rightarrow \ttbar$ production at LO.
In the $\chi^2$ definition above, $t_h$ and $t_l$ refer to the hadronically and leptonically decaying top quarks.
The values of the $\chi^2$ central-value parameters $m_{W_h}$, $m_{t_h-W_h}$, $m_{t_l}$, and $p_{T,t_h}-p_{T,t_l}$,
and the values of the width parameters $\sigma_{W_h}$, $\sigma_{t_h-W_h}$, $\sigma_{t_l}$, and $\sigma_{\mathrm{diff}p_T}$
are obtained from Gaussian fits to the distributions of relevant reconstructed variables, using signal MC events for
which the lepton, the reconstructed neutrino and the three jets are matched to the lepton, neutrino and quarks from the hard scattering process.
All possible jet permutations are tested, and the one with the lowest $\chi^2$ is used; if there are two solutions for the neutrino longitudinal momentum,
the one that yields the lowest $\chi^2$ is kept.
Any $b$-tagged jet can only be assigned to either of the $b$-quarks produced by the decay of a top quark, reducing the number of permutations to test.
 
The estimate of the reconstructed \ttbar invariant mass by the $\chi^2$ algorithm is improved via in-situ calibrations,
applied in both data and simulated samples, that rely on the fact that the masses of the top quark and $W$ boson are known to high accuracy.
First, in each event, the momenta of the two jets assigned to the hadronic $W$ decay are scaled by a correction factor $\alpha = (80.4\mathrm{~GeV})/m_{jj}$
so that their corrected invariant mass is equal to the $W$-boson mass.
Then the momenta of the third jet assigned to the hadronic top-quark decay is scaled by a factor $\beta$ that is based on the constraints
that the total invariant mass of this jet and the two in-situ calibrated jets must yield the top-quark mass.
The scale factor $\beta$ is obtained as the positive solution to the quadratic equation $a \beta^2 + b \beta + c = 0$, where $a = m_b^2$, $b = m_{jjb} - \alpha m_{jj} - m_b^2$,
$c = 173.3\mathrm{~GeV} - \alpha m_{jj}$, and $m_b$ is the reconstructed mass of the $b$-candidate jet.
The scale factors $\alpha$ and $\beta$ obtained with this method are in the ranges 0.90--1.00 and 0.88--1.10, respectively,
and the jet energy corrections thus within ranges consistent with the uncertainty in the jet energy scale
and the jet energy resolution for the jet collection used in this search~\cite{JETM-2018-05}.
No such scaling is applied to the jet assigned to the leptonically decaying top quark to avoid an over-correction of the jet momenta beyond the range compatible with
the jet energy resolution and the uncertainty in the jet energy scale due to the poorer \met resolution.
This kinematic scaling improves the \mttbar resolution\footnote{
The experimental resolution of the reconstructed \ttbar invariant mass, \mttbar, is defined as the width of a Gaussian fit to the distribution $(\mttbar - \mttFSR)/\mttFSR$
in pure-$S$ MC samples, where \mttFSR denotes the combined invariant mass of the six \ttbar decay products at the parton level after the emission of FSR.
}
for the resolved topology by about 12\%.
The \mttbar resolution for events in the resolved category varies between 12\% for $\mttbar = 400$~GeV and 10\% for $\mttbar = 1000$~GeV.
 
\subsubsection{Merged topology}
 
For the merged topology, the hadronically decaying top quark can be straightforwardly identified with the selected large-$VR$ jet that passes the requirements
in Section~\ref{sec:objects}.
If there is more than one large-$VR$ jet passing these requirements, the one with the highest transverse momentum is taken as proxy for the hadronically decaying top quark.
The leptonically decaying top quark is reconstructed from the selected jet identified with the $b$-jet from the leptonic top-quark decay,
the selected lepton, and the reconstructed neutrino (Section~\ref{subsec:selection_1l}).
The \mttbar resolution for events in the Merged  signal region is around 10\% for $\mttbar > 600$~GeV.
 
In Figure~\ref{fig:SI_reco:1L}, the $S+I$ distributions in the reconstructed \ttbar invariant mass are shown
for a single pseudo-scalar $A$ with mass 500~GeV and scalar with mass 800~GeV and different values of \tanb.
The corresponding parton-level distributions can be found in Figure~\ref{fig:SI_truth}.
The peak-dip structure of the signal-plus-interference pattern is clearly visible in the \mttbar spectrum.
 
\begin{figure}[t]
\centering
\subfloat[]{
\includegraphics[width=0.5\textwidth]{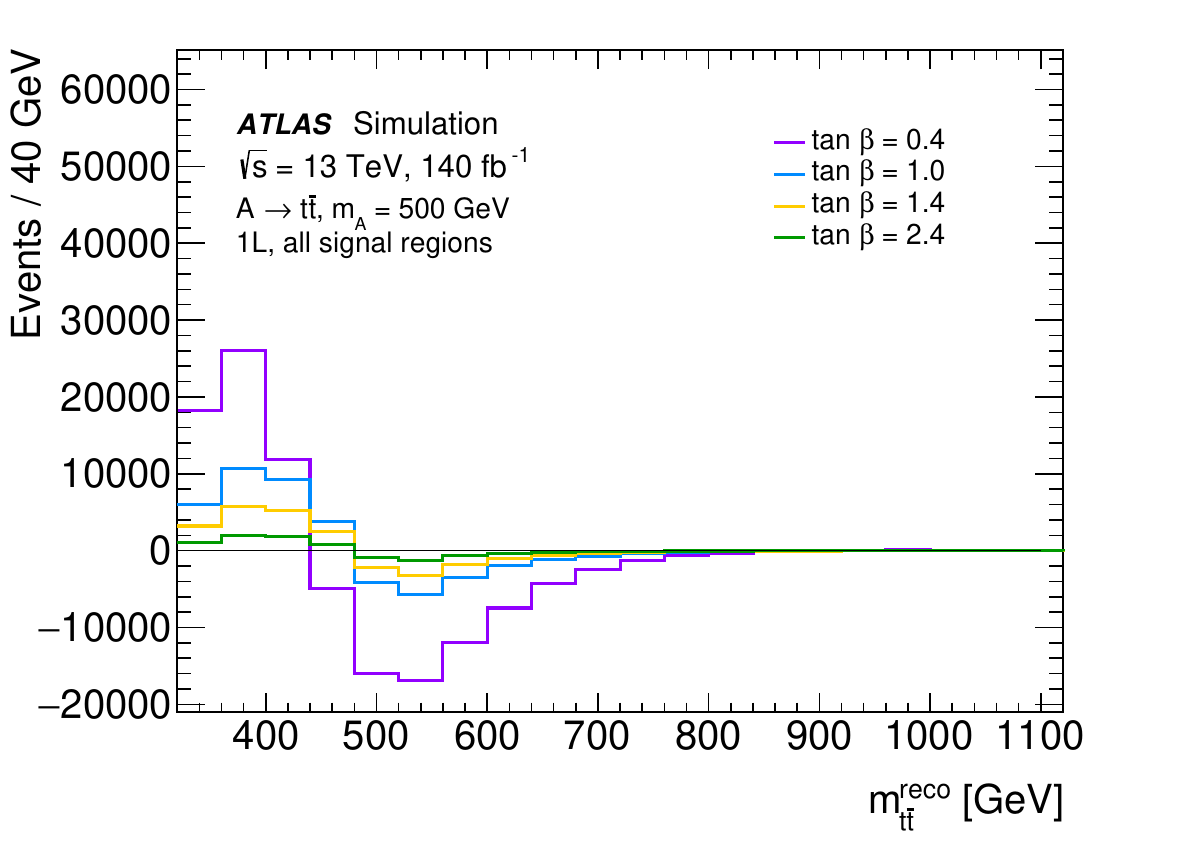}
}
\subfloat[]{
\includegraphics[width=0.5\textwidth]{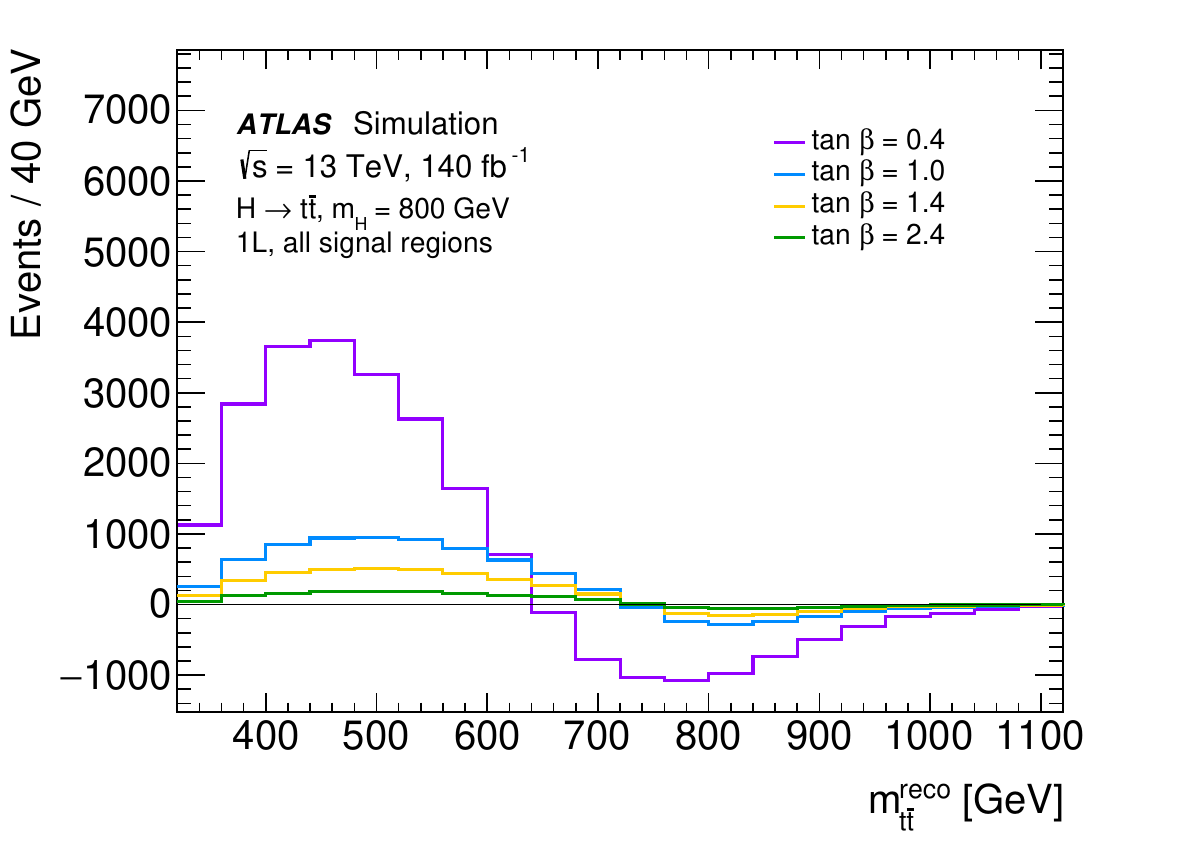}
}\\
\caption{Signal-plus-interference distributions in \mttbar\ after the signal selection of the 1-lepton channel for (a) a single pseudo-scalar $A$ with mass $m_A = 500$~GeV
and (b) a single scalar $H$ with mass $m_H = 800$~GeV for various values of \tanb.
Events from the Resolved $2b$, Resolved $1b$, and Merged signal regions are included.
\label{fig:SI_reco:1L}}
\end{figure}

\subsection{Observables in the 2-lepton channel}
 
The invariant mass of the $\ttbar$ system cannot be unambiguously reconstructed due to the presence of the two neutrinos from the two leptonically decaying top quarks.
Therefore, instead of the $\ttbar$ invariant mass, the invariant mass of the two selected $b$-jets (Section~\ref{sec:selection}) and the two charged leptons, \mllbb,
is used as a discriminating variable.
In addition, the azimuthal angle between the two leptons, \dPhi, is calculated to define the five orthogonal signal regions of the 2-lepton channel,
as discussed in Section~\ref{sec:selection}.


\section{Estimation of background contributions from data}

\subsection{$W$+jets background in the 1-lepton channel}
\label{subsec:wjets_1l}
 
Scale factors derived from data are applied before the final fit to correct the normalisation of the $W$+jets background
in the \SHERPA MC simulation samples for possible mis-modelling of the cross-section times acceptance.
The scale factors are determined by comparing the measured $W$-boson charge asymmetry in data~\cite{STDM-2017-13} with that predicted by the simulation.
The total number of $W$+jets events in data in a given signal region, $N_{W^{+}}+N_{W^{-}}$, is given by
\begin{equation}
N_{W^{+}}+N_{W^{-}} = \frac{r_{\mathrm{MC}}+1}{r_{\mathrm{MC}}-1} (D_{\mathrm{corr+}} - D_{\mathrm{corr-}}),
\label{eq:wjets_ca}
\end{equation}
where $r_{\mathrm{MC}}$ denotes the ratio of the number of $W$+jets events with a positively charged lepton to that with a negatively charged lepton
obtained from the \SHERPA MC simulation and $D_{\mathrm{corr+(-)}}$ refers to the number of observed data events with a positively (negatively) charged lepton
in the same signal region.
This method relies on the fact that the charge asymmetry obtained in MC simulation is in excellent agreement with the value measured in data~\cite{STDM-2015-03}.
Contributions to $D_{\mathrm{corr+(-)}}$ from charge-asymmetric processes such as single-top, $WZ$, and $\ttbar+W$ production are estimated from MC simulation
and are subtracted from the data samples.
Contributions from charge-symmetric processes such as \ttbar production (and the signal process itself) cancel out in the difference on the right-hand side of Eq.~\ref{eq:wjets_ca}.
The final scale factor, $C_A$, for a given signal region is then calculated as the ratio of $N_{W^{+}}+N_{W^{-}}$ evaluated from data to that predicted from the \SHERPA MC simulation.
The scale factors are evaluated separately for the $e$+jets and $\mu$+jets events in the Resolved $2b$, Resolved $1b$, and Merged regions, respectively,
without splitting the resolved regions into angular bins.
Their values agree within their statistical uncertainties and with the scale factors obtained inclusively for all $e$+jets and $\mu$+jets events across all signal regions.
The latter, $C_A = 1.125 \pm 0.031$, is used to correct the normalisation of the generated $W$+jets samples.
An additional conservative 20\% uncertainty is assigned on the $W$+jets normalisation after the $C_A$ correction to account for potential residual mis-modellings of the
relative contributions from $W$-boson production in association with heavy-flavour jets~\cite{EXOT-2015-04}.

\subsection{Multijet background in the 1-lepton channel}
\label{subsec:multijet_1l}
 
The multijet background in events satisfying the resolved- or merged-selection criteria consists of events with a non-prompt lepton
or a jet misreconstructed as a lepton that satisfies the lepton identification and isolation criteria applied in this search.
These are referred to as \textit{tight} lepton requirements in the following.
In the $\mu$+jets channel, muons arising from semileptonic decays of hadrons inside jets constitute the main source of this background.
In the $e$+jets channel, additional multijet background arises from events containing jets with a large electromagnetic component, for example from $\pi\rightarrow\gamma\gamma$ decays,
or photons mis-identified as isolated electrons.
The normalisation, \mttreco~shape, as well as statistical and systematic uncertainties associated with the multijet background are estimated from data using the \textit{matrix method}.
The matrix method used in this search is based on the one used in previous ATLAS \ttbar resonance searches and measurements~\cite{EXOT-2015-04,TOPQ-2012-18}.
 
The matrix method relies on an alternative, looser lepton definition that is based on a set of loose identification criteria~\cite{PERF-2017-01} without any additional isolation requirements applied.
The number of multijet events in a given signal region can then be estimated by solving a set of two equations describing the composition of events in two regions:
the signal region with its tight lepton requirements and a region defined by applying the same criteria as in the signal region, except that the loose lepton requirements are applied.
The number of events with leptons satisfying the loose lepton identification criteria, $N_L$, is defined as
\begin{equation}
N_{\mathrm{L}} = N_{\mathrm{prompt}} + N_{\mathrm{multijet}},
\label{eq:multijet_loose}
\end{equation}
where $N_{\mathrm{prompt}}$ and $N_{\mathrm{multijet}}$ denote the number of events with prompt leptons and events with leptons from other sources, respectively,
satisfying the loose identification criteria.
The number of events with leptons satisfying the tight lepton identification criteria applied in the signal regions, $N_T$, can be written as
\begin{equation}
N_{\mathrm{T}} = \epsilon \times N_{\mathrm{prompt}} + f \times N_{\mathrm{multijet}},
\label{eq:multijet_tight}
\end{equation}
where the \textit{real rate} $\epsilon$ is an estimator for the probability of a prompt lepton passing the loose lepton identification criteria to also pass the tight critieria.
Similarly, the \textit{fake rate} $f$ is an estimator for the probability that a lepton from other sources that passes the loose criteria also passes the tight criteria.
The fake efficiency $f$ is estimated from a control region, defined by applying the same selection criteria as for the resolved signal region,
but with the missing transverse momentum and transverse mass requirements inverted.
Contributions from processes leading to prompt leptons are subtracted from the data in this multijet enriched control region using MC simulation.
The real efficiency $\epsilon$ is determined using SM \ttbar MC samples.
Solving Eqs.~(\ref{eq:multijet_loose}) and~(\ref{eq:multijet_tight}) for $N_{\mathrm{prompt}}$ and $N_{\mathrm{multijet}}$ yields the multijet contribution in the given signal region.
The multijet estimate is performed separately for the $e$+jets and $\mu$+jets channels.
 
Good modelling of the shape of kinematic distributions, in particular \mttreco, is achieved by parameterising the real and fake rates as functions of
the transverse momentum of the lepton and a calorimeter-based isolation variable.
In addition, the rates are derived separately for the two cases $\Delta R(\ell,j)>0.4$ and $\Delta R(\ell,j)\leq 0.4$, where $j$ denotes the nearest selected jet.
The fake rates for electrons vary from 5\% to 86\%, with the largest values occurring at high electron \pT, with low nearby calorimeter activity.
This behaviour is explained by the track-based lepton isolation criterion that uses a \pT-dependent cone and leads to a looser isolation requirement at higher \pT.
The fake rates for muons vary from 9\% to 84\%, with the largest values occurring at low muon \pT, with low nearby calorimeter activity,
which is typical for soft muons arising from semileptonic decays of hadrons inside jets.
 
The modelling is validated in separate dedicated validation regions, where only one of the \met or $\met+\mwt$ requirements is inverted.
These validation regions contain a background composition that is more similar to that in the signal region while still having an enhanced multijet contribution.
A conservative 50\% normalisation uncertainty is assigned to the multijet background based on the modelling performance in the validation regions.
 
\subsection{$Z$+jets background in the 2-lepton channel}
\label{subsec:zjets_2l}
 
Data-driven corrections are applied to the \SHERPA MC simulation samples for the $Z$+jets background in the 2-lepton channel before the final fit to correct
the overall normalisation of this background component as well as the modelling of relevant kinematic variables.
The corrections are derived in the $Z$+jets enriched \CRZ\,
where a poor modelling of background prediction obtained with the \SHERPA MC simulated samples is observed for two kinematic variables in particular:
the transverse momentum of the di-lepton system, \ptll, and the invariant mass \mllbb.
In both cases, the mis-modelling exhibits a linear trend, with good agreement observed for small values of these variables
and a linear increase in the difference between data and MC prediction towards higher values.
A reweighting procedure in \mllbb is used to correct the \SHERPA MC samples for the $Z$+jets background to the data in \CRZ.
Each simulated event is assigned a multiplicative correction weight based on its \mllbb:
\[
w = a \cdot (1 - b \cdot \mllbb).
\]
The correction factors $a$ and $b$ are derived from a maximum likelihood fit of the \mllbb distributions, taking into account only the statistical uncertainty
in the MC predictions. The following values are obtained for the correction factors:
\[
a = 1.190 \pm 0.003, \quad b = (-1.7 \pm 0.1) \cdot 10^{-4}\textrm{ GeV}^{-1}.
\]
Good agreement between the data and the SM expectation is observed for all relevant kinematic variables after applying the reweighting correction.
A systematic uncertainty is assigned to the reweighting procedure based on a comparison of the nominal reweighting with an equivalent reweighting approach based on \ptll.
It amounts to a 3\% variation of the correction factor $a$.
In the final fit, an additional conservative 30\% normalisation uncertainty is applied to the $Z$+jets background
to cover any residual mismodellings in the production of heavy-flavour jets~\cite{STDM-2017-38}.
 
\subsection{Fakes background in the 2-lepton channel}
\label{subsec:fakes_2l}
 
The fakes background is estimated using MC simulation and the modelling is validated in the $\mu\mu$ and $e\mu$ channels of the
fake-enriched control region \CRF\ (Section~\ref{subsec:selection_1l}).
First, the modelling of background events arising from fake muons is studied in various representative kinematic variables in the $\mu\mu$ channel,
such as the leading and sub-leading muon transverse momenta, the di-muon invariant mass or the $W$-boson transverse masses for the leading and sub-leading muons.
The MC-based predictions are found to be in good agreement with the data within a conservative 30\% normalisation uncertainty in the MC-based fake-muon prediction.
Second, the modelling of background events with fake electrons is studied in the $e\mu$ channel, using an equivalent set of kinematic variables.
Again, good agreement between the MC-based background prediction and the data is found within a conservative 30\% normalisation uncertainty.


\section{Systematic uncertainties}
\label{sec:unc}

The modelling of signal and all background events is affected by various experimental uncertainties related to the reconstruction, identification, and calibration
of object and event properties. In addition, uncertainties related to the theoretical modelling of the simulated backgrounds, most importantly SM \ttbar production,
as well as the data-driven background estimates and corrections are taken into account.
Some of the uncertainties affect both the shape and the normalisation of the \mttbar and \mllbb spectra, while others affect the normalisation only.
The sources of the largest systematic uncertainties in the analysis are related to the modelling of the SM \ttbar background processes,
followed by uncertainties in the jet energy scale and resolution.

\subsection{Modelling uncertainties}
\label{subsec:unc_mod}
 
A range of modelling uncertainties is estimated for the dominant and irreducible background from SM \ttbar\ production.
These uncertainties can largely be grouped into three categories:
\begin{enumerate}
\item Uncertainties in the higher-order prediction to which the NLO MC sample is reweighted, including uncertainties in the reweighting method itself;
\item Uncertainties obtained from a comparison to alternative SM \ttbar\ MC samples;
\item Uncertainties in the NLO+PS prediction obtained from the nominal SM \ttbar\ MC sample without a corresponding uncertainty in the higher-order prediction.
\end{enumerate}
 
Category~1 comprises uncertainties related to scale variations in the fixed-order calculations, to the choice of PDF set,
and to uncertainties in the EW contribution, which is not present in the NLO+PS prediction.
The scale uncertainties are estimated separately for the (anti-)top \pT\ and \mttbar distributions by varying independently the renormalisation and factorisation scales up and down by a factor of two.
The PDF uncertainty is estimated as the envelope of the intra-PDF uncertainties of the LUXQED PDF set~\cite{Manohar:2016nzj},
which is used to obtain the NNLO-QCD+NLO-EW predictions (Section~\ref{sec:datamc}).
The uncertainty in the EW contribution is estimated by comparing the nominal spectra, obtained with the LUXQED PDF set,
to spectra obtained with the NNPDF3.0QED PDF set.
This variation has been found to significantly alter the EW part of the prediction, as the two PDF sets rely on a different treatment of the photon PDF.
An additional uncertainty is derived to describe possible differences between the parton-level top-quark definitions in the theory and NLO MC predictions,
which are due to the fact that the latter includes any number of real emissions from the final-state top quarks, while the NNLO prediction includes at most two real emissions.
The resulting uncertainty in the higher-order prediction is estimated by comparing the nominal SM \ttbar\ spectra
to those obtained by reweighting an alternative NLO+PS MC sample created with the same generators and settings as the nominal NLO sample
but without decaying the top quarks and forbidding PS emissions after the first one, thus obtaining a parton-level NLO+PS prediction coherent with the NNLO prediction.
An uncertainty in the reweighting method itself is derived by comparing the nominal reweighted SM \ttbar\ sample
to a sample obtained through an alternative reweighting that is first applied twice on the (anti-)top quark \pT and then in \mttbar.
 
Uncertainties in Category~2 are estimated by comparing the nominal SM \ttbar\ prediction to alternative predictions obtained from alternative SM \ttbar\ MC samples.
All alternative MC samples are reweighted to the same higher-order predictions as the nominal \POWHEG[v2]+\PYTHIA[8.230] MC sample.
This is done since none of these systematic uncertainties in the NLO+PS prediction are meant to affect the parton-level variables used for the reweighting,
but have a non-negligible effect on the final observables due to different correlations between these kinematical variables,
as well as other properties of the event that affect acceptance and shapes of the kinematic variables of interest.
The uncertainties in Category~2 are therefore referred to as \textit{residual uncertainties}.
The uncertainty due to the choice of the PS and hadronisation model is estimated by comparing the nominal predictions to those obtained from a sample generated at NLO in QCD with \POWHEGBOX[v2]
with the same PDF set and \hdamp value as for the nominal sample but interfaced with \HERWIG[7.713]~\cite{Bahr:2008pv,Bellm:2015jjp},
using the H7UE set of tuned parameters~\cite{Bellm:2015jjp} and the \MMHT[lo] PDF set~\cite{Harland-Lang:2014zoa}.
The uncertainty related to the ME-PS matching is estimated by comparing the predictions obtained with the nominal sample to those obtained from an alternative sample
generated with the same generator settings as the nominal sample but setting the $p_{\mathrm{T}}^{\mathrm{hard}}$ parameter in \PYTHIA to 1 instead of 0~\cite{ATL-PHYS-PUB-2023-029}.
This parameter regulates the definition of the vetoed region of the showering to avoid holes or overlaps in the phase space filled by \POWHEG and \PYTHIA.
This estimate of the uncertainty follows the description in Ref.~\cite{pThard}
and replaces the comparison with an alternative sample generated with \MGNLO that was used in previous ATLAS searches in \ttbar final states~\cite{EXOT-2016-04,TOPQ-2012-18,EXOT-2015-04}.
An additional uncertainty, referred to as \textit{lineshape} uncertainty,
related to the treatment of \ttbar spin correlations in different generators is estimated by comparing the predictions obtained with the nominal sample
to those obtained with an alternative sample generated with the same setup as the nominal sample but using \MADSPIN to decay the top quarks.
An uncertainty related to the choice of the \hdamp parameter is estimated by comparing the predictions of the nominal sample to those obtained with an alternative sample
with the \hdamp parameter increased by a factor of 1.5 compared to its nominal value.
In all four cases, the resulting one-sided uncertainties are assumed to be symmetric in their relative impact on the \mttbar and \mllbb spectra
to obtain effective \textit{up} and \textit{down} variations compared to the nominal predictions.
Variations in the ISR are estimated by varying the factorisation and renormalisation scales independently up and down by a factor of two.
Similarly, the uncertainty related to FSR is assessed by varying the renormalisation scale for final-state parton-shower emissions up and down by a factor of two.
In both cases, the variations are obtained from internal weights of the NLO+PS generator and the resulting alternative samples are reweighted to the higher-order predictions.
 
Category~3 comprises uncertainties in the NLO+PS prediction without a corresponding uncertainty in the higher-order predictions.
These uncertainties are kept without \textit{reducing} them, i.e. without reweighting them separately to the higher-order predictions.
Instead, the nominal reweighting is applied on each of the alternative MC predictions.
The main uncertainty of Category~3 is that related to the variations of the renormalisation scale in the ISR parton shower,
obtained via the corresponding systematic variation in the A14 tune, and accessed through internal generator weights.
The uncertainty in the \mttbar\ and \mllbb\ spectra arising from the uncertainty in the top quark mass is evaluated
by comparing the spectra obtained using the nominal sample to those generated with top quark masses of 170.0 and 175.0 GeV,
and multiplying the difference by 0.7 to approximate a one standard deviation uncertainty, corresponding to the $\pm 0.76$~GeV uncertainty in the top-quark mass world average~\cite{ATLAS:2014wva}.
This uncertainty is assigned to Category~3 to avoid reducing its impact by reweighting the alternative samples to the same higher-order prediction estimated for the central \mtop value of 173.3~GeV.
 
Additionally, the uncertainty in the SM \ttbar cross-section is applied as a pure normalisation uncertainty.
It is calculated by summing in quadrature the uncertainties related to scale, PDF+$\alpha_s$, and $m_{\mathrm{top}}$ variations with respect to the nominal value (Section~\ref{sec:datamc}).
This yields a $^{+5.6\%}_{-6.1\%}$ variation.
 
The uncertainty due to the presence of a hypothetical \ttbar bound state ("toponium"), which is not included in the MC simulation of the SM \ttbar background,
is expected to be negligible in this analysis as its effect is limited to the narrow kinematic region with $\mttbar < 350$~GeV~\cite{Sumino:2010bv}.
This region does not contribute significantly to the sensitivity of the search, even for the smallest tested value of $m_{A/H}=400$~GeV.
 
The main uncertainty in the modelling of single-top quark production, is determined from a comparison of the diagram removal and the alternative diagram subtraction scheme
for the treatment of interference effects and overlap between SM \ttbar and $tW$ production.
It is estimated by comparing the nominal $tW$ samples, generated with the diagram removal scheme,
to a set of alternative samples obtained with the diagram subtraction scheme~\cite{Frixione:2008yi,ATL-PHYS-PUB-2016-020}
using the same generator settings as for the nominal single-top samples.
As in the case of the SM \ttbar background, an uncertainty related to the choice of the PS and hadronisation model is estimated for the $tW$ background component
by comparing the nominal prediction to that obtained with the same ME generator settings but interfaced with \HERWIG[7.713],
using the H7UE set of tuned parameters and the \MMHT[lo] PDF set.
The uncertainty related to the ME-PS matching is estimated by comparing the predictions obtained with the nominal sample to those obtained from an alternative sample
generated with the same generator settings as the nominal sample but setting the $p_{\mathrm{T}}^{\mathrm{hard}}$ parameter in \PYTHIA to 1 instead of 0.
Modelling uncertainties related to the choice of the renormalisation and factorisation scales, the levels of ISR and FSR,
and the choice of PDF set have also been evaluated but are found to be negligible.
The theoretical uncertainties in the cross-sections for $tW$-, $t$-, and $s$-channel production (Section~\ref{sec:datamc}) are applied as pure normalisation uncertainties in the respective components.
They are estimated by summing in quadrature the scale, PDF, and $\alpha_s$ uncertainties in the nominal cross-sections
and amount to $\pm5.4\%$ for $tW$ production, $^{+4.3\%}_{-3.7\%}$ for $t$-channel production, and $^{+4.4\%}_{-4.1\%}$ for $s$-channel production.
 
Both shape and normalisation uncertainties are taken into account for the background from $W$+jets production in the 1-lepton channel.
In addition to the  $\pm 20\%$ normalisation assigned on top of the data-driven $C_A$ correction (Section~\ref{subsec:wjets_1l}),
uncertainties related to the choice of scales and PDF set are taken into account for the MC modelling of this background component.
The latter affect both the shape and normalisation of the $W$+jets background spectra but are found to be negligible and pruned from the final likelihood fit.
In the 2-lepton channel, no separate uncertainty is assigned to the background from $W$+jets production as it is included in the fakes background component.
The uncertainty in the fakes background component is described below.
 
The main systematic uncertainty related to the $Z$+jets background in the 2-lepton channel is the uncertainty related to the data-driven reweighting described in Section~\ref{subsec:zjets_2l}.
It affects both the shape and normalisation of this background component.
An additional conservative 30\% normalisation uncertainty is assigned to cover any residual mismodellings in the production of heavy-flavour jets (Section~\ref{subsec:zjets_2l}).
In the 1-lepton channel, where $Z$+jets production constitutes an even smaller background component,
a conservative 30\% normalisation uncertainty is applied, which covers both the $\pm 5\%$ uncertainty in the $Z$+jets production cross-section
and acceptance uncertainties related to possible mismodellings of higher jet multiplicities.
 
In the case of the remaining small backgrounds, only normalisation uncertainties are applied.
A conservative 50\% normalisation uncertainty is assigned to the data-driven estimate of the multijet background in the 1-lepton channel (Section~\ref{subsec:multijet_1l}).
The normalisation uncertainty in the fakes background in the 2-lepton channel amounts to 30\% (Section~\ref{subsec:fakes_2l}).
For the backgrounds from $\ttbar Z$, $\ttbar W$, and $\ttbar h$ production, the uncertainties in the respective higher-order cross-sections (Section~\ref{sec:datamc})
are taken into account. They amount to $^{+10.4\%}_{-12.0\%}$, $^{+13.3\%}_{-12.0\%}$, and $^{+6.8\%}_{-9.8\%}$, respectively.
A conservative 50\% normalisation uncertainty is applied to the diboson background to take into account any possible mismodelling in the
production of additional jets~\cite{Alwall:2007fs} and heavy-flavour jets~\cite{HIGG-2017-03}.
 
Modelling uncertainties are also taken into account for the signal ($S$) and signal-plus-interference ($S+I$) spectra.
Uncertainties due to the choices of renormalisation and factorisation scales are estimated by varying the two scales independently up and down by a factors of 2.0 and 0.5.
Uncertainties related to the choice of PDF set and $\alpha_s$ are derived based on the \PDFforLHC[15] prescription.
The uncertainty in the top-quark mass is taken into account by comparing the $S$ or $S+I$ spectra obtained
for $m_t = 173.3$~GeV to spectra obtained for values of $m_t$ varied by $\pm 0.76$~GeV from the nominal value.
No uncertainty is applied to the signal MC reweighting as the agreement between generated test samples and reweighted samples
has been found to be within statistical uncertainties.
 
\subsection{Experimental uncertainties}
 
The dominant experimental uncertainties affecting the \mttbar and \mllbb spectra are the uncertainties in the jet energy scale (JES) and resolution (JER).
These are evaluated from a combination of simulations, test-beam data, and in situ measurements.
Additional contributions from jet flavour composition, punch-through, single-particle response, calorimeter response to different jet flavours and pile-up are taken into account.
The jet flavour uncertainties, which were among the dominant JES uncertainties in the previous analysis~\cite{EXOT-2016-04},
were significantly reduced compared to those given in Ref.~\cite{JETM-2018-05}.
This was achieved by re-calibrating to remove the dependence on the MC hadronisation model~\cite{ATL-PHYS-PUB-2022-021},
and considering uncertainties related to quark, gluon, charm and bottom quarks derived from comparisons between models which agree with previous measurements,
and also constraining these for the flavour mixture where the JES is measured in data with \textit{in situ} techniques.
The JER uncertainties are obtained with an in situ measurement of the jet response in di-jet events~\cite{JETM-2018-05}.
Uncertainties related to the jet mass scale (JMS), which are propagated to the large-$VR$ jet uncertainties,
are derived using the $R_{\mathrm{trk}}$ method~\cite{JETM-2018-02},
which compares the ratio of the mass of jets reconstructed from calorimeter clusters to that of jets reconstructed from tracks in data and MC simulation.
An uncertainty is assigned to the efficiency of the JVT requirement on jets~\cite{PERF-2014-03}.
 
Further experimental uncertainties in the correction factors for the $b$-tagging efficiencies~\cite{FTAG-2018-01},
as well as the rates of mis-tagging $c$-\cite{FTAG-2020-08}, and light-flavour jets~\cite{FTAG-2019-02}
are applied to the simulated event samples by looking at dedicated flavour-enriched samples in data.
An additional $b$-tagging uncertainty is applied for high-momentum jets ($\pT > 400$~GeV) to account for uncertainties in the modelling of the track reconstruction in high-$\pT$ environments.
It is calculated from simulated events by considering variations in the quantities affecting the $b$-tagging performance such as the impact parameter resolution,
the percentage of poorly measured tracks, the description of the detector material, and the track multiplicity per jet.
The dominant effect on the uncertainty when extrapolating to high jet \pT is related to the different tagging efficiency when smearing the track impact parameters
based on the resolution measured in data and simulation.
 
Smaller experimental uncertainties are related to the efficiencies of the lepton identification, isolation, and reconstruction,
as well as the lepton energy scale and resolution~\cite{EGAM-2018-01,MUON-2018-03} and to the scale and resolution of the track soft-term in the \met calculation~\cite{ATLAS-CONF-2018-023}.
Variations in the reweighting applied to simulated samples to match the expected mean number of interactions observed in each bunch crossing in data are included.
They cover the uncertainty in the ratio between the predicted and measured inelastic cross-section.
A constant 0.83\% normalisation uncertainty is applied to all signal and background samples, except multijet and $W$+jets, which are estimated from data.
It accounts for the uncertainty in the integrated luminosity (Section~\ref{sec:datamc}).

\subsection{Uncertainty correlation scheme}
\label{sec:uncert_corr}
 
All experimental uncertainties are treated as fully correlated across samples and the signal regions of the 1- and 2-lepton channels in the final profile-likelihood fit (Section~\ref{sec:stats}).
The modelling uncertainties are treated as uncorrelated between all signal and background samples, with the exception of the uncertainty in the top-quark mass $m_{\mathrm{top}}$,
which is treated as correlated between signal, signal-plus-interference, and SM \ttbar samples.
Additionally, the uncertainties in the SM \ttbar background related to the modelling of the parton shower and hadronisation, the PS-ME matching, and the choice of the \hdamp parameter
are treated as uncorrelated between all 11+5 signal regions of the 1- and 2-lepton channels.
These uncertainties are obtained by comparing the nominal to an alternative SM \ttbar sample in each case (\textit{two-point systematics}),
an approach that typically yields conservative uncertainties with non-negligible constraints.
Keeping these uncertainties uncorrelated between all signal regions is a conservative approach that has been chosen to prevent constraints from propagating across regions,
thus minimising the constraints on these nuisance parameters.
Additionally, the uncertainties in the SM \ttbar background related to the inclusive \ttbar cross-section, the choice of renormalisation and factorisation scales,
and the levels of ISR/FSR cannot be assumed to be fully correlated across different kinematic regimes and are therefore treated
as uncorrelated between the Resolved and Merged signal regions of the 1-lepton channel and between the 1- and 2-lepton channels,
although they are treated as correlated across the different angular bins in \cts and \dPhi.
All other modelling nuisance parameters, such as the one related to the top-quark mass uncertainty, are treated as correlated across all signal regions.

\subsection{Uncertainty impact}
 
The relative importance of the different categories of systematic uncertainties is quantified via their post-fit impact on the observed signal strength for two representative signal hypotheses.
The results are given in Table~\ref{tab:syst}, where the fractional contribution of each category to the total systematic uncertainty in the observed signal strength is summarised.
The impact of each nuisance parameter on the signal strength is evaluated by repeating the profile likelihood fit for a given $S+I+B$ hypothesis twice,
with the nuisance parameter fixed to the value corresponding to its $\pm 1\sigma$ variation.
The impact of a group of nuisance parameters is obtained by summing in quadrature the impacts of all nuisance parameters in this category.
Similarly, the total systematic uncertainty is obtained by summing in quadrature the impacts of all nuisance parameters.
The statistical uncertainty in the signal strength is calculated based on the requirement that the sum in quadrature of the statistical uncertainty and the total systematic uncertainty,
obtained as described above, must yield the total uncertainty in the signal strength as obtained from the initial profile likelihood fit in which all nuisance parameters are allowed to float.
The \ttbar modelling systematics are found to have the largest impact for the majority of the tested signal hypotheses.
The dominant uncertainties in this irreducible backgrounds are those related to the NNLO estimate (reweighting, scales, PDFs), the ME-PS matching ($p_{\mathrm{T}}^{\mathrm{hard}}$, \hdamp),
and modelling of ISR and FSR.
The largest experimental systematics are those related to the JES and JER of the small-$R$ jets, followed by $b$-tagging related systematics.
 
Most nuisance parameters with large impact on the signal strength are only moderately constrained in the profile likelihood fit, with the notable exception of the nuisance parameters related to
the modelling of the parton shower and hadronisation for the SM \ttbar background.
The central values of most of the high-ranking nuisance parameters are changed only moderately in the fit, typically by less than half a pre-fit standard deviation.
No changes by significantly more than one pre-fit standard deviation are observed. Among the nuisance parameters with a roughly one-standard-deviation change in its central value
is one of the nuisance parameters related to the JER, which exhibits a linear slope at high \mttbar that roughly matches a small pre-fit difference between data and the MC prediction
in this region, which is corrected in the profile likelihood fit.i
 
\begin{table}[t]
\begin{center}
\caption{
Post-fit fractional contributions of different uncertainty categories to the total uncertainty in the observed signal strength $\sqrt{\mu}$,
as determined in the combined fit of the 1- and 2-lepton channels for two representative signal hypotheses:
$m_A=800$~\GeV, $\tanb=0.4$ and $m_A=m_H=500$~\GeV, $\tanb=2.0$.
The impact of each nuisance parameter on the signal strength is evaluated by repeating the profile likelihood fit for a given $S+I+B$ hypothesis twice,
with the nuisance parameter fixed to the value corresponding to its $\pm 1\sigma$ variation.
The impact of a group of nuisance parameters is then obtained by summing in quadrature the impacts of all nuisance parameters in this category.
Further details on the evaluation of the total systematic and statistical uncertainties are given in the text.
The best-fit value of the observed signal strength and its total uncertainty amount to $-0.147 \pm 0.104$ ($+0.071 \pm 0.224$)
for the hypothesis with $m_A=800$~\GeV ($m_A=m_H=500$~\GeV).
Note that the sum in quadrature of all impacts does not sum exactly to 100\% due to the rounding precision of the numbers quoted in this table.
}
\label{tab:syst}
\begin{tabular}{l|cc}
\toprule
Uncertainty component & \multicolumn{2}{c}{Fractional contribution [\%]} \\
\multirow{2}{*}{ } & {$m_A=800$~\GeV} & {$m_A=m_H=500$~\GeV} \\
& {$\tanb=0.4$}          & {$\tanb=2.0$} \\
\midrule
\textbf{Experimental}       & \textbf{30} &  \textbf{42}  \\
Small-$R$ jets (JER, JES)   & 22 &  29 \\
Large-$VR$ jets 	    & 11 &  20 \\
Flavour tagging	            & 13 &  17 \\
Leptons		            & 4 &  5 \\
Other (\MET, luminosity, pile-up, JVT)	& 10 & 14 \\
\midrule
\textbf{Modelling: SM \ttbar and signal} & \textbf{91} & \textbf{79}  \\
\ttbar NNLO	& 49 & 28 \\
\ttbar lineshape& 27 & 29 \\
\ttbar ME-PS ($p_{\mathrm{T}}^{\mathrm{hard}}$) & 36 & 30 \\
\ttbar ME-PS (\hdamp)	& 41  & 25 \\
\ttbar ISR\& FSR& 9 &  13 \\
\ttbar PS	& 29 & 41 \\
\ttbar cross-section & 21 & 31 \\
\ttbar Scales \& PDF & 21  & 16\\
$m_t$ 		&  6     & 4 \\
Signal 		&  19     & 9 \\
\midrule
\textbf{Modelling: other} & \textbf{41} & \textbf{16} \\
$W$+jets	& 11 & 8 \\
$Z$+jets	& 1 &  2 \\
Multijet 	& 27 & 10 \\
Fakes		& <1 & 1 \\
Other bkg.      & 29 & 10 \\
\midrule
\textbf{MC statistics}   & \textbf{18}  & \textbf{26} \\
\midrule
\textbf{Total systematic uncertainty} & $\mathbf{\pm 100}$ & $\mathbf{\pm 100}$ \\
\textbf{Total statistical uncertainty} & $\mathbf{<1}$ & $\mathbf{<1}$ \\
\bottomrule
\end{tabular}
\end{center}
\end{table}

\FloatBarrier


\section{Statistical data analysis}
\label{sec:stats}

The agreement between the data and the SM prediction (null hypothesis) as well as different signal hypotheses
is quantified using a profile-likelihood fit~\cite{Cowan:2010js} of the expected distributions in \mttbar (1-lepton channel) and \mllbb (2-lepton channel)
and the observed ones.
The fit is performed simultaneously in the eleven signal regions of the 1-lepton and the five signal regions of the 2-lepton channels.
The statistical and systematic uncertainties are taken into account as nuisance parameters (NPs) in the fit
with a correlation scheme as described in Section~\ref{sec:uncert_corr}.
 
The shape of the binned \mttbar or \mllbb distributions in the presence of a signal interfering with the background is parameterised in terms of the signal strength $\mu$~\cite{EXOT-2016-04}:
\begin{equation}
\mu\;S + \sqrt{\mu}\;I + B = (\mu-\sqrt{\mu})\;S + \sqrt{\mu}\;(S+I) + B.
\label{eq:LH}
\end{equation}
The terms $S$ and $S+I$ on the right-hand side of Eq.~\ref{eq:LH} denote the \mttbar or \mllbb distributions obtained from the $S$ and $S+I$ samples
for a given signal hypothesis, respectively,
while $B$ denotes the corresponding distributions for the total background expected under the SM-only hypothesis.
The fitted variable is $\sqrt{\mu}$ and the case $\mu=1$ ($\mu = 0$) corresponds to the signal hypothesis under consideration (the background-only hypothesis).
In some cases, in which interference effects are negligible, Eq.~\ref{eq:LH} reduces to the common parameterisation $\mu\;S + B$
and then $\mu$ can be interpreted as the ratio of the observed to the predicted value of the pure-signal cross-section times the branching fraction to \ttbar.
This approach relies on the simplifying assumption that, for a given signal hypothesis, the \mttbar (\mllbb) shapes for $S$ and $S+I$ in Eq.~\ref{eq:LH},
and hence the width of the interference pattern, do not change with $\mu$.
 
The likelihood used in this analysis is given by~\cite{HistFactory}:
\begin{equation} 
\mathcal{L}(\text{data}|\sqrt{\mu}, \bm{\theta}_{\mu}) =
\prod\limits_{c=1}^{N_{\text{cats}}} \mathcal{L}_c(\text{data}|\sqrt{\mu}, \bm{\theta}_{\sqrt{\mu}})
\prod\limits_{k=1}^{N_{\text{\text{cons}}}} \mathcal{F}(\tilde{\theta}_{\sqrt{\mu},k}|\theta_{\sqrt{\mu},k})
\label{eq:llh}
\end{equation}
where $\bm{\theta}$ is the vector of NPs,
$N_{\text{cats}}$ is the number of categories,
$N_{\text{\text{cons}}}$ is the number of constrained NPs,
$\tilde{\theta}_k$ is the global observable corresponding to $\theta_k$,
$c$ is the index for the categories,
$k$ is the index for the constrained NPs,
and $\mathcal{F}$ denotes a Poisson, a Gaussian or a Log-normal distribution depending on the type of uncertainty.
 
The agreement of the data with a set of signal hypotheses (search stage) is quantified using the \CLs frequentist formalism~\cite{Read:2002hq}
with the profile likelihood ratio test statistic ($q_{0}$), defined as~\cite{Cowan:2010js}:
\begin{equation} 
q_{0} = -2 \mathrm{ln} \frac{\mathcal{L}(0,\hat{\hat{\bm{\theta}}}_{0} ) }{ \mathcal{L}(\hat{\sqrt{\mu}},\hat{\bm{\theta}}_{\hat{\sqrt{\mu}}} )  },
\end{equation}
where the numerator is evaluated for the set of NPs $\hat{\hat{\bm{\theta}}}_{0}$ that maximise $\mathcal{L}$ for $\sqrt{\mu}=0$,
and the denominator is evaluated for the values $\hat{\sqrt{\mu}}$ and $\hat{\bm{\theta}}_{\hat{\sqrt{\mu}}}$ that jointly maximise the likelihood.
A best-fit value $\hat{\sqrt{\mu}}\neq 1$ indicates agreement with an interference pattern that is different from the one predicted by the signal hypothesis under consideration,
i.e. the signal hypothesis for which the $S$ and $S+I$ templates in Eq.~(\ref{eq:LH}) are obtained.
The search stage thus does not only quantify the compatibility of the data with one specific signal hypothesis,
e.g. the unique interference pattern obtained for a single scalar of mass $m_H$ and relative width $\Gamma_H/m_H$ ($\sqrt{\mu}=1$).
It also tests the compatibility of the data with the broader set of interference patterns obtained from the $S$ and $S+I$ templates for the original ($\sqrt{\mu}=1$) hypothesis
by varying the value of $\sqrt{\mu}$ according to Eq.~(\ref{eq:llh}).
A best-fit value of $\hat{\sqrt{\mu}}\neq 1$ indicates better compatibility of the data with an interference pattern that is different from the one obtained under the original hypothesis,
for example a peak-peak rather than a peak-dip structure, see e.g. Figure~\ref{fig:SI_truth_mod_ind}.
In this context, it is worth noting that $g_{A/H\ttbar}^2$ is equivalent to $\sqrt{\mu}$ according to Eq.~(\ref{eq:generic}).
Variations of $\sqrt{\mu}$ thus yield a set of interference patterns like the ones shown in Figure~\ref{fig:SI_truth_mod_ind} for variations of $g_{A/H\ttbar}^2$ for two represenative
parameter points in the generic benchmark scenario.
 
In the absence of any significant deviation from the background-only hypothesis (see Section~\ref{sec:results}),
the level at which a given signal hypothesis is excluded by the data (exclusion stage) is also quantified with the \CLs frequentist formalism
but using a different test statistic compared to the one used at the search stage.
It is defined as the simple likelihood ratio of the two values of the likelihood obtained for $\sqrt{\mu}=1$ (signal hypothesis) and $\sqrt{\mu}=0$ (background-only hypothesis):
\begin{equation} 
q_{1,0} = -2 \mathrm{ln} \frac{\mathcal{L}(1,\hat{\hat{\bm{\theta}}}_{1})}{\mathcal{L}(0,\hat{\hat{\bm{\theta}}}_{0})}.
\label{eq:plr_limits}
\end{equation}
The asymptotic formula of this test statistic, which provides an analytic approximation of its sampling distribution~\cite{Wilks:1938dza,Wald}, has been taken from Ref.~\cite{Cowan:2010js}.
Its implementation in the statistical framework and general validity of the asymptotic approximation in the context of this analysis has been verified using toy experiments.
The use of this test statistic for the calculation of exclusion regions is motivated by the fact that at the exclusion stage, the goal is to quantify the rejection of the specific benchmark scenario
under consideration, which corresponds to $\sqrt{\mu}=1$.
Values of $\sqrt{\mu}\neq1$ do not yield interference patterns compatible with the signal model under consideration because, as stated above, the width of the interference pattern,
which is determined by the $S$ and $S+I$ templates, does not change in the fit with the signal strength (or, equivalently, the coupling modifier $g_{A/H\ttbar}$),
while in a realistic model, the width is a function of the coupling to \ttbar.
In a type-II 2HDM, for example, the latter is determined by the value of \tanb (Section~\ref{sec:models}).
The choice of test statistic for the exclusion stage thus differs from that for the search stage,
where the aim is to test whether the SM hypothesis should be rejected in favour of an alternative signal hypothesis that agrees better with the data.
In this case, it is sensible to consider the wider range of interference patterns obtained for values $\sqrt{\mu}\neq1$.

A signal hypothesis is excluded at 95\% confidence level (CL) if the \CLs value, $p_{\CLs}$, for $\sqrt{\mu}=1$ is smaller than 0.05.
In this context, it should be noted that the quadratic dependence of the likelihood function on $\sqrt{\mu}$ can lead to a non-monotonic
behaviour of $p_{\CLs}$ as a function of $\sqrt{\mu}$, as shown in Figure~\ref{fig:CLsScan} for one of the signal hypotheses considered in this paper.
This means that for some signal hypotheses, the $p_{\CLs}$ function may cross the value of $p{\CLs} = 0.05$ multiple times,
leading to disjoint exclusion intervals. In these cases, no unique upper limit on the signal strength can be defined.
Additionally, the crossing point(s) for the median and the $N\sigma$ bands may vary significantly for the same signal hypothesis.
Furthermore, the behaviour of the \CLs scans varies across signal hypotheses, with some scans exhibiting only unique crossing points or even a monotonous behaviour,
and others exhibiting several crossing points in either the median and/or $N\sigma$ ($N \in \{\pm 1,\pm 2\}$) bands.
 
\begin{figure}[t]
\centering
\includegraphics[width=0.5\textwidth]{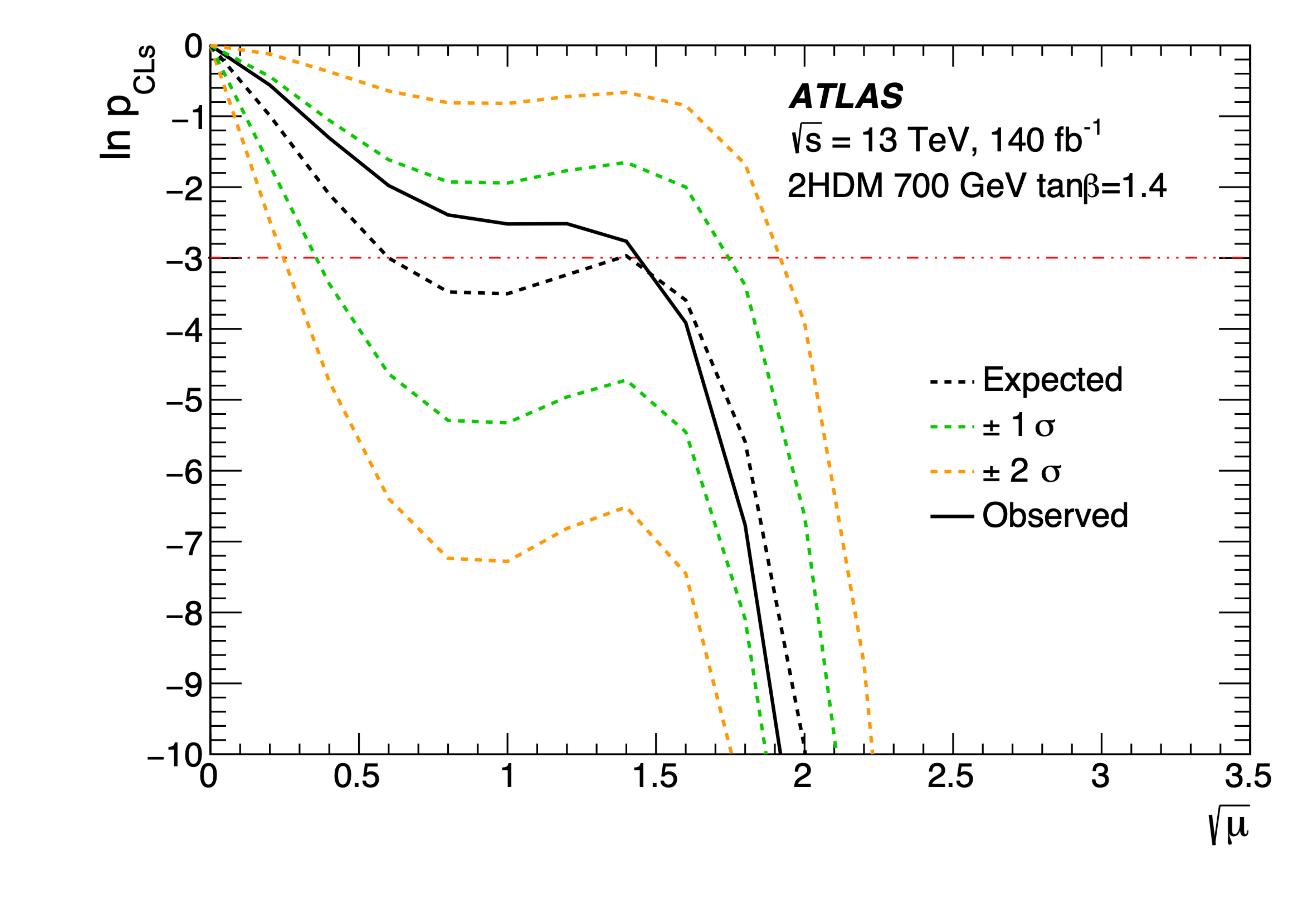}
\caption{Distribution of the logarithm of the \CLs value, $\ln(p_{\CLs})$ as a function of $\sqrt{\mu}$ for the signal hypothesis with $m_A=m_H=700$~GeV and $\tan\beta=1.4$
showing a non-monotonic behaviour for the observed and expected median, $\pm 1\sigma$, and $\pm 2\sigma$ bands.
The red horizontal line corresponds to $p_{\CLs}=0.05$. Values of $\sqrt{\mu}$ with $p_{\CLs}<0.05$ are excluded at 95\% CL.
\label{fig:CLsScan}}
\end{figure}
 
This behaviour of $p_{\CLs}$ as a function of $\sqrt{\mu}$ has implications for the calculation of two-dimensional exclusion regions in the parameter spaces of the different benchmark models.
Instead of calculating upper limits on $\sqrt{\mu}$ for selected signal hypotheses and then using an interpolation technique to obtain the contours corresponding to $\sqrt{\mu} = 1$
to define the parameter regions excluded at 95\% CL, a different approach is chosen in this paper.
The 95\% CL exclusion regions are obtained by calculating the observed (expected) values of $p_{\CLs}$ using the test statistic in Eq.~(\ref{eq:plr_limits})
for each point in a fine, uniform grid of points in the parameter plane of interest.
Each point corresponds to a different signal hypothesis, for which $S$ and $S+I$ templates have been obtained via the reweighting technique described in Section~\ref{sec:datamc}.
A linear interpolation between the values of $\ln(p_{\CLs})$ is then used to determine the observed (expected) exclusion contour corresponding to $p_{\CLs}(\sqrt{\mu}=1)=0.05$.
This approach avoids the ambiguities in the definition of upper limits on $\sqrt{\mu}$ in the presence of signal-background interference effects
that necessitate a quadratic likelihood parameterisation.
 
The non-linear dependence of the likelihood and hence \CLs value on $\sqrt{\mu}$ also has consequences for the calculation of the $N\sigma$ bands.
These are commonly understood as an indication of how frequently each signal hypothesis would be excluded under the background-only hypothesis,
with the edges of the $\pm 1\sigma$ ($\pm 2\sigma$) bands marking the range of signal hypotheses that would be excluded under the background-only hypothesis in 68\% (95\%) of equivalent searches.
Common ATLAS statistics tools to date have relied on a slightly different definition of the $N\sigma$ exclusion regions.
The latter are obtained from a set of Asimov datasets that are representative of $N\sigma$ fluctuations under the background-only hypothesis, referred to as $N\sigma$ Asimov datasets.
The exclusion limit for a given $N\sigma$ Asimov dataset defines the $N\sigma$ exclusion region, referred to as $N\sigma$ \textit{limit contour} in the following.
If the likelihood function is quadratic in the POI, as is the case for models including signal-background interference, unphysical crossings of the limit contours with the median exclusion limit
can occur. Such crossings contradict the simple frequentist interpretation of the $N\sigma$ exclusion regions outlined above.
Therefore, an alternative, more appropriate way of defining the $N\sigma$ exclusion regions, referred to as \textit{limit bands}, is chosen in the analysis presented in this paper.
These limit bands represent the hypotheses that have an exclusion rate between $\Phi(-N)$ and $\Phi(N)$,
where $\Phi(X)$ is the standard Gaussian cumulative distribution function and $\Phi(N)$ is the probability of $N^{\mathrm{th}}$ normile.
This approach to determine the $N\sigma$ exclusion regions more accurately reflects the frequentist interpretation of the $N\sigma$ exclusion regions and, by construction,
avoids unphysical crossings of the $N\sigma$ band edges and the median expected limit.
In the case of a likelihood that is linear in the POI, the limit contours approximate the limit bands very well.


\section{Results}
\label{sec:results}

 
The level of agreement between the observed data and the SM prediction is tested in a fit under the background-only hypothesis ($\mu=0$) in which only the nuisance parameters are allowed to vary.
The \mttbar distributions in the eleven signal regions of the 1-lepton channel after the fit to the full 140~\ifb dataset
are shown in Figures~\ref{fig:mttbar:postfit:final_2b} and~\ref{fig:mttbar:postfit:final_1b}.
The \mllbb\ distributions in the five signal regions of the 2-lepton channel are shown in Figure~\ref{fig:mttbar:postfit:final_2l}.
The observed \mttbar and \mllbb spectra are compatible with the post-fit expected spectra within the (constrained) uncertainty bands in all signal regions.
 
The agreement between the data and various signal hypotheses is tested at the search stage (Section~\ref{sec:stats}).
The significance of local excesses or deficits is quantified by fitting the observed data with different signal hypotheses ($S+I+B$ model, at LO precision),
predicting either a single scalar or pseudo-scalar with masses in the range 400 -- 1400~GeV and values of the relative width of 1\%, 5\%, 10\%, 20\%, 30\%, and 40\%.
No significant interference pattern is found in the data.
The most significant deviation from the background-only prediction is obtained for the interference pattern of a pseudo-scalar
with $m_A=800$~GeV, $\Gamma_A/m_A=10\%$, and a best-fit value of $\sqrt{\mu}=4.0$,
with a local significance of $2.3\sigma$.
In the absence of any significant interference patterns in the data compared to the background-only prediction,
exclusion regions are derived for the benchmark scenarios described in Section~\ref{sec:models}.
 
\begin{figure}[t]
\centering
\subfloat[]{
\includegraphics[width=0.32\textwidth]{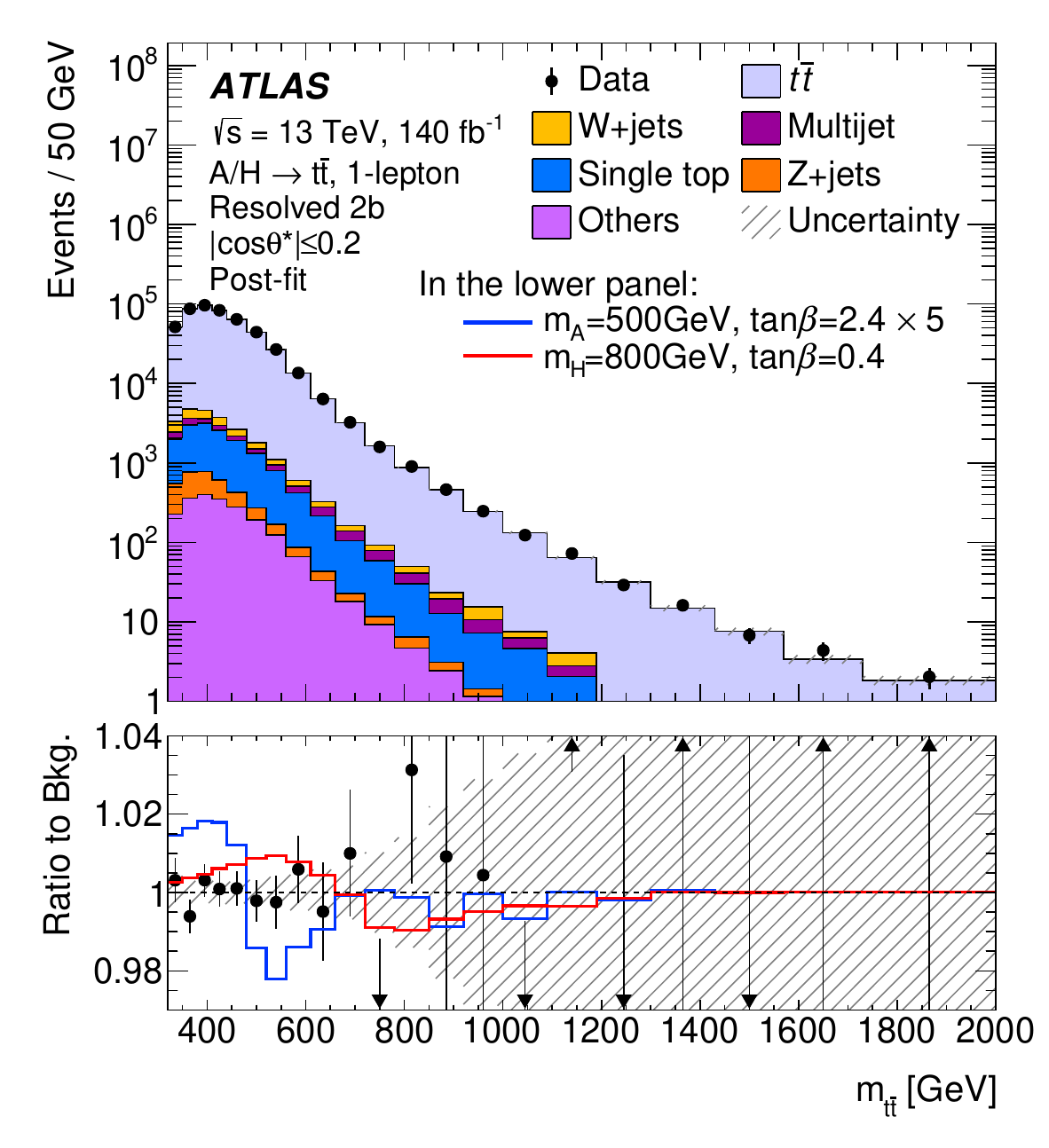}
}
\subfloat[]{
\includegraphics[width=0.32\textwidth]{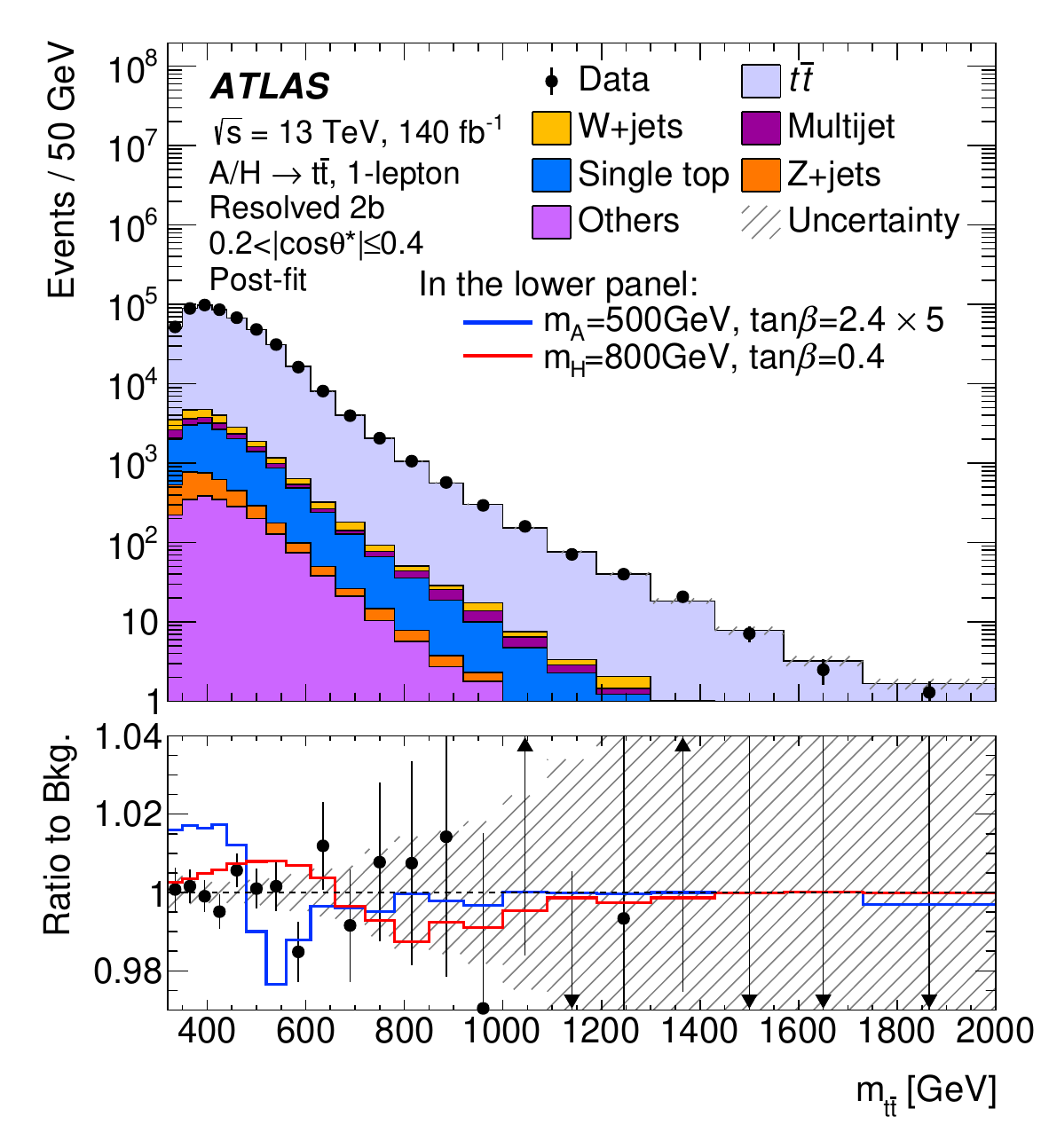}
}\\
\subfloat[]{
\includegraphics[width=0.32\textwidth]{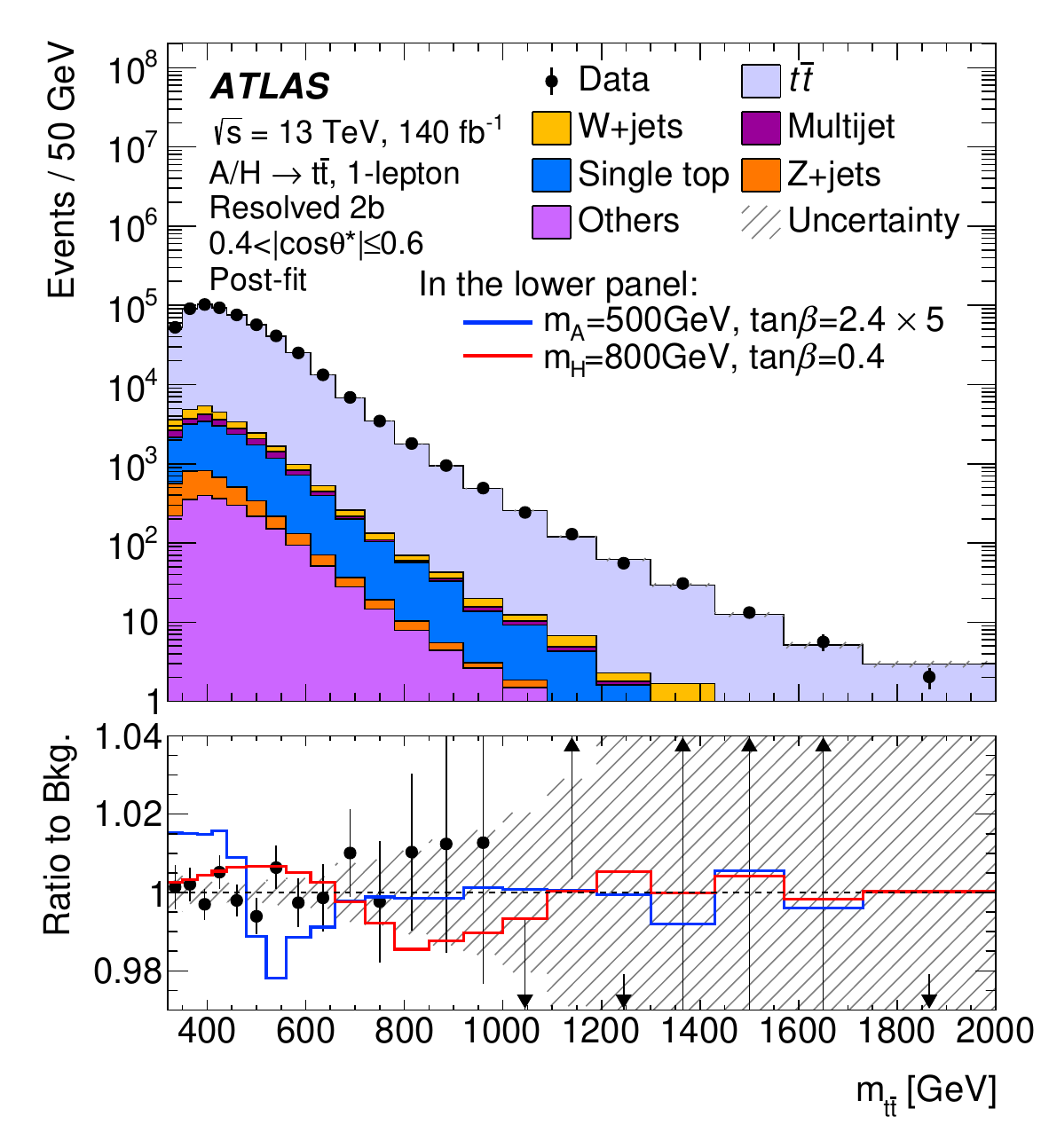}
}
\subfloat[]{
\includegraphics[width=0.32\textwidth]{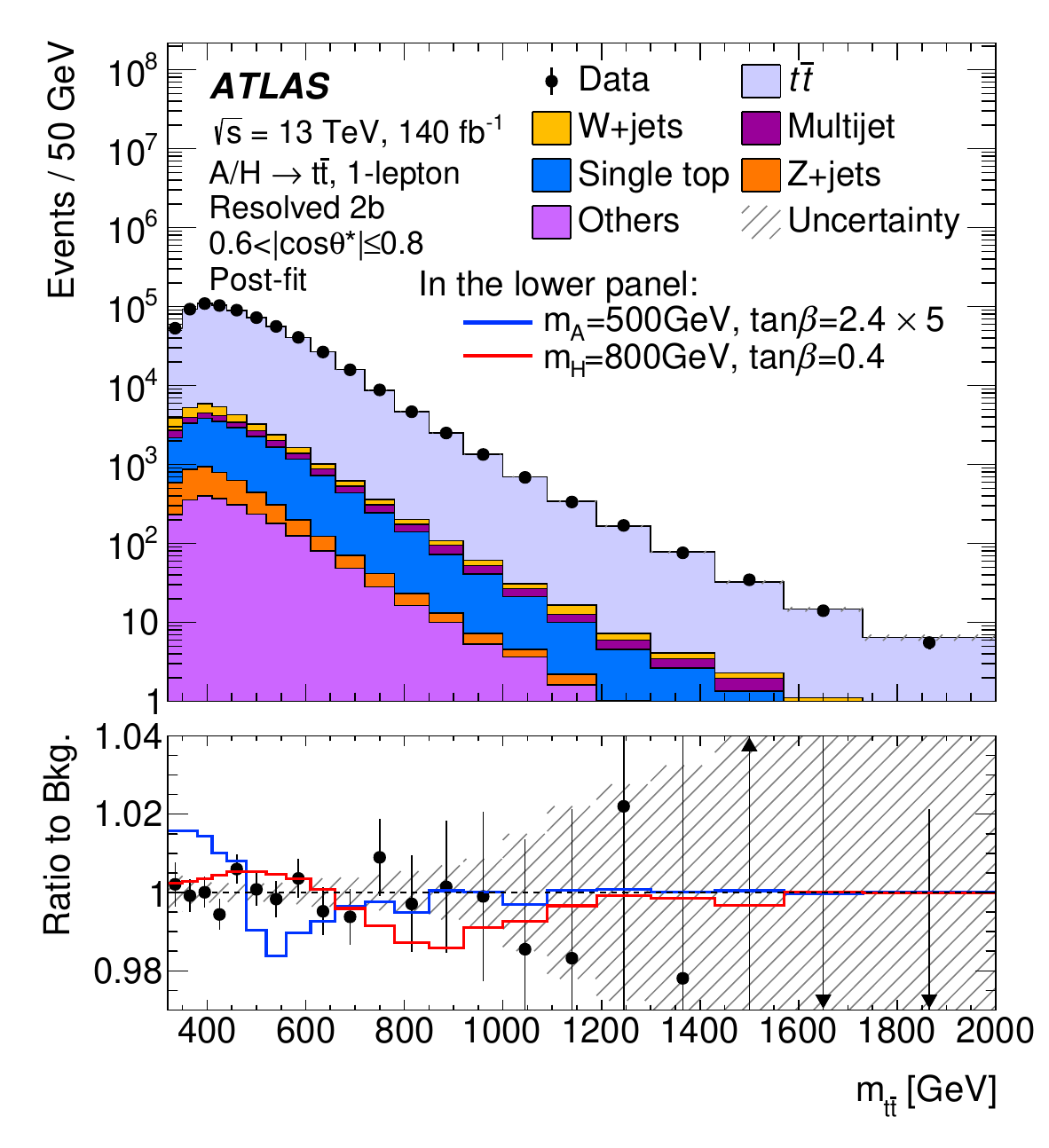}
}\\
\subfloat[]{
\includegraphics[width=0.32\textwidth]{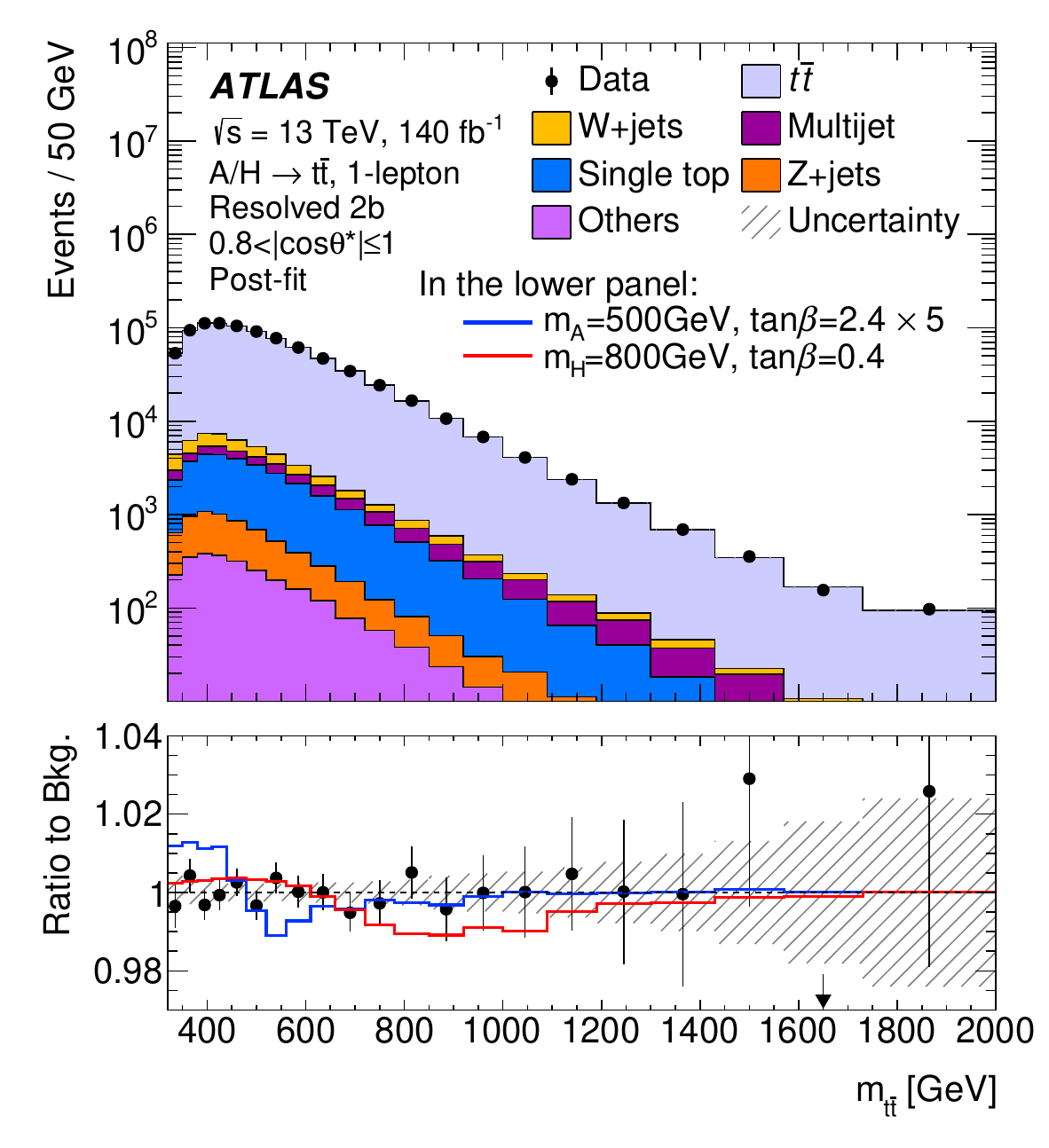}
}
\subfloat[]{
\includegraphics[width=0.32\textwidth]{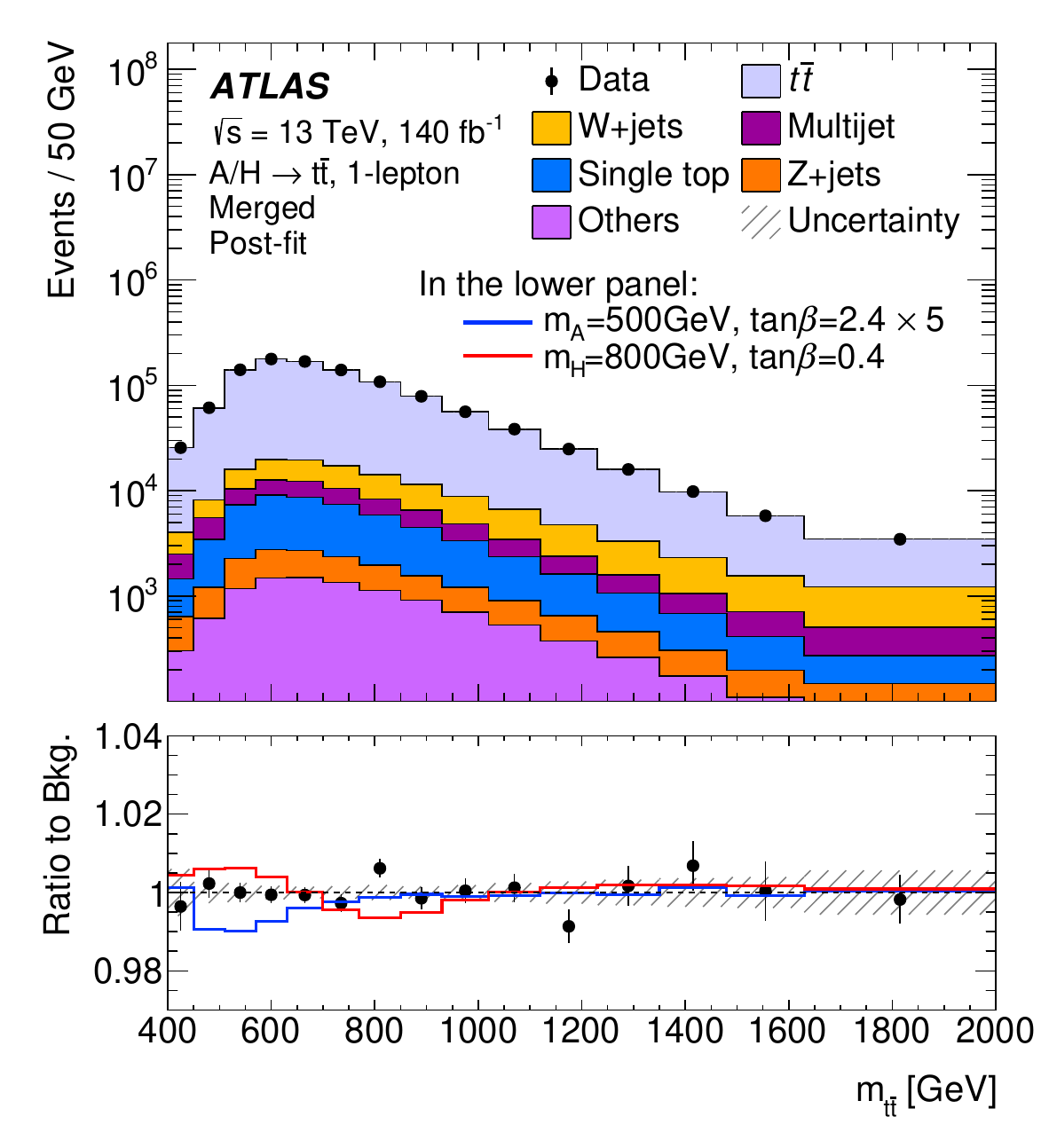}
}\\
\caption{Post-fit distributions of the reconstructed $m_{t\bar{t}}$ for the five Resolved $2b$ signal regions
with (a) $|\cos\theta^*| \leq 0.2$, (b) $0.2 < |\cos\theta^*| \leq 0.4$, (c) $0.4 < |\cos\theta^*| \leq 0.6$, (d) $0.6 < |\cos\theta^*| \leq 0.8$, and (e) $0.8 < |\cos\theta^*| \leq 1.0$,
and (f) the Merged signal region of the 1-lepton channel.
In the lower panels, the ratio of the data and the post-fit prediction are shown (data points).
The expected relative deviation from the background prediction in the presence of an interference pattern, $(S+I+B)/B$,
for two representative signal hypotheses (one of them scaled by a factor of five for better visibility) is also shown in the ratio panel.
\label{fig:mttbar:postfit:final_2b}}
\end{figure}
 
\begin{figure}[t]
\centering
\subfloat[]{
\includegraphics[width=0.32\textwidth]{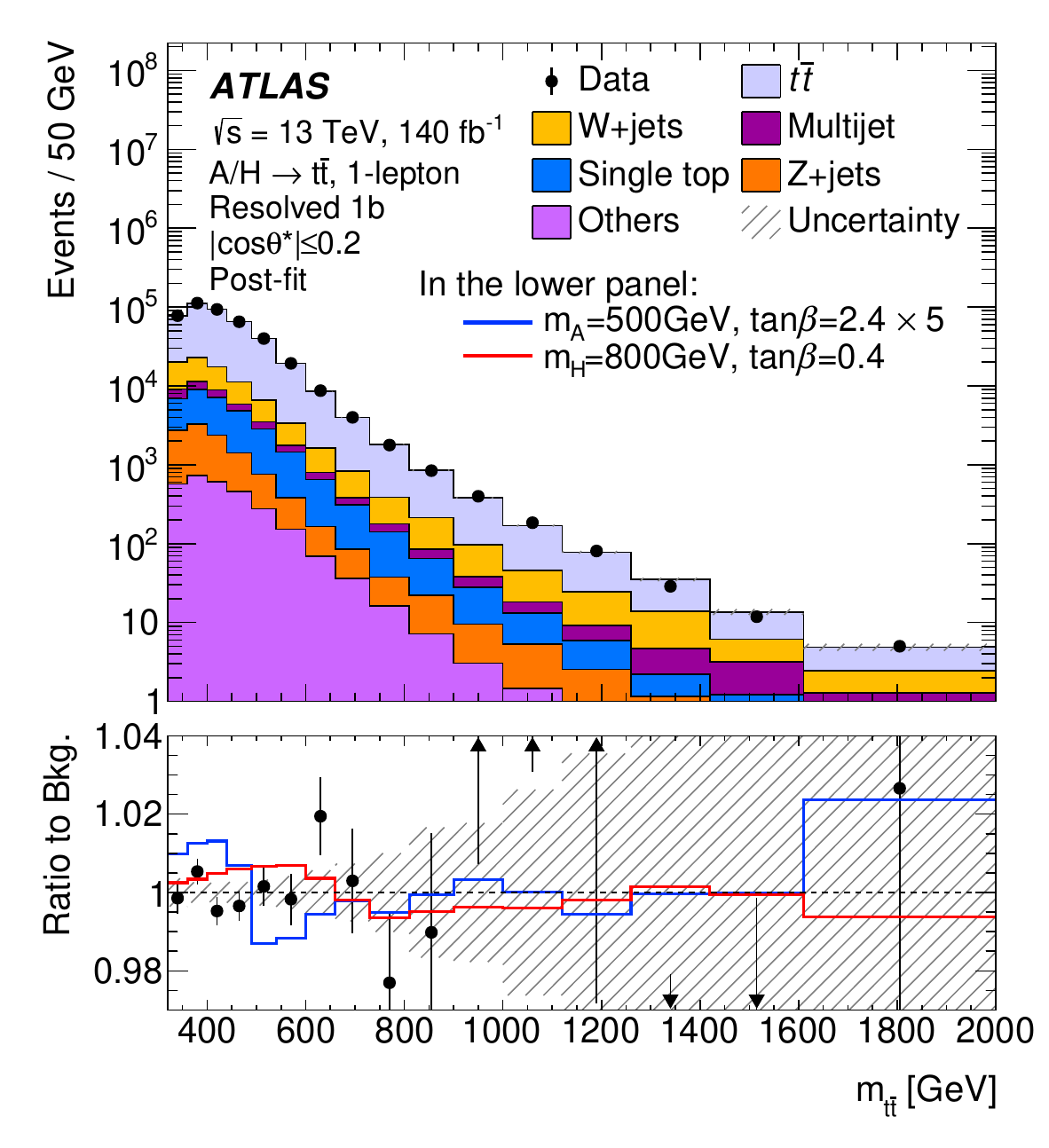}
}
\subfloat[]{
\includegraphics[width=0.32\textwidth]{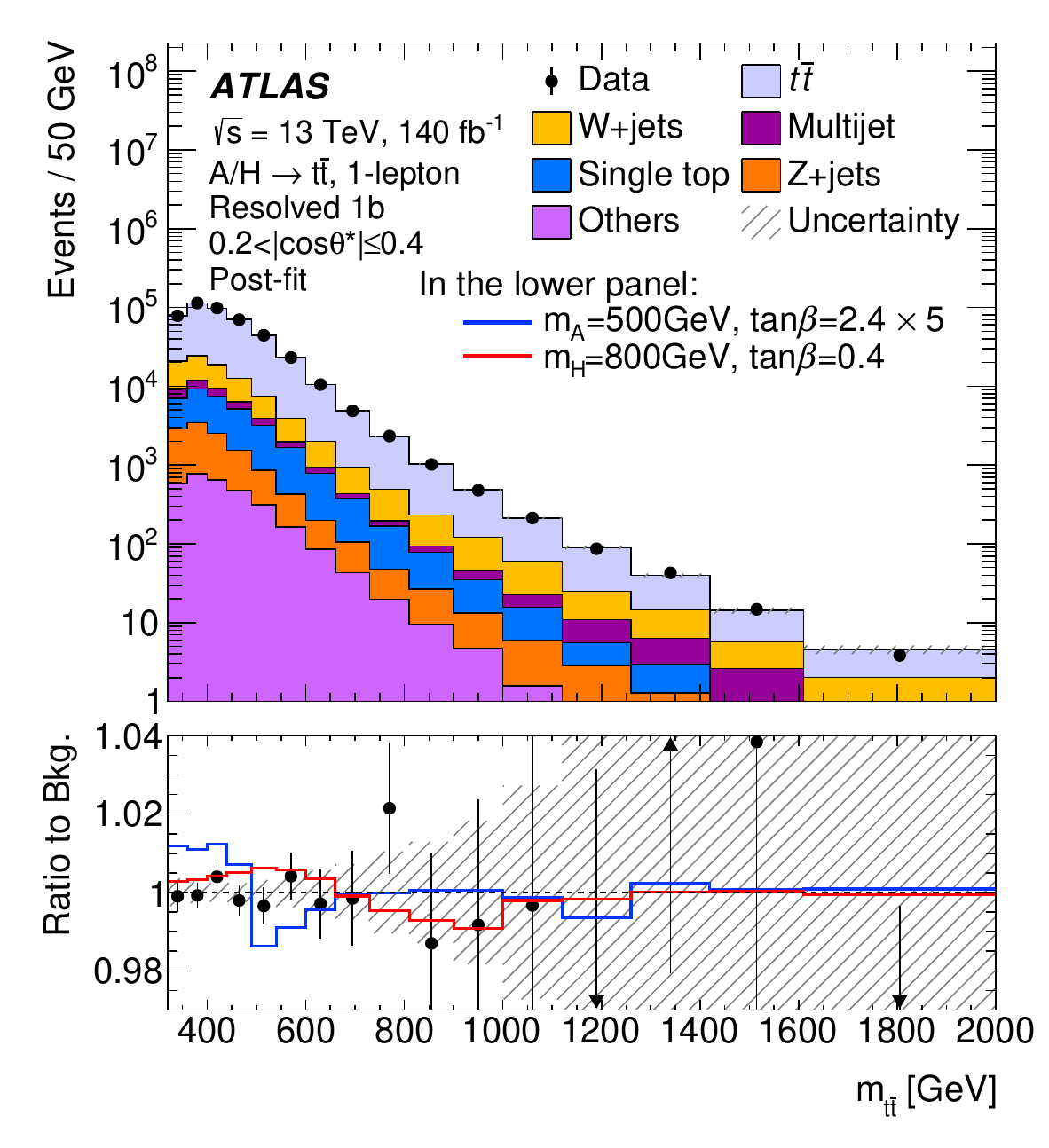}
}\\
\subfloat[]{
\includegraphics[width=0.32\textwidth]{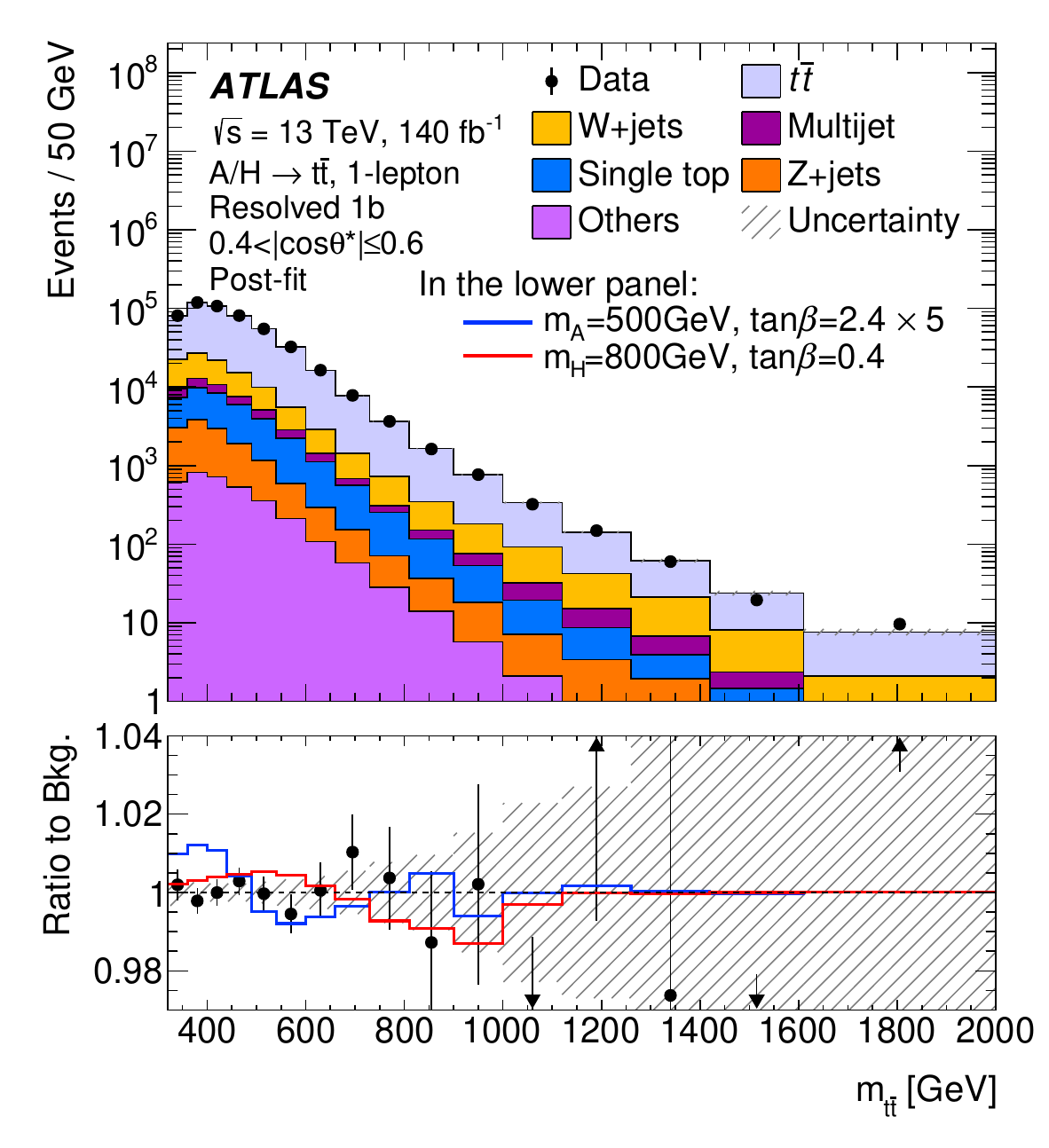}
}
\subfloat[]{
\includegraphics[width=0.32\textwidth]{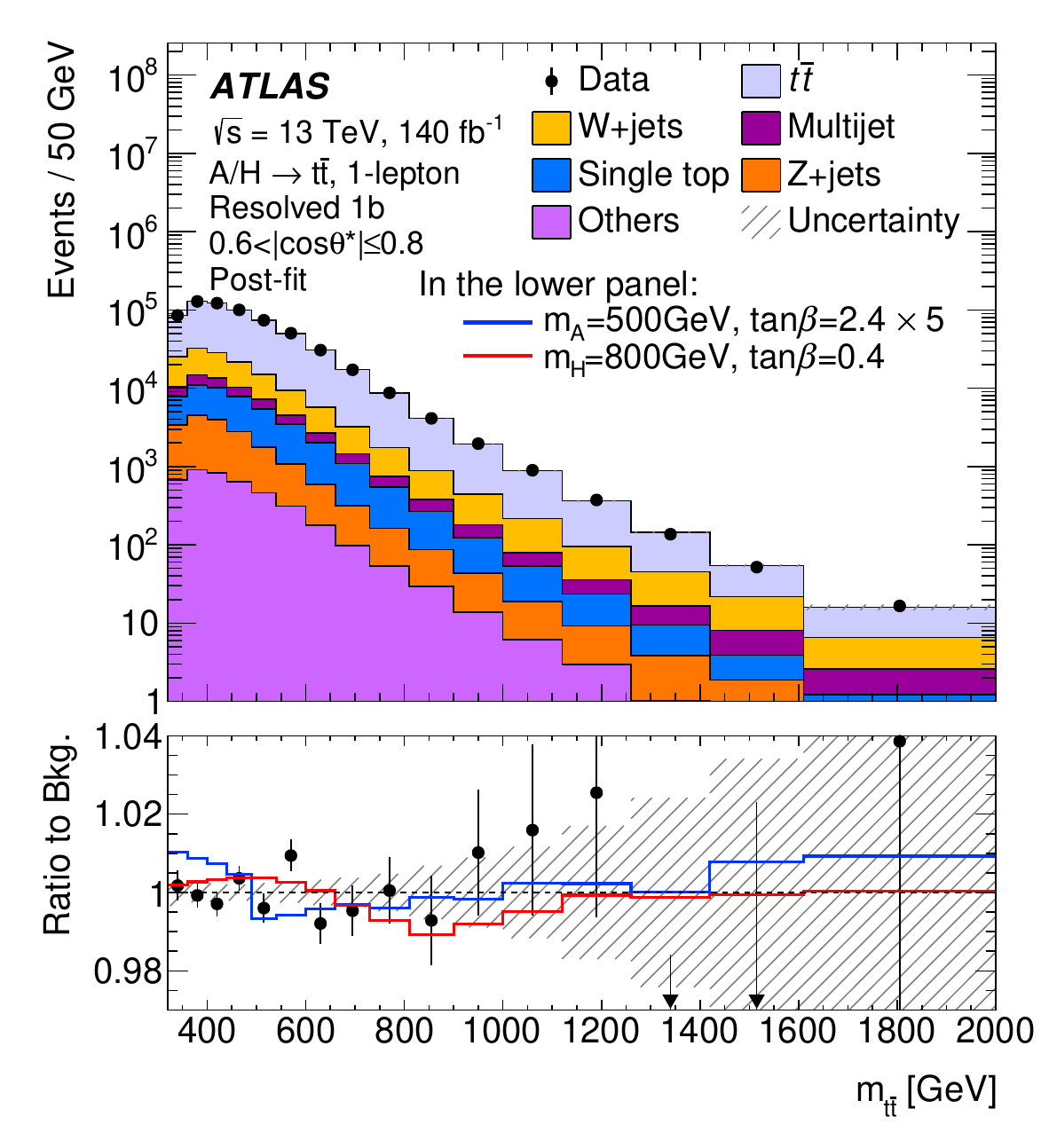}
}\\
\subfloat[]{
\includegraphics[width=0.32\textwidth]{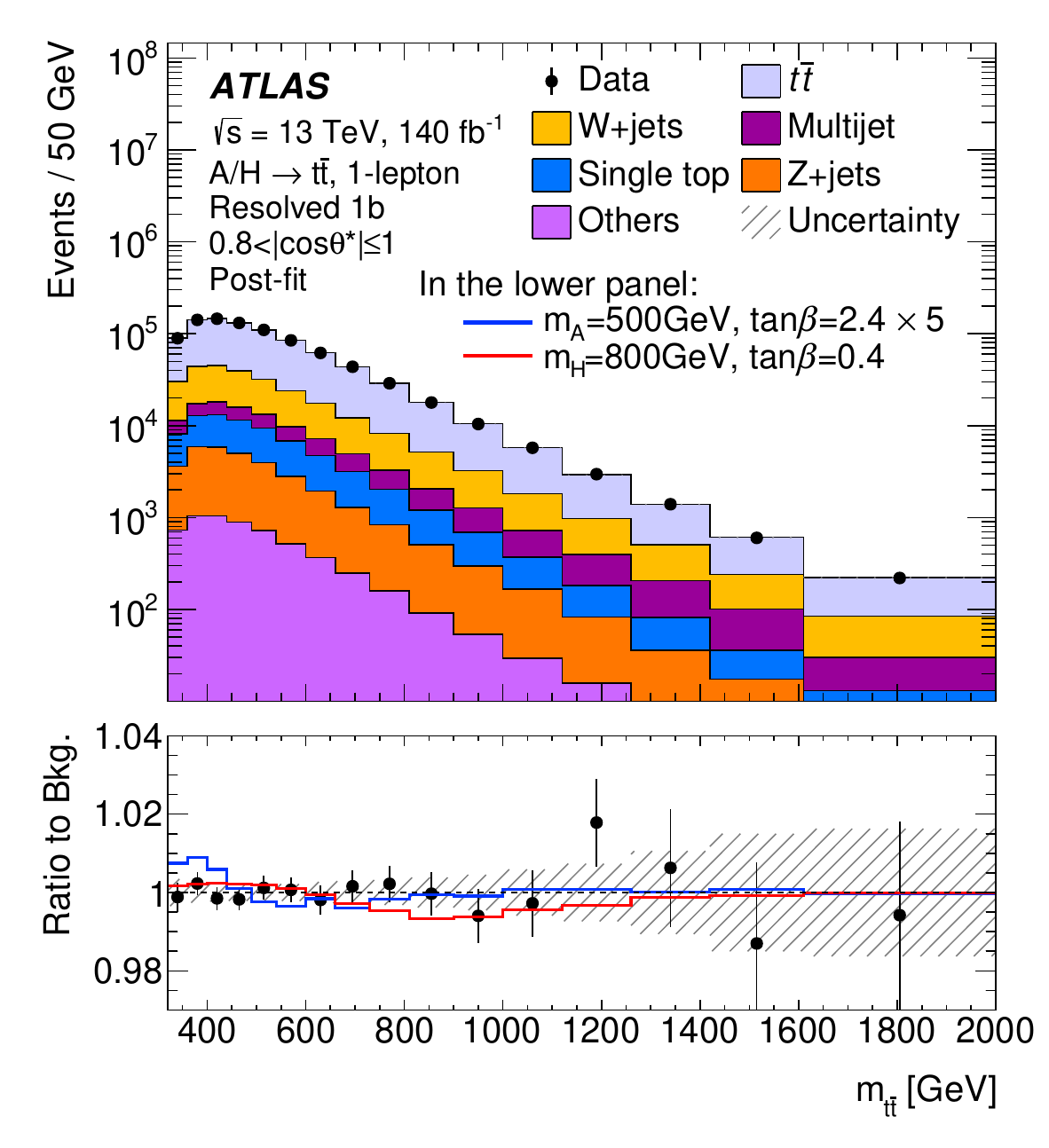}
}
\caption{Post-fit distributions of the reconstructed $m_{t\bar{t}}$ for the five Resolved $1b$ signal regions of the 1-lepton channel
with (a) $|\cos\theta^*| \leq 0.2$, (b) $0.2 < |\cos\theta^*| \leq 0.4$, (c) $0.4 < |\cos\theta^*| \leq 0.6$, (d) $0.6 < |\cos\theta^*| \leq 0.8$, and (e) $0.8 < |\cos\theta^*| \leq 1.0$.
In the lower panels, the ratio of the data and the post-fit prediction are shown (data points).
The expected relative deviation from the background prediction in the presence of an interference pattern, $(S+I+B)/B$,
for two representative signal hypotheses (one of them scaled by a factor of five for better visibility) is also shown in the ratio panel.
\label{fig:mttbar:postfit:final_1b}}
\end{figure}
 
\begin{figure}[t]
\centering
\subfloat[]{
\includegraphics[width=0.32\textwidth]{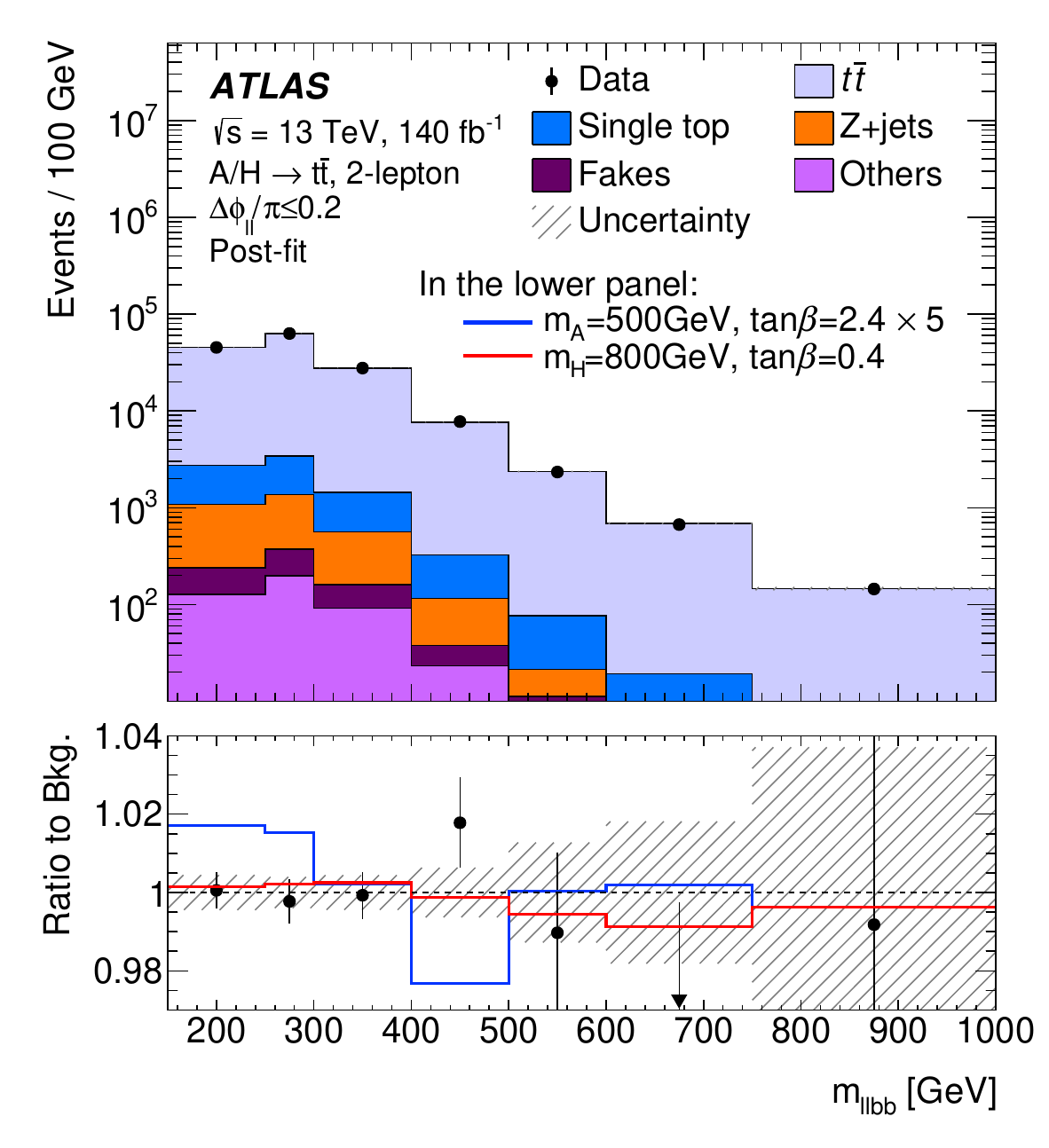}
}
\subfloat[]{
\includegraphics[width=0.32\textwidth]{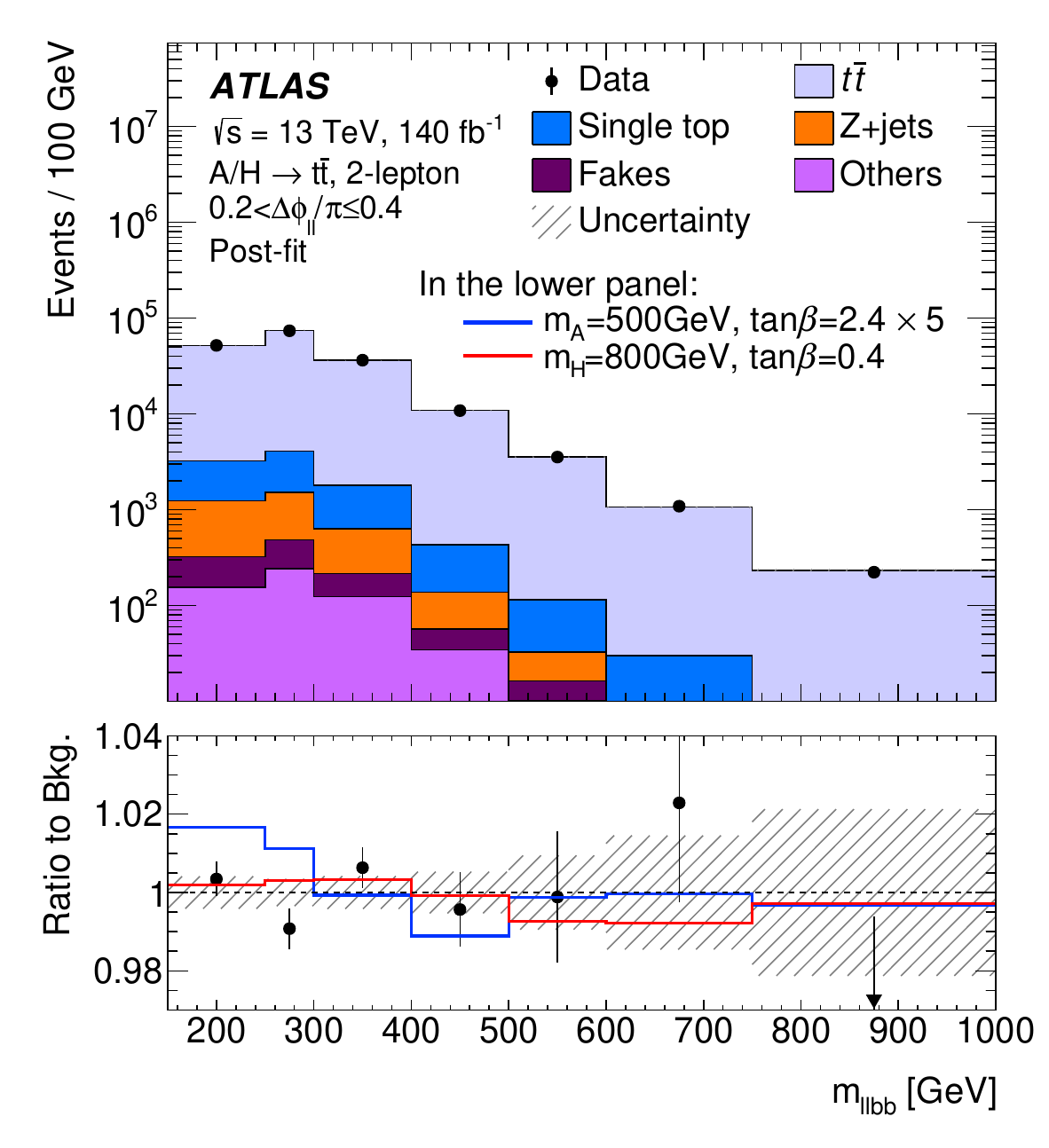}
}\\
\subfloat[]{
\includegraphics[width=0.32\textwidth]{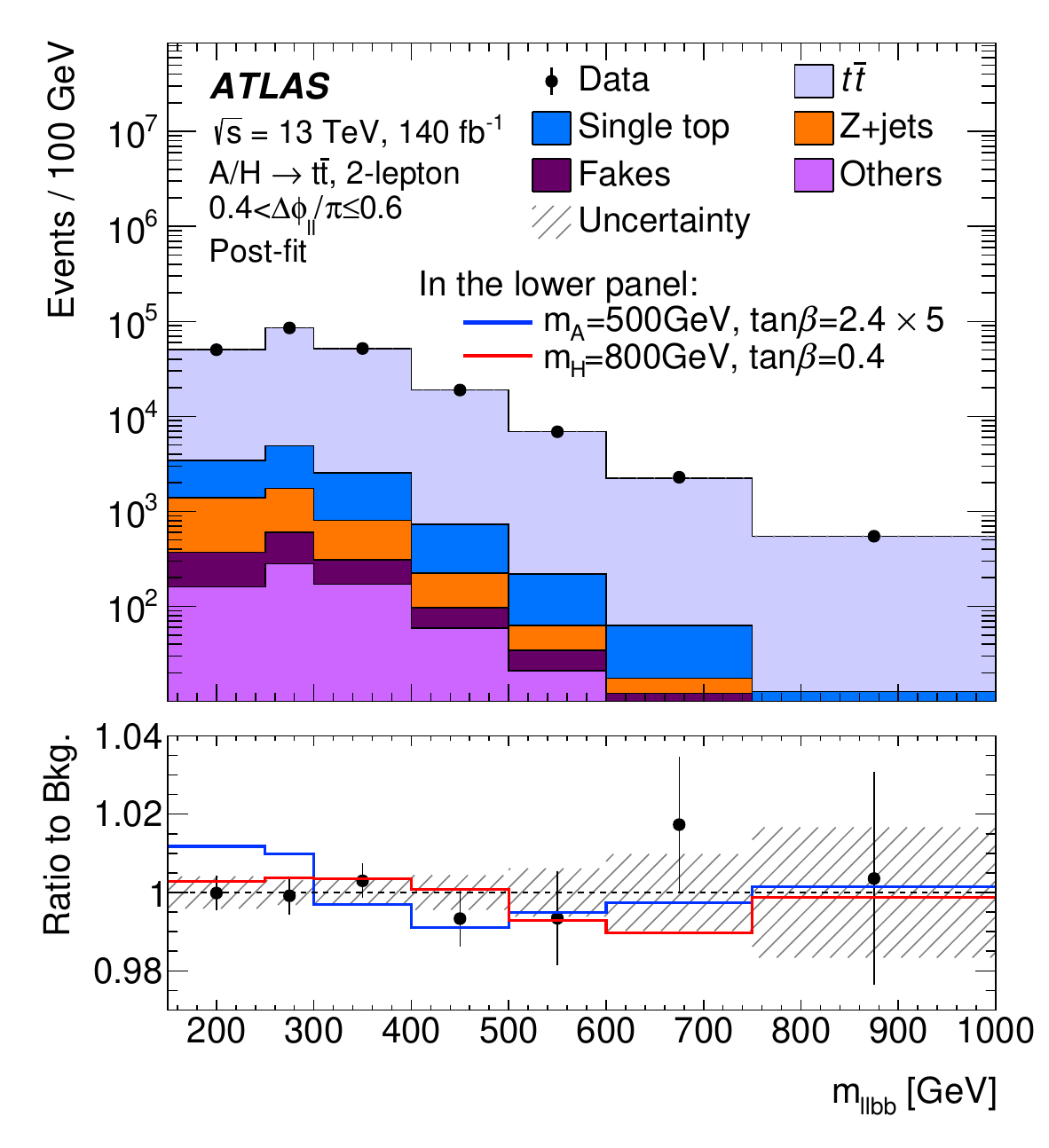}
}
\subfloat[]{
\includegraphics[width=0.32\textwidth]{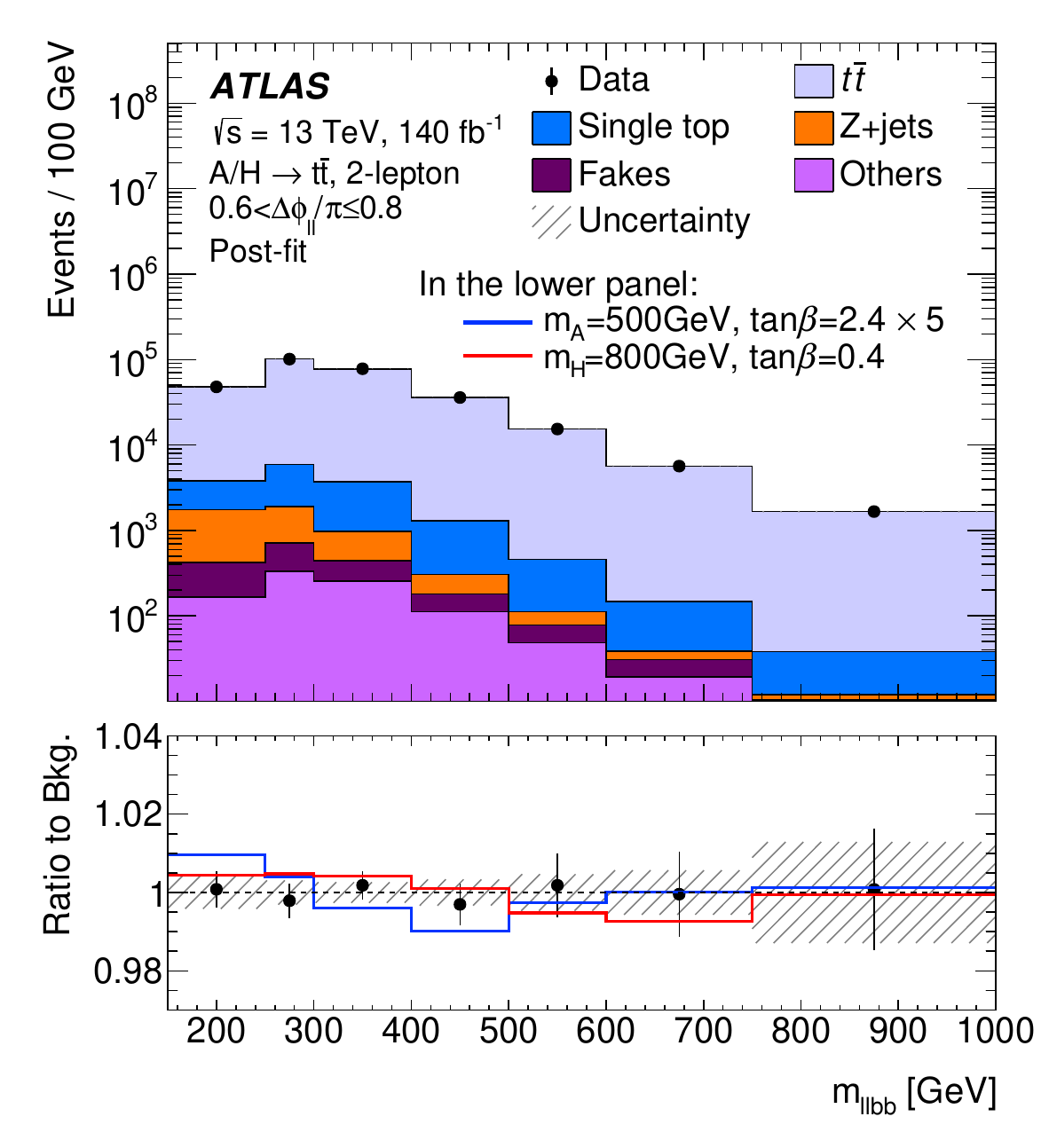}
}\\
\subfloat[]{
\includegraphics[width=0.32\textwidth]{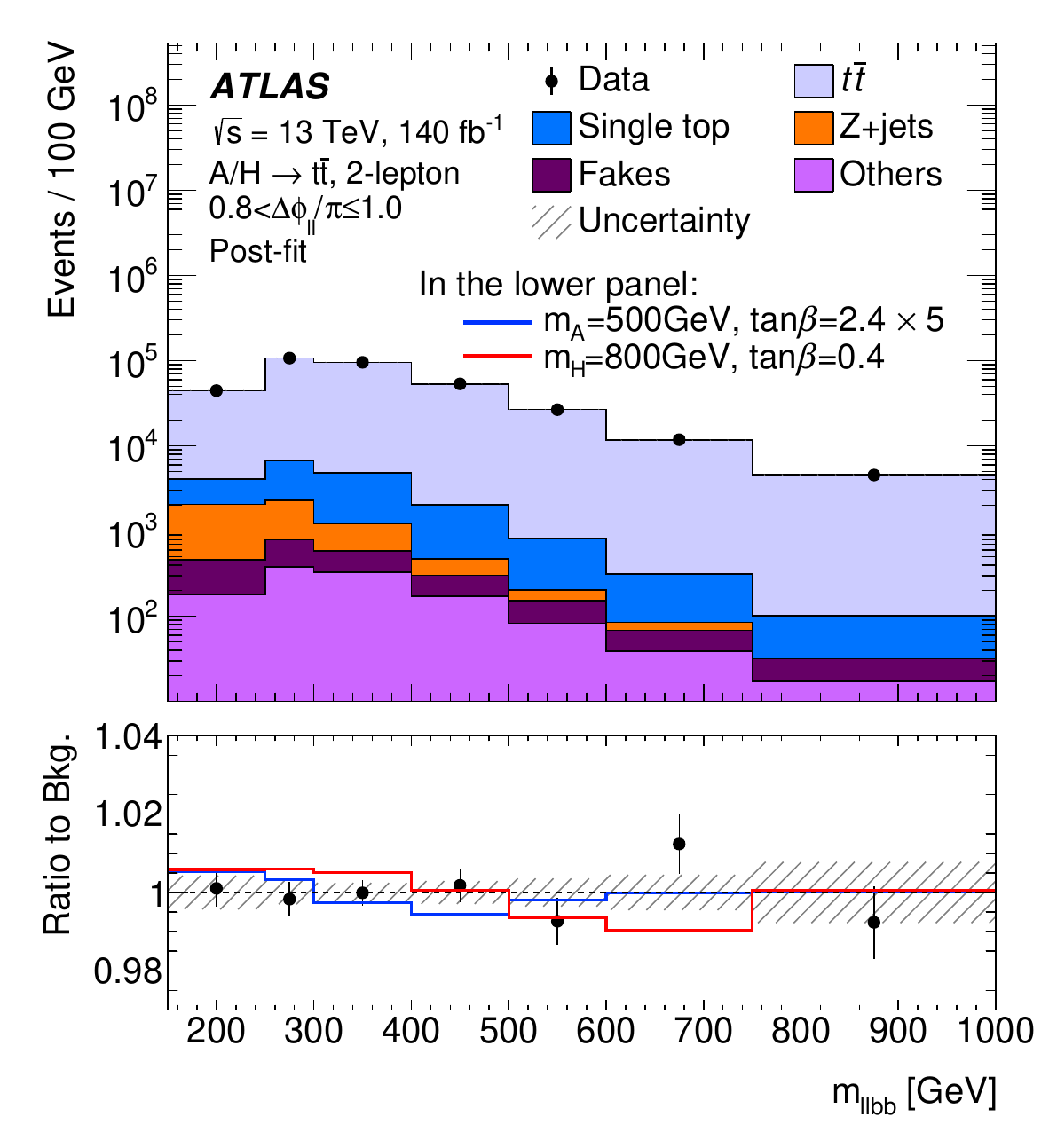}
}\\
\caption{Post-fit distributions of the reconstructed $m_{\ell\ell b b}$ for the five signal regions of the 2-lepton channel
with (a) $\Delta\phi_{\ell\ell}/\pi \leq 0.2$, (b) $0.2 < \Delta\phi_{\ell\ell}/\pi \leq 0.4$, (c) $0.4 < \Delta\phi_{\ell\ell}/\pi \leq 0.6$, (d) $0.6 < \Delta\phi_{\ell\ell}/\pi \leq 0.8$, and (e) $0.8 < \Delta\phi_{\ell\ell}/\pi \leq 1.0$.
In the lower panels, the ratio of the data and the post-fit prediction are shown (data points).
The expected relative deviation from the background prediction in the presence of an interference pattern, $(S+I+B)/B$,
for two representative signal hypotheses (one of them scaled by a factor of five for better visibility) is also shown in the ratio panel.
\label{fig:mttbar:postfit:final_2l}}
\end{figure}

The excluded region at 95\% CL in the $m_{A/H}$-$\tanb$ plane for the hypothesis of both a pseudo-scalar and a scalar with equal masses ($m_A = m_H$) in a type-II 2HDM with $\cos(\beta-\alpha)=0$
is shown in Figure~\ref{fig:limits_PHI_2hdm}. The exclusion region for the hMSSM is shown in Figure~\ref{fig:limits_PHI_hmssm}.
In both cases, the interference patterns for both $A$ and $H$ are assumed to be present in the spectra of the fitted variables.
Values of \tanb smaller than 3.49 (3.52) are observed (expected) to be excluded for $m_A = m_H = 400$\,GeV in the 2HDM.
In the hMSSM, values of \tanb smaller than 3.16 (3.37) are observed (expected) to be excluded for $m_A = 400$\,GeV.
Masses up to 1240~GeV (1210~GeV) are observed (expected) to be excluded for the lowest tested \tanb value of 0.4 in the 2HDM.
In the hMSSM, masses up to 950~GeV (830~GeV) are observed (expected) to be excluded for $\tanb=1.0$.
In both scenarios, the observed exclusion is stronger than the expected exclusion by about $2\sigma$ in the mass region $m_A = m_H \approx 850$~GeV.
This deviation is consistent with the location of the largest local excess in the data, which is found for $\mttbar \approx 850$~GeV, as mentioned above.
In the type-II 2HDM and the hMSSM, a local deficit compared to the SM expectation is predicted, hence the local excess in the data leads to a stronger-than-expected exclusion of these BSM scenarios
in this mass region.
The sensitivity of the search is dominated by the 1-lepton channel due to its larger branching ratio.
The sensitivity improvement from the statistical combination with the 2-lepton channel is largest for low values of $m_{A/H}$.
For $m_{A/H}=400$~GeV, the observed (expected) exclusion range in \tanb is 11\% (5\%) larger for the combined 1- and 2-lepton channels
compared to the exclusion obtained with the 1-lepton channel alone.
 
Constraints are also derived for the 2HDM+$a$ benchmark, specifically for benchmark scenarios 2a and 2b defined in Ref.~\cite{EXOT-2018-64}.
In these benchmark scenarios, the value of the mixing angle is chosen such that $\sint = 0.35$ and 0.7, respectively.
The masses of the additional neutral and charged Higgs bosons of the 2HDM are set to 600~GeV ($m_A = m_H = m_{H^{\pm}}$), while the mass of the DM particle is set to 1 GeV.
Exclusion limits are derived as a function of the mass of the pseudo-scalar mediator, $m_a$ and \tanb (Figure~\ref{fig:limits_2d_2hdma}).
They only show a moderate dependence on $m_a$, which results from a decrease in the production cross-section of the mediator $a$ with increasing $m_a$
combined with an increase of the branching ratio for the decay $A\rightarrow \ttbar$. The latter is related to the fact that the branching ratio for the decay $A\rightarrow ah$ decreases with $m_a$.
Values of \tanb up to almost 1.1 (0.9) are excluded in the probed $m_a$ range for the benchmark scenario with $\sint=0.35$ (0.70).
The lower sensitivity in the case of the scenario with the larger $a-A$ mixing parameter \sint is due to the fact that the branching ratio for the decay $A\rightarrow \ttbar$
is smaller for larger $a-A$ mixing as the mixing increases the branching ratio for the invisible decay $A\rightarrow \chi\chi$.
The constraints on the 2HDM+$a$ can be compared with the constraints from other searches on the same dataset, summarised in Ref.~\cite{EXOT-2018-64}, which represent the most stringent constraints
on the 2HDM+$a$ to date. The search presented in this paper provides the most stringent expected constraints on \tanb for $m_a \gtrsim 400$~GeV for the benchmark scenarios under consideration,
while the observed constraints are slightly weaker than those obtained from a search for charged Higgs bosons decaying into a top and a bottom quark ($tbH^{\pm}(tb)$), which provides the strongest
observed exclusion among all individual searches in this region.
In this context, it is worth noting that the cross-section for $tbH^{\pm}$ production in the 2HDM+$a$ is known at NLO in QCD~\cite{EXOT-2018-64},
while, as pointed out in Section~\ref{sec:datamc}, the cross-section for $pp \rightarrow a/A/H \rightarrow \ttbar$ production is only known at LO accuracy,
which leads to more conservative constraints from this search.
 
For the generic benchmark model, the constraints on the coupling modifiers for the pseudo-scalar $A$ and the
scalar $H$ to \ttbar are shown separately as a function of the (pseudo-)scalar mass in Figures~\ref{fig:limits_2d_mod_ind_a} and~\ref{fig:limits_2d_mod_ind_h}, respectively.
In these cases, only the interference pattern for either $A$ or $H$ is assumed to be present in the spectra of the fitted variables.
The constraints are derived for different, fixed values of the relative total width $\Gamma_{A/H}/m_{A/H}$.
Given that the partial width $\Gamma(A/H\rightarrow \ttbar)$ is proportional to $g_{A/H\ttbar}$, it can exceed the total width in some regions of the shown phase space.
These unphysical regions are marked by hatched lines.
The constraints on the coupling modifier $g_{A/Ht\bar{t}}$ for signal hypotheses with a single pseudo-scalar are more stringent than for those
with a single scalar of the same width due to the fact that the production cross-sections are generally larger for the pseudo-scalar compared to the scalar case for the same mass and width.
Further differences in the exclusion regions arise due to differences in the interference patterns of scalars and pseudo-scalars. In the former case, the peak is generally narrower and located
closer to the dip.
The observed constraints on the coupling strength are slightly weaker than expected for $m_{A/H}\approx 850$~GeV,
especially in the scenarios with $\Gamma_{A/H}/m_{A/H} = 5\%$ and $\Gamma_{A/H}/m_{A/H}=10\%$, consistent with the small narrow excess of events in data
compared to the SM expectation observed in this region. This weaker-than-expected exclusion can be explained by the fact that for $g_{A/H\ttbar}>1$
the pure-signal component $S$, which scales like $g_{A/H\ttbar}^4$ (Eq.~(\ref{eq:generic})), is enhanced over the interference component, which scales like $g_{A/H\ttbar}^2$.
This means that larger values of $g_{A/H\ttbar}$ correspond to interference patterns with a shallower dip or, for very large couplings, even a peak-peak instead of a peak-dip structure.
Such patterns are more compatible with a narrow excess in the data than the interference patterns with a more pronounced dip obtained for smaller values of $g_{A/H\ttbar}$.
A small "island" occurs in the observed exclusion contour for the scenario with a single pseudo-scalar with total width $\Gamma_{A}=5\%$ in the region around $m_A=800$~GeV,
for which coupling values around $g_{A\ttbar}\approx 0.9$ are observed to be excluded but slightly larger values of $g_{A\ttbar}$ are not (Figure~\ref{fig:limits_A_5}).
This "island" can be explained by the fact that \CLs value for this choice of $m_A$ and $\Gamma_{A}$ exhibits a local minimum around $g_{A\ttbar}\approx 0.9$ when expressed as a function of
$g_{A\ttbar}$, similar to the scenario shown in Figure~\ref{fig:CLsScan}, where $\sqrt{\mu}$ can be read as the equivalent of $g_{A/Ht\bar{t}}^2$ (compare also Eqs.~(\ref{eq:generic})
and~(\ref{eq:LH})).
This local minimum causes the \CLs value for a narrow range of $g_{A/Ht\bar{t}}$ to fall below the exclusion threshold of 0.05.
Similarly, the "hole" in the observed exclusion contour for the scenario with a single pseudo-scalar with total width $\Gamma_{A}=10\%$ in the region around $m_A=800$~GeV,
for which coupling values in a narrow range around $g_{A\ttbar}\approx 1.8$ are not excluded (Figure~\ref{fig:limits_A_10}),
can be explained by a local maximum in the \CLs scan around $g_{A\ttbar}\approx 1.8$.
 
\begin{figure}[t]
\centering
\subfloat[]{
\includegraphics[width=0.5\textwidth]{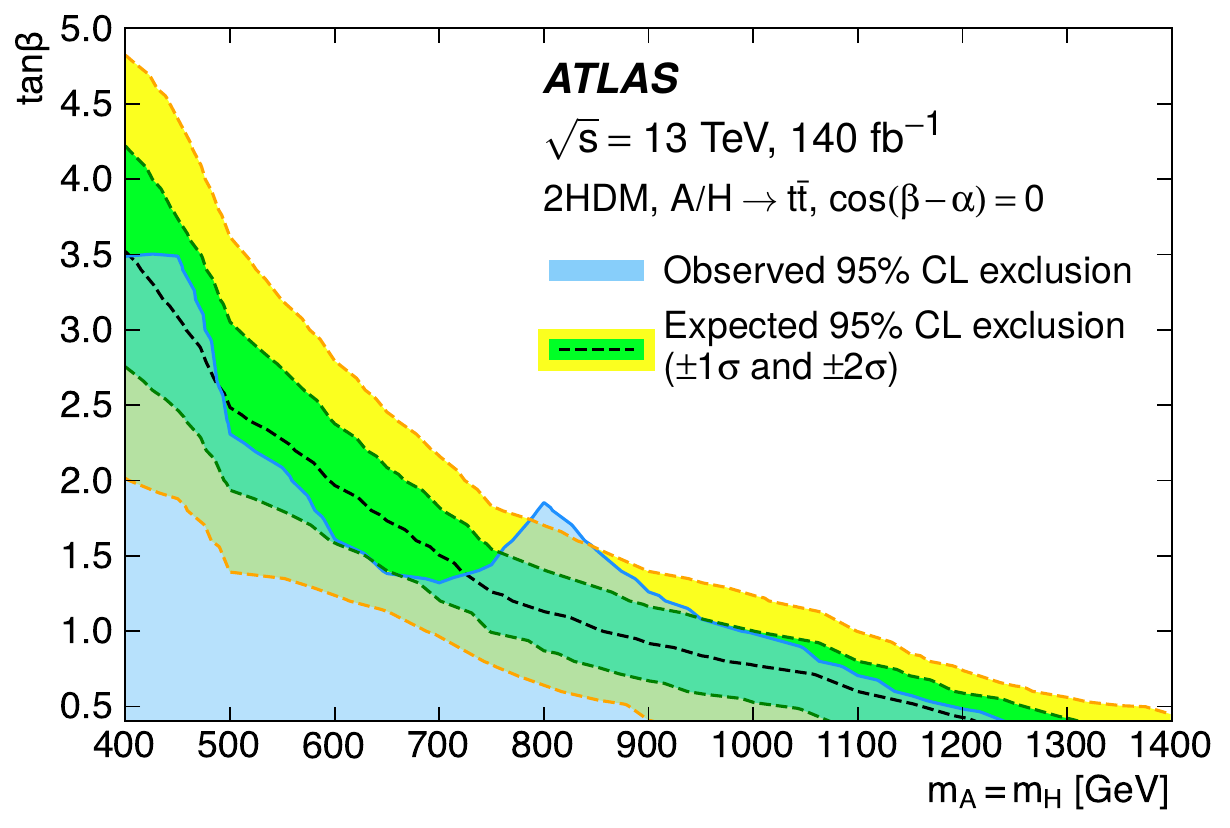}
\label{fig:limits_PHI_2hdm}
}
\subfloat[]{
\includegraphics[width=0.5\textwidth]{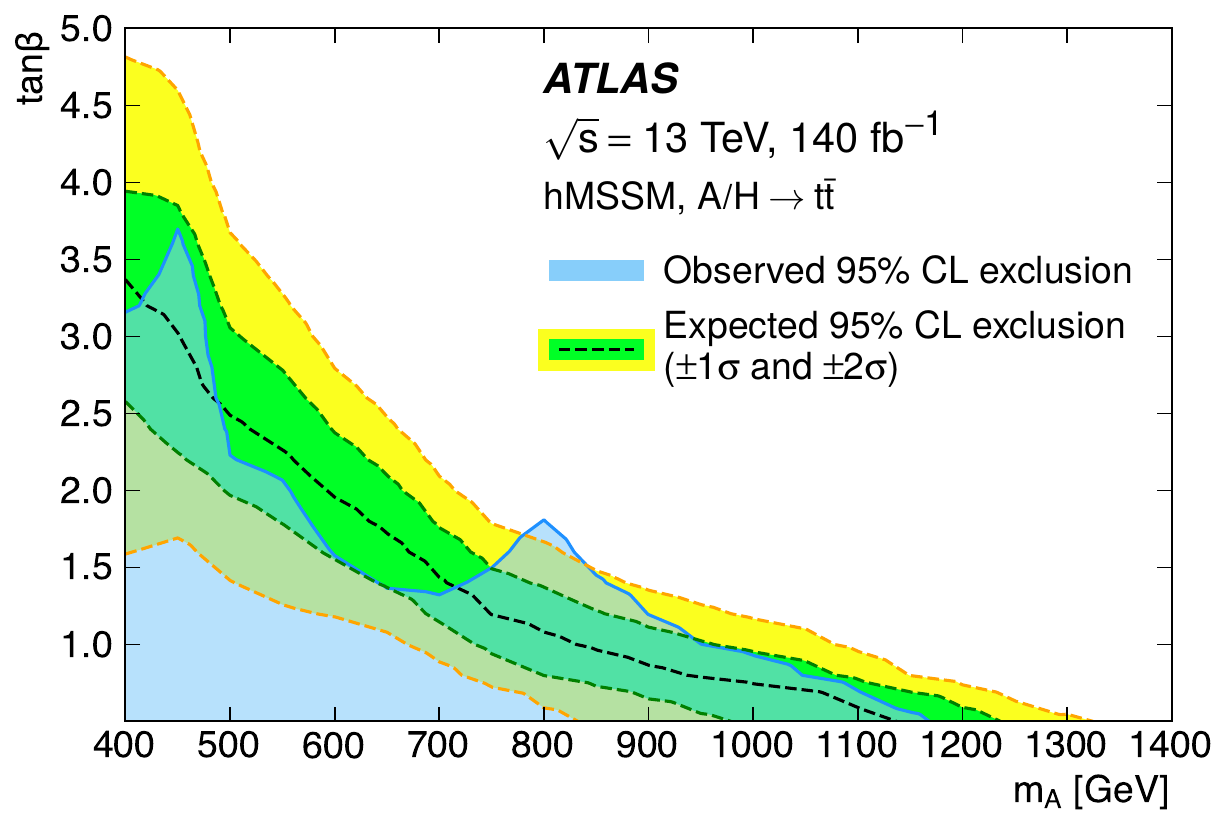}
\label{fig:limits_PHI_hmssm}
}
\caption{Observed and expected exclusion contours in the $m_{A/H}-\tan\beta$ plane for (a) a type-II 2HDM in the alignment limit ($\cos(\beta-\alpha)=0$)
with mass-degenerate pseudo-scalar and scalar states, $m_A = m_H$ and (b) the hMSSM.
The observed exclusion regions are indicated by the shaded area.
The boundary of the expected exclusion region under the background-only hypothesis is marked by the dashed line.
The surrounding shaded bands correspond to the $\pm 1$ and $\pm 2$ standard deviation $(\pm 1\sigma, \pm 2\sigma)$ uncertainty.
}
\end{figure}

\begin{figure}[t]
\centering
\subfloat[]{
\includegraphics[width=0.5\textwidth]{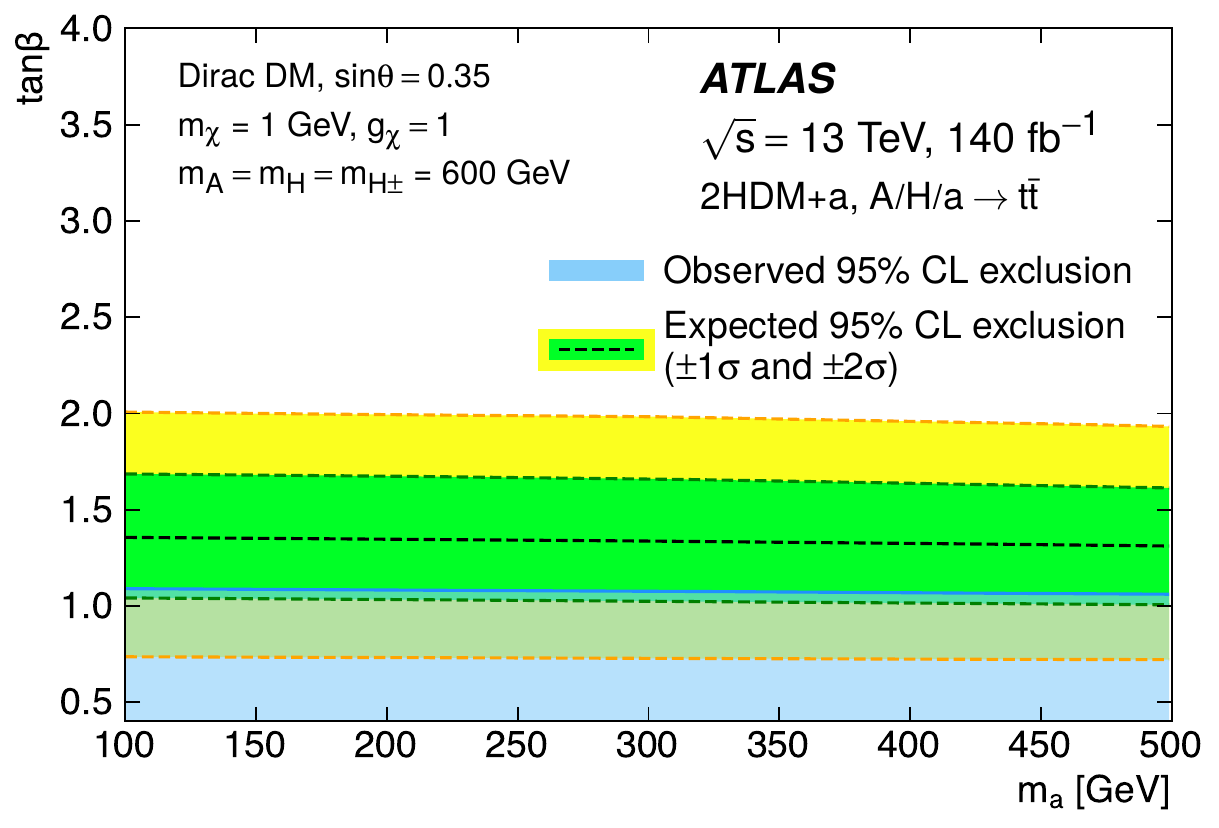}
}
\subfloat[]{
\includegraphics[width=0.5\textwidth]{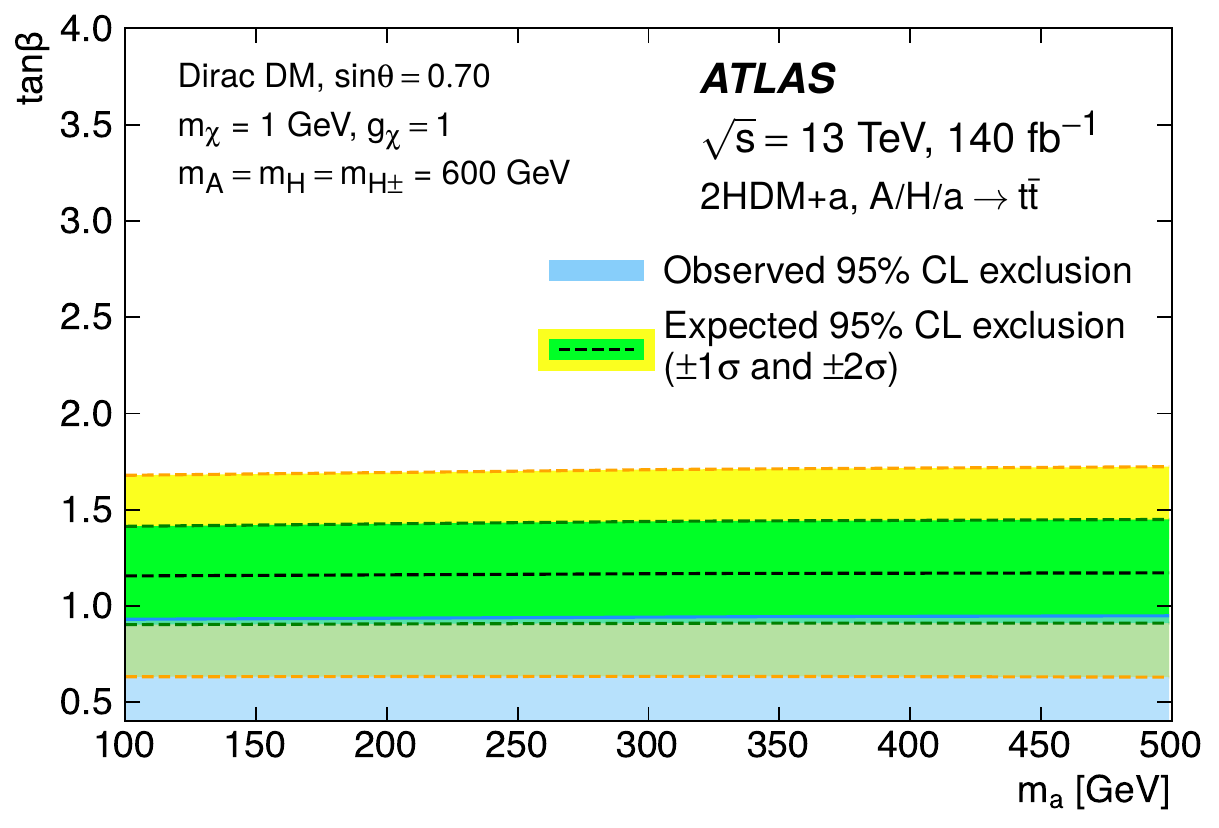}
}
\caption{Observed and expected exclusion contours in the $m_{a}-\tan\beta$ plane for the 2HDM+$a$ with $m_A = m_H = 600$~GeV for (a) $\sin\theta = 0.35$ and (b) $\sin\theta = 0.70$.
The model settings correspond to those of benchmark scenarios 2a and 2b defined in Ref.~\cite{EXOT-2018-64}, respectively.
The observed exclusion regions are indicated by the shaded area.
The boundary of the expected exclusion region under the background-only hypothesis is marked by the dashed line.
The surrounding shaded bands correspond to the $\pm 1$ and $\pm 2$ standard deviation $(\pm 1\sigma, \pm 2\sigma)$ uncertainty.
}
\label{fig:limits_2d_2hdma}
\end{figure}

\begin{figure}[t]
\centering
\subfloat[]{
\includegraphics[width=0.5\textwidth]{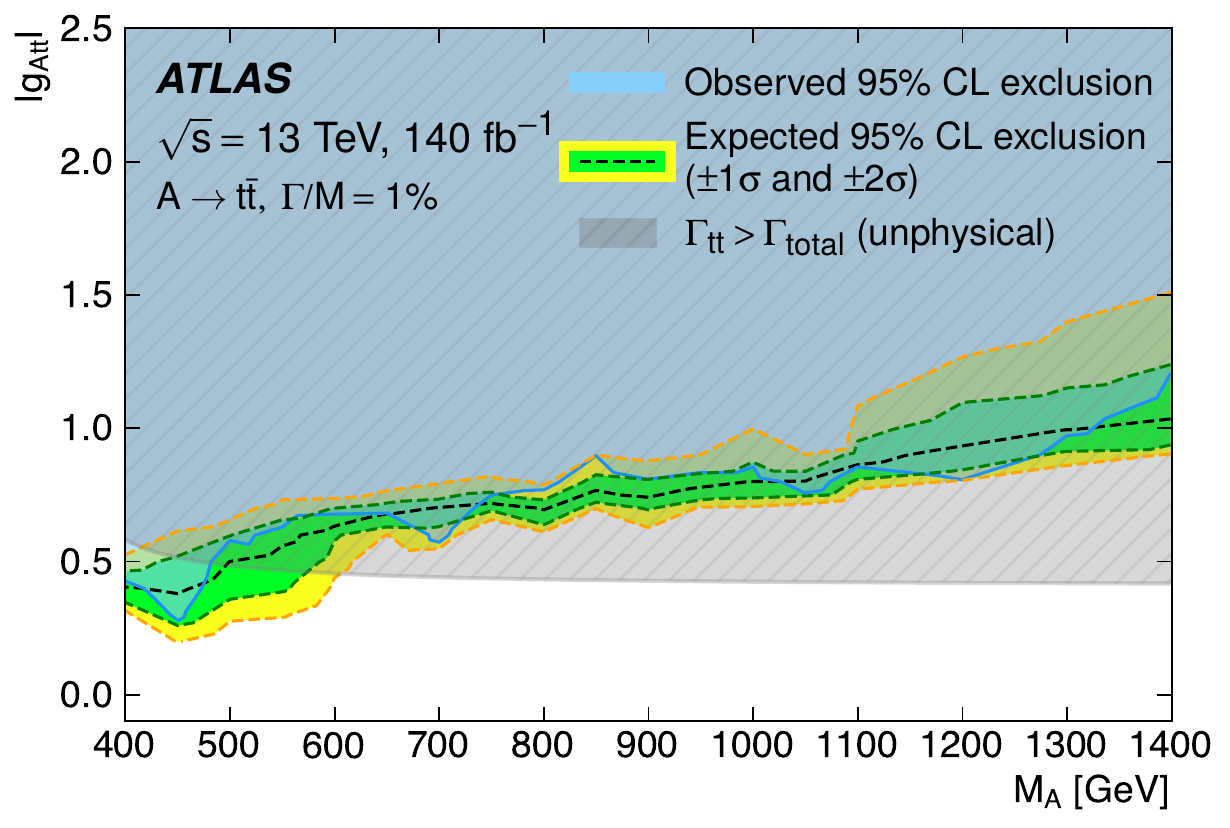}
}
\subfloat[]{
\includegraphics[width=0.5\textwidth]{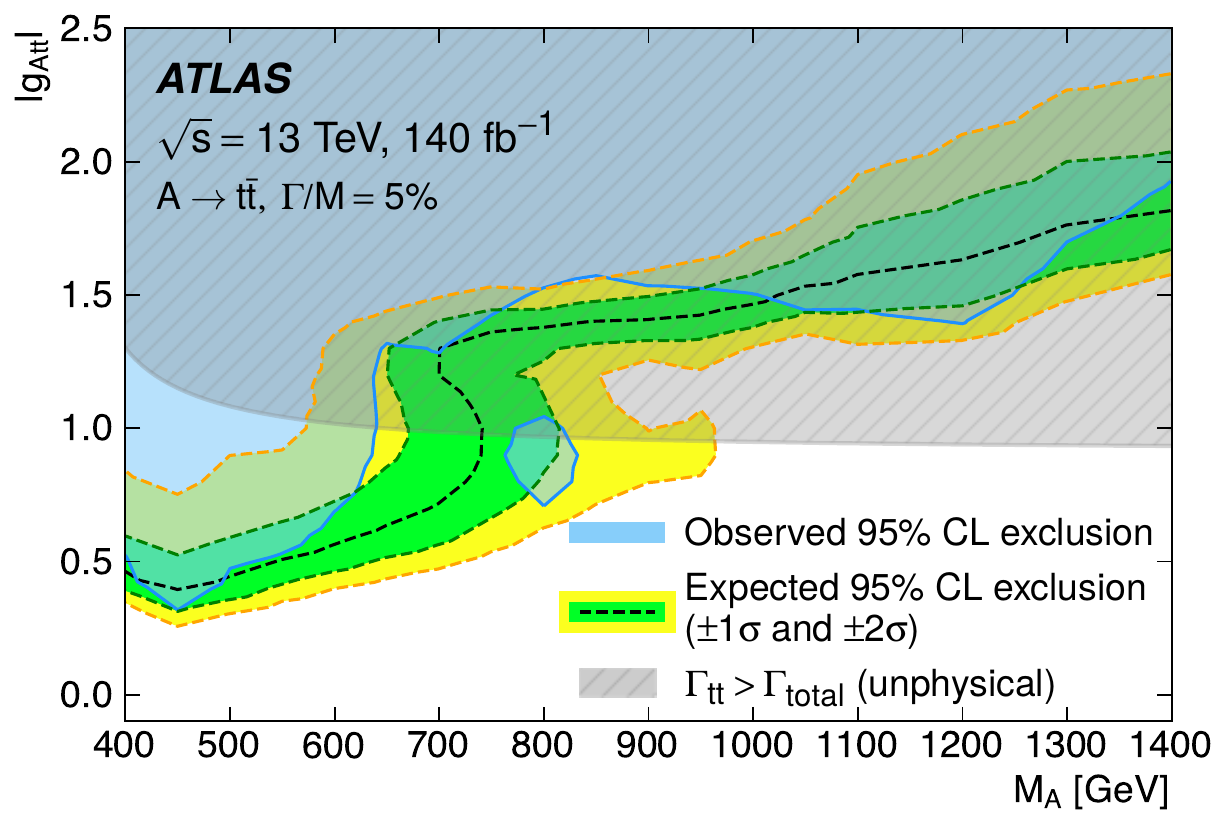}
\label{fig:limits_A_5}
}\\
\subfloat[]{
\includegraphics[width=0.5\textwidth]{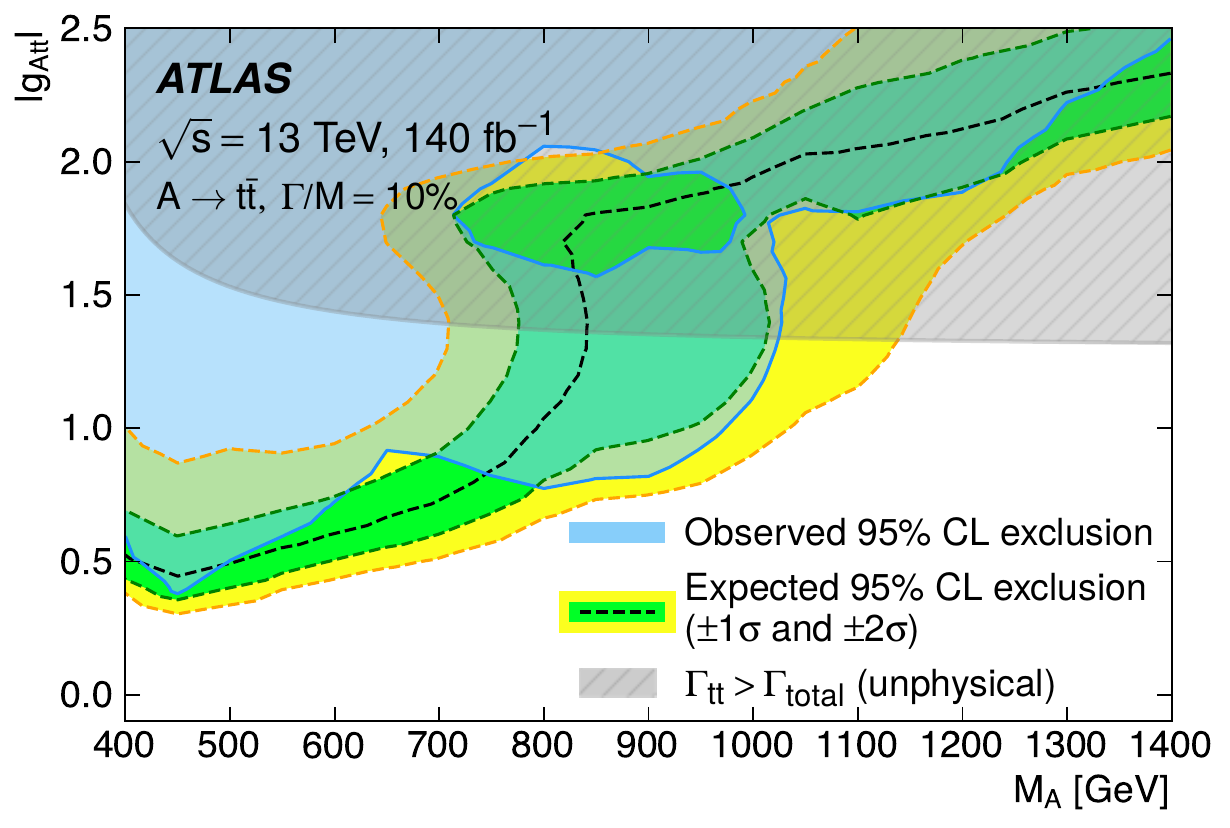}
\label{fig:limits_A_10}
}
\subfloat[]{
\includegraphics[width=0.5\textwidth]{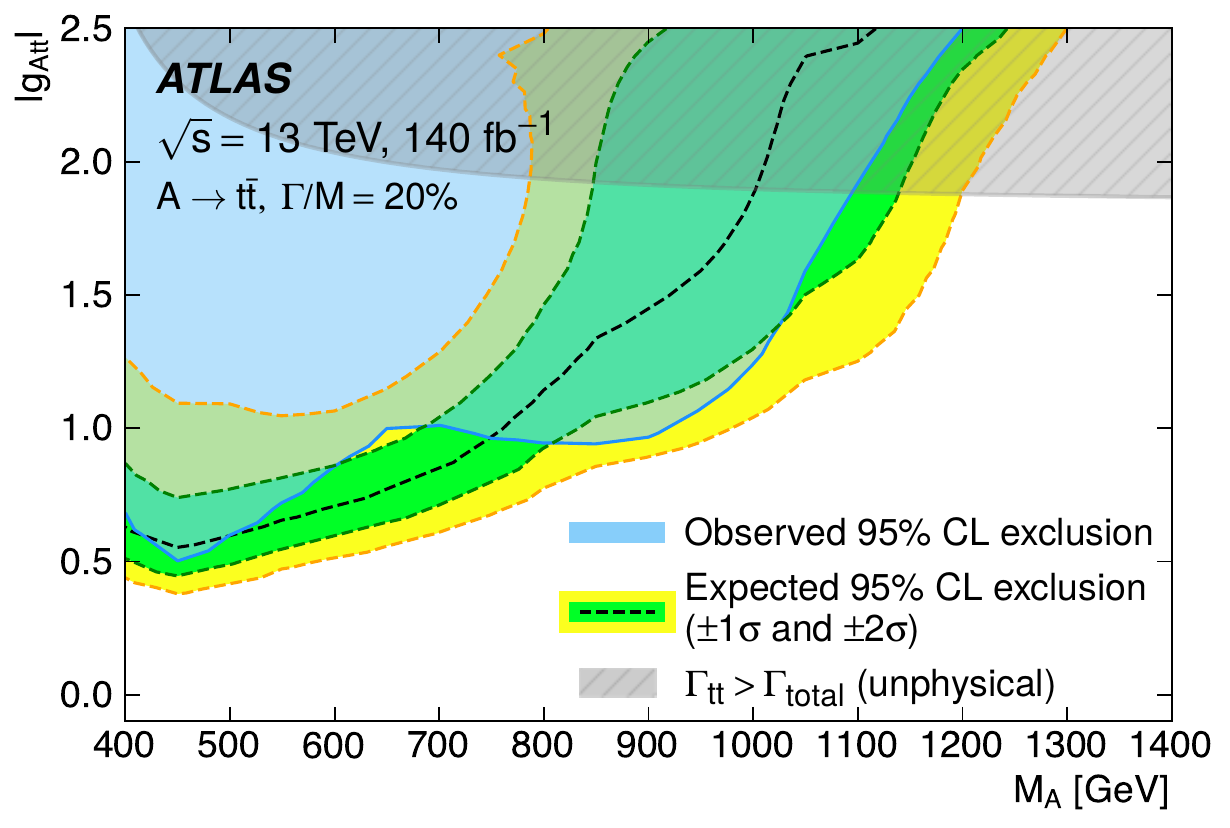}
}\\
\subfloat[]{
\includegraphics[width=0.5\textwidth]{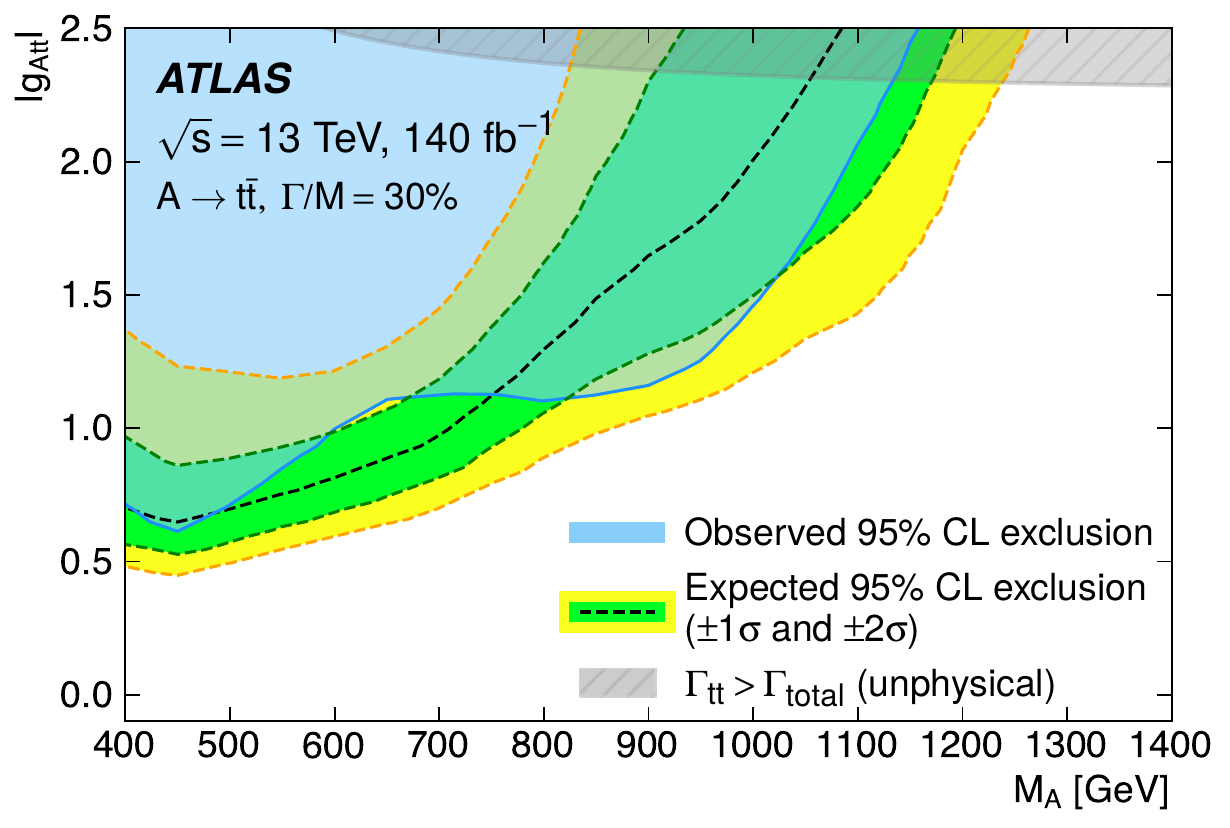}
}
\subfloat[]{
\includegraphics[width=0.5\textwidth]{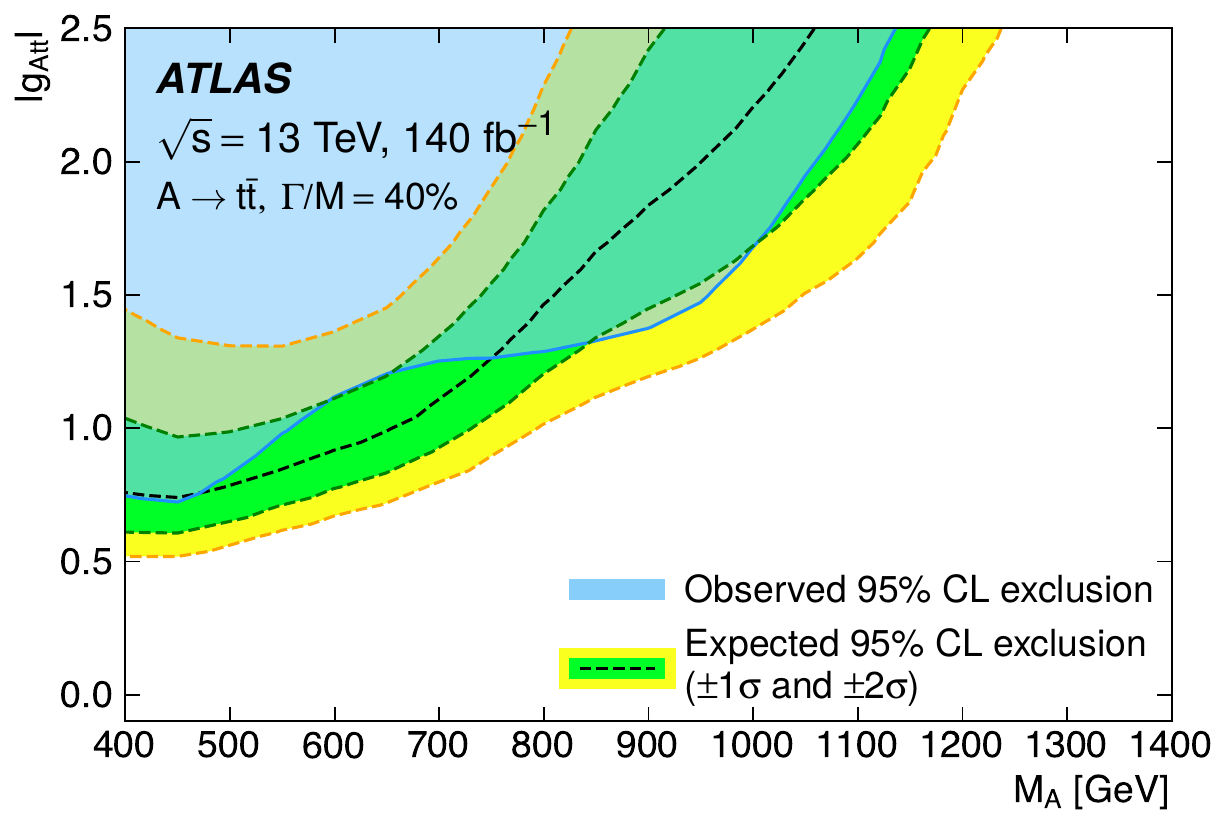}
}\\
\caption{Constraints on the coupling strength modifier $g_{At\bar{t}}$ as a function of $m_A$ for different values of the relative width of the pseudo-scalar $A$:
(a) 1\%, (b) 5\%, (c) 10\%, (d) 20\%, (e) 30\%, (f) 40\%.
The observed exclusion regions are indicated by the shaded area.
The boundary of the expected exclusion region under the background-only hypothesis is marked by the dashed line.
The surrounding shaded bands correspond to the $\pm 1$ and $\pm 2$ standard deviation $(\pm 1\sigma, \pm 2\sigma)$ uncertainty.
The hatched area indicates the unphysical region of phase space where the partial width $\Gamma(A\rightarrow t\bar{t})$ is larger than the total width of $A$.
}
\label{fig:limits_2d_mod_ind_a}
\end{figure}

\begin{figure}[t]
\centering
\subfloat[]{
\includegraphics[width=0.5\textwidth]{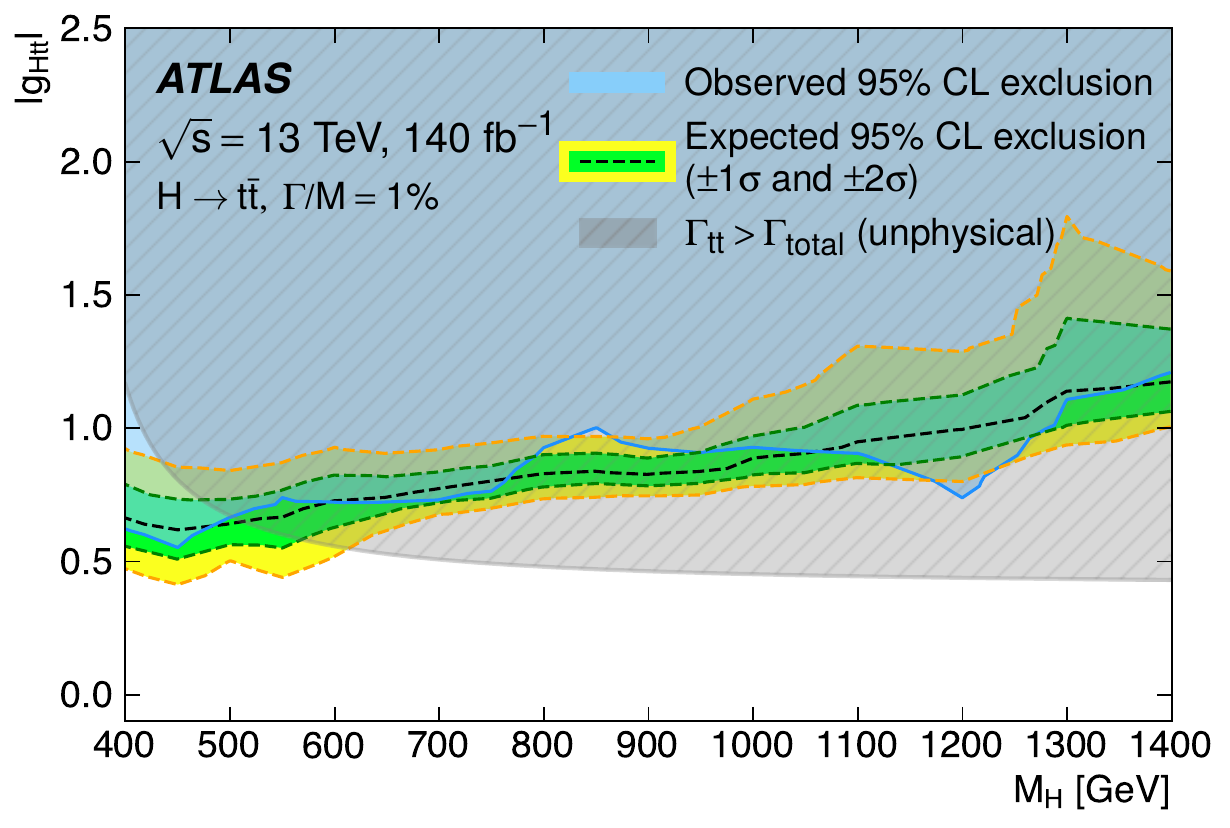}
}
\subfloat[]{
\includegraphics[width=0.5\textwidth]{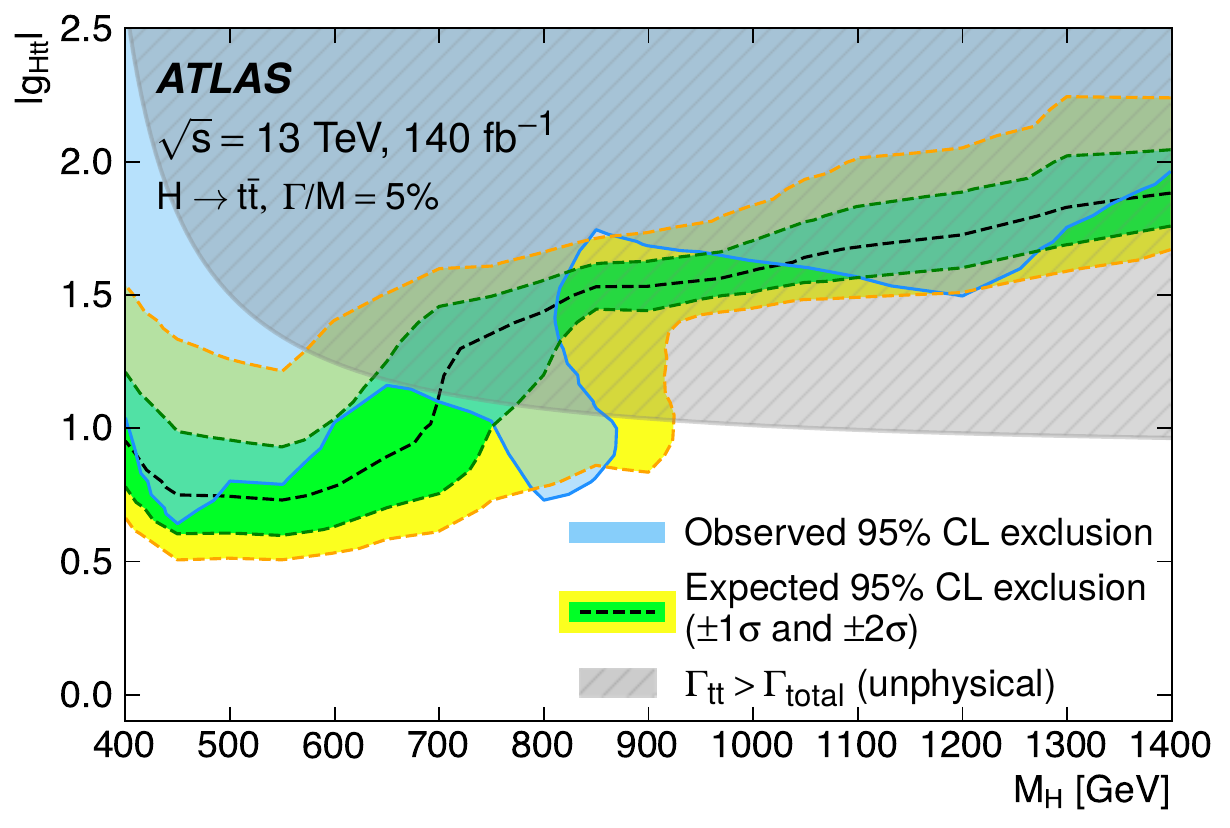}
}\\
\subfloat[]{
\includegraphics[width=0.5\textwidth]{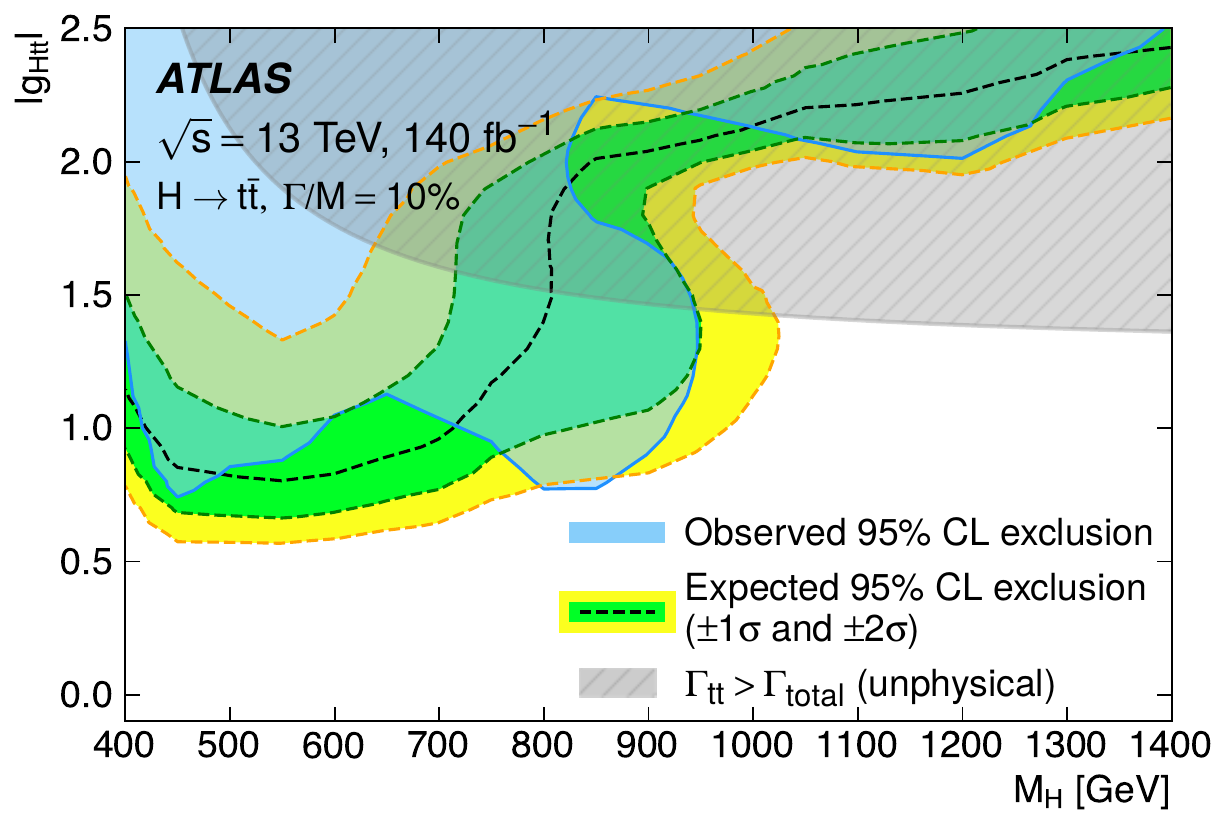}
}
\subfloat[]{
\includegraphics[width=0.5\textwidth]{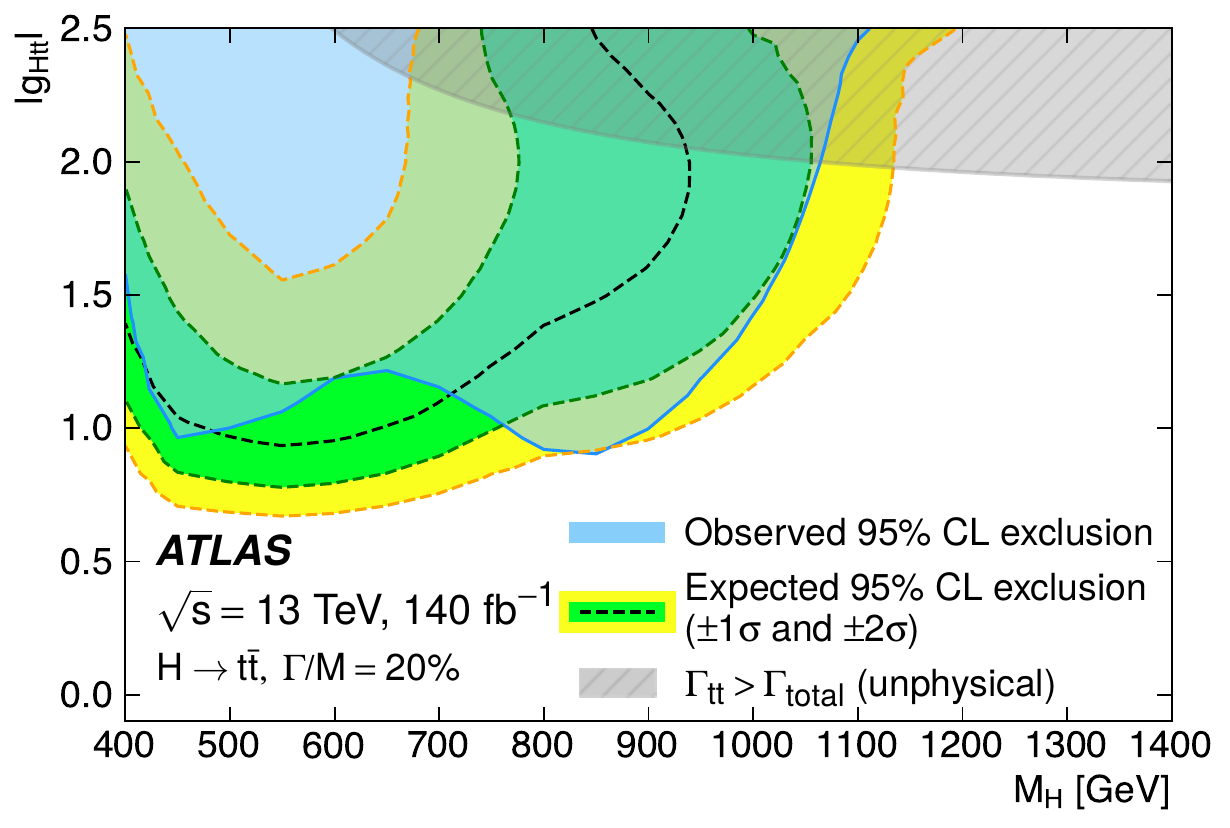}
}\\
\subfloat[]{
\includegraphics[width=0.5\textwidth]{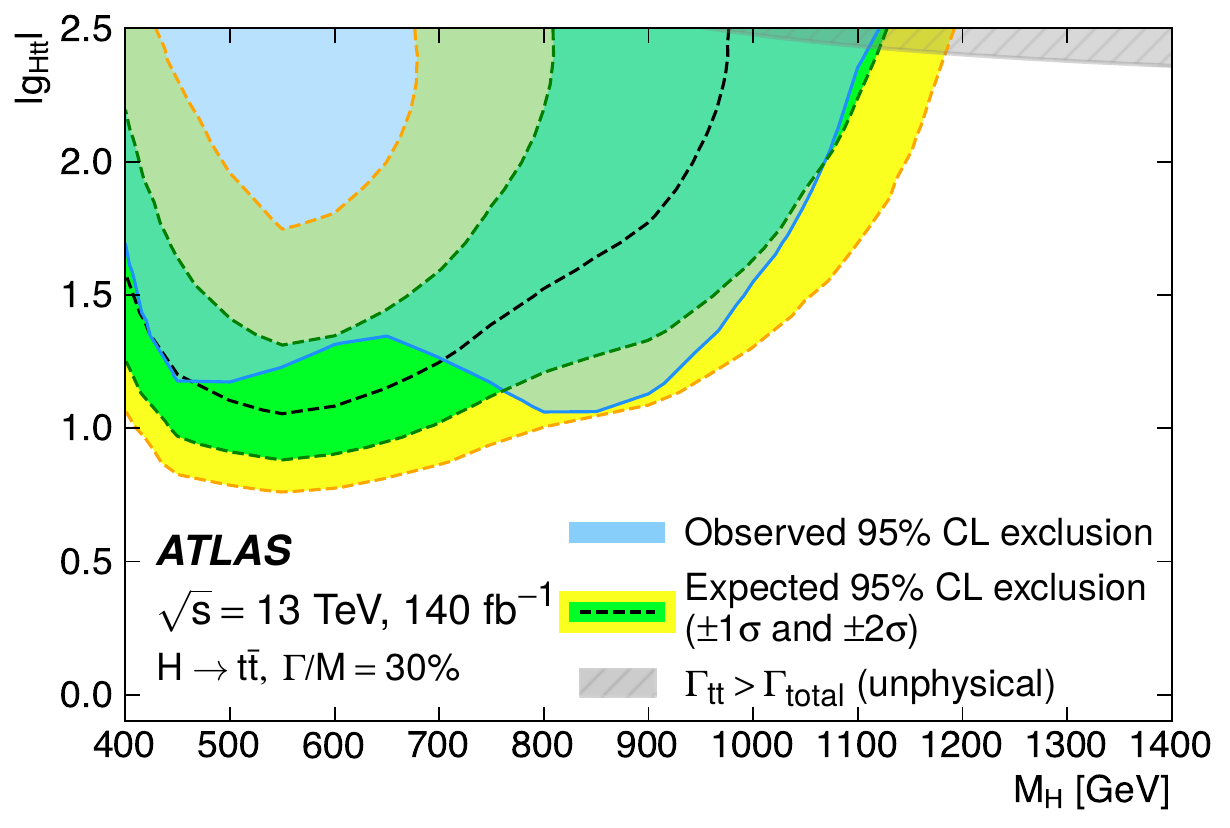}
}
\subfloat[]{
\includegraphics[width=0.5\textwidth]{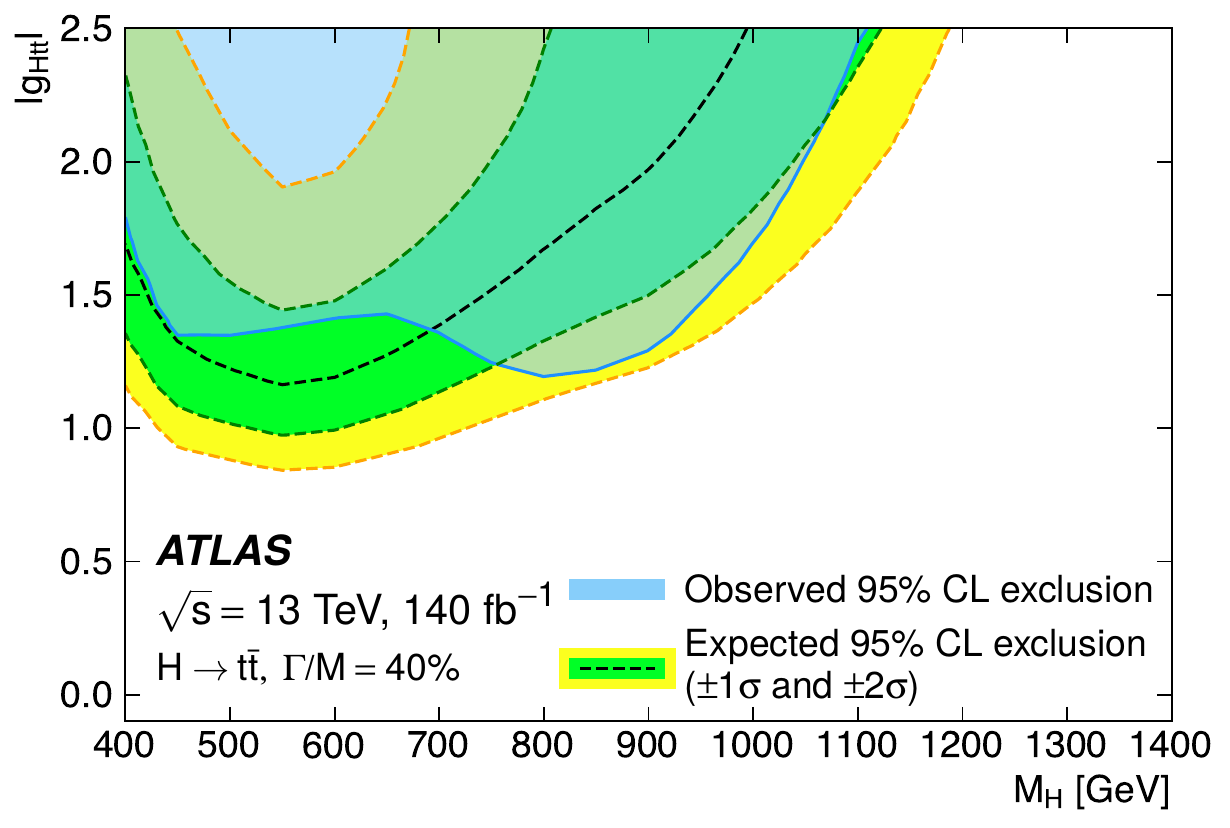}
}\\
\caption{Constraints on the coupling strength modifier $g_{Ht\bar{t}}$ as a function of $m_H$ for different values of the relative width of the scalar $H$:
(a) 1\%, (b) 5\%, (c) 10\%, (d) 20\%, (e) 30\%, (f) 40\%.
The observed exclusion regions are indicated by the shaded area.
The boundary of the expected exclusion region under the background-only hypothesis is marked by the dashed line.
The surrounding shaded bands correspond to the $\pm 1$ and $\pm 2$ standard deviation $(\pm 1\sigma, \pm 2\sigma)$ uncertainty.
The hatched area indicates the unphysical region of phase space where the partial width $\Gamma(H\rightarrow t\bar{t})$ is larger than the total width of $H$.
}
\label{fig:limits_2d_mod_ind_h}
\end{figure}


\FloatBarrier
 
\section{Conclusion}

A search for massive pseudo-scalar and scalar resonances decaying into \ttbar has been conducted on 140~\ifb of $pp$ collision data at $\sqrt{s}=13$~TeV
recorded by the ATLAS experiment at the LHC.
The non-negligible interference between the signal and the main background from SM \ttbar production is taken into account.
The search targets events with a pair of top quarks whose decay leads to a semileptonic or dileptonic final state.
For the semileptonic decays, separate analysis strategies targeting resolved and merged hadronic top-quark decays are used.
The agreement between the data and the SM prediction is quantified in distributions of the \ttbar invariant mass (1-lepton final states)
and the invariant mass of the two leptons and two $b$-jets (2-lepton final states) in several orthogonal signal regions, which are combined in a final likelihood fit.
 
No significant deviation from the SM prediction is observed, and exclusion regions at 95\% CL are derived for several representative benchmark models
that predict new scalar and pseudo-scalar states decaying into \ttbar,
such as a generic type-II 2HDM, the hMSSM, and the 2HDM+$a$ benchmark for LHC DM searches.
Additionally, the search results are interpreted in a more model-independent scenario in which only the interference pattern of a single scalar or pseudo-scalar
of a given mass and width is considered and upper limits are derived on the coupling of this particle to \ttbar.
In the 2HDM, values of \tanb smaller than 3.49 are excluded for $m_A = m_H = 400$\,GeV,
and mass values up to 1240\,GeV are excluded for the lowest tested \tanb value of 0.4.
In the hMSSM, values of \tanb smaller than 3.16 are excluded for $m_A = 400$\,GeV
and mass values up to 950~GeV are excluded for \tanb=1.0.
The search presented in this paper provides the most stringent constraints on the 2HDM and hMSSM parameter space in the region of high $m_A$ and low \tanb to date,
surpassing previous constraints from \ttbar interference and $\ttbar\ttbar$ searches on Run-2 data.
In the 2HDM+$a$, values of \tanb below 1.1 (0.9) are excluded across the probed range of the mediator mass $m_a$ for a benchmark scenario with low (high) $a-A$ mixing.
The search presented in this paper is the first to consider the more complex interference patterns arising in the presence of two pseudo-scalars and a scalar
and the first \ttbar interference search to set constraints on pseudo-scalar mediators to dark matter.


\section*{Acknowledgements}


We thank CERN for the very successful operation of the LHC and its injectors, as well as the support staff at
CERN and at our institutions worldwide without whom ATLAS could not be operated efficiently.
 
The crucial computing support from all WLCG partners is acknowledged gratefully, in particular from CERN, the ATLAS Tier-1 facilities at TRIUMF/SFU (Canada), NDGF (Denmark, Norway, Sweden), CC-IN2P3 (France), KIT/GridKA (Germany), INFN-CNAF (Italy), NL-T1 (Netherlands), PIC (Spain), RAL (UK) and BNL (USA), the Tier-2 facilities worldwide and large non-WLCG resource providers. Major contributors of computing resources are listed in Ref.~\cite{ATL-SOFT-PUB-2023-001}.
 
We gratefully acknowledge the support of ANPCyT, Argentina; YerPhI, Armenia; ARC, Australia; BMWFW and FWF, Austria; ANAS, Azerbaijan; CNPq and FAPESP, Brazil; NSERC, NRC and CFI, Canada; CERN; ANID, Chile; CAS, MOST and NSFC, China; Minciencias, Colombia; MEYS CR, Czech Republic; DNRF and DNSRC, Denmark; IN2P3-CNRS and CEA-DRF/IRFU, France; SRNSFG, Georgia; BMBF, HGF and MPG, Germany; GSRI, Greece; RGC and Hong Kong SAR, China; ISF and Benoziyo Center, Israel; INFN, Italy; MEXT and JSPS, Japan; CNRST, Morocco; NWO, Netherlands; RCN, Norway; MNiSW, Poland; FCT, Portugal; MNE/IFA, Romania; MESTD, Serbia; MSSR, Slovakia; ARRS and MIZ\v{S}, Slovenia; DSI/NRF, South Africa; MICINN, Spain; SRC and Wallenberg Foundation, Sweden; SERI, SNSF and Cantons of Bern and Geneva, Switzerland; MOST, Taipei; TENMAK, T\"urkiye; STFC, United Kingdom; DOE and NSF, United States of America.
 
Individual groups and members have received support from BCKDF, CANARIE, CRC and DRAC, Canada; PRIMUS 21/SCI/017, CERN-CZ and FORTE, Czech Republic; COST, ERC, ERDF, Horizon 2020, ICSC-NextGenerationEU and Marie Sk{\l}odowska-Curie Actions, European Union; Investissements d'Avenir Labex, Investissements d'Avenir Idex and ANR, France; DFG and AvH Foundation, Germany; Herakleitos, Thales and Aristeia programmes co-financed by EU-ESF and the Greek NSRF, Greece; BSF-NSF and MINERVA, Israel; Norwegian Financial Mechanism 2014-2021, Norway; NCN and NAWA, Poland; La Caixa Banking Foundation, CERCA Programme Generalitat de Catalunya and PROMETEO and GenT Programmes Generalitat Valenciana, Spain; G\"{o}ran Gustafssons Stiftelse, Sweden; The Royal Society and Leverhulme Trust, United Kingdom.
 
In addition, individual members wish to acknowledge support from CERN: European Organization for Nuclear Research (CERN PJAS); Chile: Agencia Nacional de Investigaci\'on y Desarrollo (FONDECYT 1190886, FONDECYT 1210400, FONDECYT 1230812, FONDECYT 1230987); China: Chinese Ministry of Science and Technology (MOST-2023YFA1605700), National Natural Science Foundation of China (NSFC - 12175119, NSFC 12275265, NSFC-12075060); Czech Republic: Czech Science Foundation (GACR - 24-11373S), Ministry of Education Youth and Sports (FORTE CZ.02.01.01/00/22\_008/0004632), PRIMUS Research Programme (PRIMUS/21/SCI/017); EU: H2020 European Research Council (ERC - 101002463); European Union: European Research Council (ERC - 948254, ERC 101089007), Horizon 2020 Framework Programme (MUCCA - CHIST-ERA-19-XAI-00), European Union, Future Artificial Intelligence Research (FAIR-NextGenerationEU PE00000013), Italian Center for High Performance Computing, Big Data and Quantum Computing (ICSC, NextGenerationEU); France: Agence Nationale de la Recherche (ANR-20-CE31-0013, ANR-21-CE31-0013, ANR-21-CE31-0022), Investissements d'Avenir Labex (ANR-11-LABX-0012); Germany: Baden-Württemberg Stiftung (BW Stiftung-Postdoc Eliteprogramme), Deutsche Forschungsgemeinschaft (DFG - 469666862, DFG - CR 312/5-2); Italy: Istituto Nazionale di Fisica Nucleare (ICSC, NextGenerationEU); Japan: Japan Society for the Promotion of Science (JSPS KAKENHI JP21H05085, JSPS KAKENHI JP22H01227, JSPS KAKENHI JP22H04944, JSPS KAKENHI JP22KK0227); Netherlands: Netherlands Organisation for Scientific Research (NWO Veni 2020 - VI.Veni.202.179); Norway: Research Council of Norway (RCN-314472); Poland: Polish National Agency for Academic Exchange (PPN/PPO/2020/1/00002/U/00001), Polish National Science Centre (NCN 2021/42/E/ST2/00350, NCN OPUS nr 2022/47/B/ST2/03059, NCN UMO-2019/34/E/ST2/00393, UMO-2020/37/B/ST2/01043, UMO-2021/40/C/ST2/00187, UMO-2022/47/O/ST2/00148, UMO-2023/49/B/ST2/04085); Slovenia: Slovenian Research Agency (ARIS grant J1-3010); Spain: Generalitat Valenciana (Artemisa, FEDER, IDIFEDER/2018/048), Ministry of Science and Innovation (MCIN \& NextGenEU PCI2022-135018-2, MICIN \& FEDER PID2021-125273NB, RYC2019-028510-I, RYC2020-030254-I, RYC2021-031273-I, RYC2022-038164-I), PROMETEO and GenT Programmes Generalitat Valenciana (CIDEGENT/2019/023, CIDEGENT/2019/027); Sweden: Swedish Research Council (Swedish Research Council 2023-04654, VR 2018-00482, VR 2022-03845, VR 2022-04683, VR 2023-03403, VR grant 2021-03651), Knut and Alice Wallenberg Foundation (KAW 2018.0157, KAW 2018.0458, KAW 2019.0447, KAW 2022.0358); Switzerland: Swiss National Science Foundation (SNSF - PCEFP2\_194658); United Kingdom: Leverhulme Trust (Leverhulme Trust RPG-2020-004), Royal Society (NIF-R1-231091); United States of America: U.S. Department of Energy (ECA DE-AC02-76SF00515), Neubauer Family Foundation.


\printbibliography
 
\clearpage
 
\begin{flushleft}
\hypersetup{urlcolor=black}
{\Large The ATLAS Collaboration}

\bigskip

\AtlasOrcid[0000-0002-6665-4934]{G.~Aad}$^\textrm{\scriptsize 104}$,
\AtlasOrcid[0000-0001-7616-1554]{E.~Aakvaag}$^\textrm{\scriptsize 17}$,
\AtlasOrcid[0000-0002-5888-2734]{B.~Abbott}$^\textrm{\scriptsize 123}$,
\AtlasOrcid[0000-0002-0287-5869]{S.~Abdelhameed}$^\textrm{\scriptsize 119a}$,
\AtlasOrcid[0000-0002-1002-1652]{K.~Abeling}$^\textrm{\scriptsize 56}$,
\AtlasOrcid[0000-0001-5763-2760]{N.J.~Abicht}$^\textrm{\scriptsize 50}$,
\AtlasOrcid[0000-0002-8496-9294]{S.H.~Abidi}$^\textrm{\scriptsize 30}$,
\AtlasOrcid[0009-0003-6578-220X]{M.~Aboelela}$^\textrm{\scriptsize 45}$,
\AtlasOrcid[0000-0002-9987-2292]{A.~Aboulhorma}$^\textrm{\scriptsize 36e}$,
\AtlasOrcid[0000-0001-5329-6640]{H.~Abramowicz}$^\textrm{\scriptsize 154}$
% ...
\bigskip
\\

% ...
$^{17}$Department for Physics and Technology, University of Bergen, Bergen; Norway.\\
% ...
$^{30}$Physics Department, Brookhaven National Laboratory, Upton NY; United States of America.\\
% ...
$^{36}$$^{(a)}$Facult\'e des Sciences Ain Chock, R\'eseau Universitaire de Physique des Hautes Energies - Universit\'e Hassan II, Casablanca;$^{(b)}$Facult\'{e} des Sciences, Universit\'{e} Ibn-Tofail, K\'{e}nitra;$^{(c)}$Facult\'e des Sciences Semlalia, Universit\'e Cadi Ayyad, LPHEA-Marrakech;$^{(d)}$LPMR, Facult\'e des Sciences, Universit\'e Mohamed Premier, Oujda;$^{(e)}$Facult\'e des sciences, Universit\'e Mohammed V, Rabat;$^{(f)}$Institute of Applied Physics, Mohammed VI Polytechnic University, Ben Guerir; Morocco.\\
% ...
$^{45}$Physics Department, Southern Methodist University, Dallas TX; United States of America.\\
% ...
$^{50}$Fakult\"{a}t Physik , Technische Universit{\"a}t Dortmund, Dortmund; Germany.\\
% ...
$^{56}$II. Physikalisches Institut, Georg-August-Universit\"{a}t G\"ottingen, G\"ottingen; Germany.\\
% ...
$^{104}$CPPM, Aix-Marseille Universit\'e, CNRS/IN2P3, Marseille; France.\\
% ...
$^{119}$$^{(a)}$New York University Abu Dhabi, Abu Dhabi;$^{(b)}$United Arab Emirates University, Al Ain; United Arab Emirates.\\
% ...
$^{123}$Homer L. Dodge Department of Physics and Astronomy, University of Oklahoma, Norman OK; United States of America.\\
% ...
$^{154}$Raymond and Beverly Sackler School of Physics and Astronomy, Tel Aviv University, Tel Aviv; Israel.\\
% ...
$^{a}$ Also Affiliated with an institute covered by a cooperation agreement with CERN.\\
$^{b}$ Also at An-Najah National University, Nablus; Palestine.\\
$^{c}$ Also at Borough of Manhattan Community College, City University of New York, New York NY; United States of America.\\
$^{d}$ Also at Center for Interdisciplinary Research and Innovation (CIRI-AUTH), Thessaloniki; Greece.\\
$^{e}$ Also at Centro Studi e Ricerche Enrico Fermi; Italy.\\
$^{f}$ Also at CERN, Geneva; Switzerland.\\
% ...

\end{flushleft}


\end{document}